\documentclass[letterpaper, 10 pt, conference]{ieeeconf}  % Comment this line out if you need a4paper

\IEEEoverridecommandlockouts                              % This command is only needed if 
\title{\LARGE \bf
Robust offset-free constrained Model Predictive Control with Long Short-Term Memory Networks - Extended version
}

\author{Irene Schimperna, and Lalo Magni% <-this % stops a space
\thanks{I. Schimperna and L. Magni are with Department of Civil and Architecture Engineering, University of Pavia, via Ferrata 3, Pavia, 27100, Italy (e-mails: irene.schimperna01@universitadipavia.it, lalo.magni@unipv.it)}%
}

\usepackage{graphicx} % Required for inserting images

\usepackage{amsmath}
\usepackage{amssymb}
\usepackage{tikz}
\usetikzlibrary{shapes,arrows}
\usepackage{comment}
\newcommand{\norm}[1]{\left\lVert#1\right\rVert}
\newcommand{\abs}[1]{\left\lvert#1\right\rvert}

\newtheorem{assumption}{Assumption}
\newtheorem{theorem}{Theorem}

\newtheorem{lemma}{Lemma}
\newtheorem{corollary}{Corollary}

\newtheorem{remark}{Remark}

\usepackage[sorting=none]{biblatex}
\begin{document}

\maketitle
\thispagestyle{empty}
\pagestyle{empty}

%%%%%%%%%%%%%%%%%%%%%%%%%%%%%%%%%%%%%%%%%%%%%%%%%%%%%%%%%%%%%%%%%%%%%%%%%%%%%%%%
\begin{abstract}
This paper develops a control scheme, based on the use of Long Short-Term Memory neural network models and Nonlinear Model Predictive Control, which guarantees recursive feasibility with slow time variant set-points and disturbances, input and output constraints and unmeasurable state. Moreover, if the set-point and the disturbance are asymptotically constant, offset-free tracking is guaranteed.
Offset-free tracking is obtained by augmenting the model with a disturbance, to be estimated together with the states of the Long Short-Term Memory network model by a properly designed observer.
Satisfaction of the output constraints in presence of observer estimation error, time variant set-points and disturbances is obtained using a constraint tightening approach.


\end{abstract}

\section{Introduction} \label{sec:introduction}
Model Predictive Control (MPC) \cite{rawlings2019mpc_book} is an optimization-based control method, that consists in solving at each sample time instant a Finite Horizon Optimal Control Problem (FHOCP) and then applying only the first element of the optimal control sequence. 
In the last 30 years many theoretical results have been proposed in order to derive MPC control algorithm with guaranteed properties both for linear and nonlinear systems. Stabilization of a given steady state equilibrium is a largely solved problem, with approaches based on terminal ingredients (terminal set/cost) \cite{Chen1998NMPCStability}, \cite{Magni2001Automatica} or sufficiently long prediction horizon \cite{DenicolaoIEEETAC98NMPCStability}, \cite{BocciaSCL2014}, \cite{RebleAutomatica2012}, \cite{KolerIEEETAC2023} in the nominal case, i.e. when the model equation and state are exactly known. Starting from this simple case, several results were developed in order to consider more realistic situations that involves disturbances, model uncertainties, time variant references a-priori unknown and unmeasurable states. The development of MPC control algorithms able to tackle these aspects is made particularly difficult when state/output constraints, that are one of the most important characteristic of MPC, are considered. In fact, recursive feasibility is heavily affected by all these aspects. In particular, to take into account model uncertainties and disturbances several algorithms were proposed both in a bounded deterministic context \cite{mayne2000constrained_mpc} and in a stochastic one \cite{hewing2019cautious_mpc}, \cite{Farina2015StocasticOutputFeedback}, \cite{Aken2018stochasticMPC}. For linear systems robustness is typically achieved using tube based robust MPC schemes \cite{Langson2004TubeMPC}. For nonlinear systems it is possible to rely on constraint tightening approaches, for example based on the Lipschitz constant of the system under control \cite{limon2002iss_mpc}, \cite{Pin2009IEEETACRobustMPC}, or on min-max approaches \cite{raimondo2009minmax}. 
Concerning the tracking problem most of the attention was devoted to compensate exogenous signals (reference and/or disturbance signals) generated by a neutrally stable ecosystem \cite{Magni2001Automatica_outputfeedback}, \cite{kohler2022constrained_output_regulation} and in particular constant signals \cite{MagniIJSS2005}, \cite{BettiTAC2013}, and to guarantee recursive feasibility with set-point changes by simultaneously optimizing an artificial reference \cite{LimonTAC2018}, \cite{KolerAutomantica2020}, \cite{GaluppiniNOLCOS23}. 
In \cite{kohler2020reference_generic} an offline procedure to compute a parameterized terminal cost has been developed for both set-point and trajectory stabilization.
Moreover, to guarantee offset-free tracking at steady state despite the presence of persistent disturbances and model-plant mismatch, a well-known method consists in augmenting the model with a properly designed constant disturbance, to be estimated by the observer together with the states of the model. Sufficient conditions that the disturbance model, the observer and the MPC have to satisfy in order to guarantee zero error at steady state are analysed for linear models in \cite{pannocchia2003offset_free_mpc}, \cite{maeder2009linear_offset_free} and for nonlinear models in \cite{morari2012nonlinear_offset_free}. 
The case of unmeasurable states has been typically solved by the introduction of an observer. The presence of constraints however calls for the design of a robust state-feedback MPC and for a properly designed observer. In most of the literature about output MPC the observer estimation error is regarded as a generic disturbance \cite{roset2008robustness}.
Few exceptions are \cite{kohler2019simple_robust_mpc}, that proposes a constraint tightening technique based on the evolution of the maximum observer estimation error, and \cite{copp2017simultaneous}, that uses a min-max optimization problem to simultaneously compute a Moving Horizon Estimation and a robust MPC. 
The properties of any MPC algorithm are also affected by the quality of the model that can range from linear to nonlinear, physical based to black-box identification, fine dimension to infinite dimension.  

In this paper, a Neural Network (NN) model has been considered for its flexibility and modularity and possibility to be learnt directly from the data without the need of analysing the physics of the system.
The idea of developing models inspired by human brain neurons dates back to 1957 with the Rosenblatt perceptron \cite{rosenblatt1958perceptron}, and also some interesting theoretical results concerning the modelling capabilities of NN, like the universal approximation theorem \cite{cybenko1989approximation}, have been known for years. Despite this, the use of NN had a great increase in recent years, thanks to the availability of huge amount of data and of cheap hardware specialized for parallel computations (GPUs), that allows to perform computationally intensive training procedures that were infeasible until some years ago. NN are now known to be effective in a large variety of contexts and applications \cite{abiodun2018ann_applications_survey},
like image \cite{krizhevsky2012imagenet}, speech \cite{graves2013speech} and handwriting recognition \cite{graves2008handwriting}, prediction \cite{zhang2003time_series} and forecasting \cite{zhang1998forecasting}, \cite{hippert2001forecasting}.

For MPC, a class of NN that can be used to provide input-output models of the system under control are Recurrent Neural Networks (RNN) \cite{bonassi2022rnn}. In fact RNN have the same mathematical structure of a discrete-time dynamical system, and for this reason they can be very effective in describing the dynamic nature of the system under control.
In the family of RNN, two architectures that have shown remarkable performances in several tasks are Long Short-Term Memory (LSTM) networks \cite{hochreiter1997lstm} and Gated Recurrent Units (GRU) \cite{cho2014gru}. Recently, in the control field, a new architecture with guaranteed stability properties called Recurrent Equilibrium Networks (REN) has been proposed \cite{Revay2023recurrent-equilibrium-networks}.
In \cite{terzi2021mpc_lstm}, LSTM neural networks have been analysed from a stability point of view and applied as models for MPC. 
In particular, conditions on LSTM model’s parameters guaranteeing the Input-to-State Stability (ISS) \cite{jiang2001iss} and the Incremental Input-to-State Stability ($\delta$ISS) \cite{tran2016delta_iss} are designed. 
Then, assuming that the trained model exactly represents the dynamic of the system and relying on $\delta$ISS, an observer guaranteeing that the estimated state asymptotically converges to the true value is proposed. Based on the LSTM model and on the state observer, a stabilizing MPC control algorithm solving the tracking problem for constant state and input reference values and in presence of input constraints is proposed and analysed.

In this paper, the algorithm proposed in \cite{terzi2021mpc_lstm} has been modified in order to cope with time variant set-points and disturbances, output constraints and to guarantee zero-error regulation for asymptotically constant set-points and disturbances. In this respect, the $\delta$ISS LSTM model is employed in the design of an MPC scheme that guarantees recursive feasibility with a-priori unknown slow time variant bounded reference signals,  unknown disturbances and input and output constraints by means of a properly designed robust MPC inspired by \cite{kohler2019simple_robust_mpc} and a state observer properly designed for the LSTM model completed with a disturbance model that satisfies the sufficient conditions for offset-free stated in \cite{morari2012nonlinear_offset_free}. 
Further to the state-feedback MPC and the observer, the control scheme includes 
a Reference calculator, that computes state and input set-points for the MPC at every sample time instant on the base of the current value of the output reference signal and of the disturbance estimation.
A preliminary version of this zero-error control scheme for unconstrained systems was presented in \cite{schimperna2023IFACNOLCOS}. Remarkably, the results derived in this paper are useful not only in combination with LSTM models but for any $\delta$ISS model like \cite{Revay2023recurrent-equilibrium-networks} and \cite{Bonassi2021StabilityGRU}. 

The paper is organized as follows. After the notation and some preliminary definitions, in Section \ref{sec:problem_control} the control problem is described, and all the components of the proposed control algorithm are introduced and analysed. In Section \ref{sec:stability_offset_free} the main results of the paper are reported, concerning the 
feasibility, stability and offset-free analysis of the control scheme. In Section \ref{sec:numerical_example} the proposed control algorithm is tested on a benchmark pH neutralization process \cite{henson1994ph}. Finally, Section \ref{sec:conclusion} draws the conclusions of the work. All the proofs are reported in Appendix \ref{sec:appendix_proof}.

\subsection{Notation}
Considering a vector $v$, $v_{(j)}$ is its $j$-th component, $v^\top$ is its transpose, $\|v\|$ is its 2-norm, $\|v\|_\infty$ is its infinity-norm and $\|v\|^2_A = v^\top A v$ is the squared norm weighted with matrix $A$. %and $\|v\|_A = \sqrt{v^\top A v}$. 
$|v|$ is the vector containing the absolute values of the elements of $v$. Inequalities between vectors are considered element by element.
Given two vectors $v$ and $w$, $v \circ w$ is their element-wise product. 
Considering a matrix $M$, $M_{(ij)}$ is its element in position $ij$, 
$M_{(j*)}$ is its $j$-th row, 
$\rho(M)$ is its spectral radius, i.e. the maximum absolute value of its eigenvalues, and $\|M\|$ and $\|M\|_{\infty}$ are its induced 2-norm and $\infty$-norm. $\lambda_{max}(M)$ and $\lambda_{min}(M)$ are respectively the maximum and minimum eigenvalues of the symmetric matrix $M$.
Given a positive definite matrix $A$, $A^{1/2}$ is the unique positive definite matrix $B$ such that $B B = A$.
$\mathbf{0}_{m,n}$ is the $m \times n$ null matrix, $I_n$ is the $n \times n$ identity matrix and $\mathbf{1}_{n}$ is the vector of ones of $n$ elements.
$sat(v,v_{max})$ denotes the saturation operator between $-v_{max}$ and $+v_{max}$, that is applied element by element when $v$ is a vector.

\subsection{Definitions}
\textit{Definition ($\mathcal{K}$-function):} A continuous function $\gamma: \mathbb{R}_{\geq 0} \to \mathbb{R}_{\geq 0}$ is a class $\mathcal{K}$ function if $\gamma(s) > 0$ for all $s>0$, it is strictly increasing and $\gamma(0) = 0$.

\textit{Definition ($\mathcal{K}_{\infty}$-function):} A continuous function $\gamma: \mathbb{R}_{\geq 0} \to \mathbb{R}_{\geq 0}$ is a class $\mathcal{K}_{\infty}$-function if it is a class $\mathcal{K}$ function and $\gamma(s) \to \infty$ for $s \to \infty$.

\textit{Definition ($\mathcal{KL}$-function):} A continuous function $\beta: \mathbb{R}_{\geq 0} \times \mathbb{Z}_{\geq 0} \to \mathbb{R}_{\geq 0}$ is a class $\mathcal{KL}$-function if $\beta(s,k)$ is a class $\mathcal{K}$ function with respect to $s$ for all $k$, it is strictly decreasing in $k$ for all $s > 0$, and $\beta (s,k) \to 0$ as $k \to \infty$ for all $s>0$.
\hfill $\square$

The notions of Input-to-State practical Stability (ISpS) and $\delta$ISS are now introduced for the generic discrete-time dynamical system $x_{k+1} = f(x_k, u_k)$, with $x \in \mathbb{R}^{n_x}$ and $u \in \mathbb{R}^{n_u}$.
The stability notion are stated in the sets $\mathcal{X} \subseteq \mathbb{R}^{n_x}$ and $\mathcal{U} \subseteq \mathbb{R}^{n_u}$, with the set $\mathcal{X}$ assumed to be positive invariant for the dynamical system, i.e. for any $u \in \mathcal{U}$, it holds that $x \in \mathcal{X} \implies f(x,u) \in \mathcal{X}$.

\textit{Definition (ISpS, \cite{raimondo2009minmax}):}  The dynamical system $x_{k+1} = f(x_k, u_k)$ is ISpS in the sets $\mathcal{X}$ and $\mathcal{U}$ if there exist functions $\beta \in \mathcal{KL}$ and $\gamma \in \mathcal{K}_{\infty}$ and a constant $c > 0$ such that for any $k \in \mathbb{Z}_{\geq 0}$, any initial condition $x_{0} \in \mathcal{X}$, any input sequence $u_{0}, u_{1}, ..., u_{k-1}$ with $u_{h} \in \mathcal{U}$ for all $h = 1,..,k-1$, it holds that:
\begin{equation}
    \norm{x_k} \leq \beta(\|x_0\|, k) + \gamma  \left( \max_{0 \leq h < k} \| u_{h} \| \right) + c
\end{equation}

\textit{Definition ($\delta$ISS):} The dynamical system $x_{k+1} = f(x_k, u_k)$ is $\delta$ISS in the sets $\mathcal{X}$ and $\mathcal{U}$ if there exist functions $\beta \in \mathcal{KL}$ and $\gamma \in \mathcal{K}_{\infty}$ such that for any $k \in \mathbb{Z}_{\geq 0}$, any pair of initial conditions $x_{\mathrm{a},0} \in \mathcal{X}$ and $x_{\mathrm{b},0} \in \mathcal{X}$, any pair of input sequences $u_{\mathrm{a},0}, u_{\mathrm{a},1}, ..., u_{\mathrm{a},k-1}$ and $u_{\mathrm{b},0}, u_{\mathrm{b},1}, ..., u_{\mathrm{b},k-1}$, with $u_{\mathrm{a},h}, u_{\mathrm{b},h} \in \mathcal{U}$ for all $h = 1,..,k-1$, it holds that:
\begin{equation}    \label{eq:def_delta_iss}
\begin{split}
    \| x_{\mathrm{a},k} - x_{\mathrm{b},k} \| &\leq \beta(\|x_{\mathrm{a},0} - x_{\mathrm{b},0}\|, k) \\
    & + \gamma \left( \max_{0 \leq h < k} \| u_{\mathrm{a},h} - u_{\mathrm{b},h} \| \right)
\end{split}
\end{equation}
where $x_{\mathrm{a},k}$ and $x_{\mathrm{b},k}$ satisfy the dynamics equation $x_{k+1} = f(x_k, u_k)$ respectively with the inputs $u_{\mathrm{a}}$ and $u_{\mathrm{b}}$ and initial states $x_{\mathrm{a},0}$ and $x_{\mathrm{b},0}$.

\textit{Definition (exponential $\delta$ISS):} The dynamical system $x_{k+1} = f(x_k, u_k)$ is exponentially $\delta$ISS in the sets $\mathcal{X}$ and $\mathcal{U}$ if it it $\delta$ISS in the sets $\mathcal{X}$ and $\mathcal{U}$ and the function $\beta$ is exponential with respect to the second argument, i.e. there exist $\mu > 0$ and $\lambda \in (0,1)$ such that 
\begin{equation*}
    \beta(s,k) \leq \mu s \lambda^k
\end{equation*}

\section{Problem formulation and control algorithm} \label{sec:problem_control}
In this work a nonlinear plant with unknown dynamics, input $u_\phi \in \mathbb{R}^m$, output $y_{\phi} \in \mathbb{R}^p$ and a possible unknown bounded asymptotically constant disturbance $d_{\phi}$ is considered as system under control.
The plant input is saturated:
\begin{equation}    \label{eq:input_saturation_phi}
    u_\phi \in \mathcal{U}_\phi = \{ u_\phi \in \mathbb{R}^m : u_{\phi,min} \leq u_\phi \leq u_{\phi,max} \} %\subseteq \bar{\mathcal{U}}
\end{equation}
with $u_{\phi,min}, u_{\phi,max} \in \mathbb{R}^m$, and the output has to be limited:
\begin{equation}    \label{eq:output_constraint_phi}
    y_\phi \in \mathcal{Y}_\phi = \{ y_\phi \in \mathbb{R}^p : y_{\phi,min} \leq y_\phi \leq y_{\phi,max} \}
\end{equation}
with $y_{\phi,min}, y_{\phi,max} \in \mathbb{R}^p$.

\begin{assumption}  \label{ass:input_equal_output}
    The plant has the same number of inputs and outputs, i.e. $m = p$.
    %\hfill $\square$
\end{assumption}

\begin{remark}
    Assumption \ref{ass:input_equal_output} is typical of nonlinear offset-free MPC schemes \cite{morari2012nonlinear_offset_free}, and is required to have a unique input combination associated to each output at equilibrium. However, if $m > p$ the proposed algorithm can still be applied by selecting a subset of $p$ inputs to be used for control and leaving the other $m-p$ inputs constant.
    \hfill $\square$
\end{remark}

\tikzstyle{block} = [draw, rectangle, 
    minimum height=2em, minimum width=3em]
\tikzstyle{normalization} = [draw, rectangle, 
    minimum height=1em, minimum width=1em]
\tikzstyle{input} = [coordinate]
\tikzstyle{output} = [coordinate]
\tikzstyle{pinstyle} = [pin edge={to-,thin,black}]
\tikzstyle{sum} = [draw, circle, node distance=0.6cm]

\begin{figure}
\centering
\begin{tikzpicture}[auto, node distance=1.5cm,>=latex']
    % We start by placing the blocks
    \node [input, name=input] {};
    \node [normalization, right of=input, node distance=0.6cm] (n_setpoint) {N};
    \node [block, right of=n_setpoint, align=center, node distance=1.8cm] (ref_calc) {\small Reference\\ \small calculator};
    \node [block, right of=ref_calc, node distance=2.1cm, align=center] (mpc) {\small LSTM\\ \small based\\ \small MPC};
    \node [normalization, right of=mpc, node distance=1.3cm] (n_input) {N$^{-1}$};
    \node [block, right of=n_input, pin={[pinstyle]above:$d_\phi$},
            node distance=1.5cm] (plant) {\small plant};
    \draw [draw, ->] (mpc) -- node[name=u] {$u$} (n_input);
    \draw [->] (n_input) -- node[name=u_phi] {$u_\phi$} (plant);
    \node [output, right of=plant, node distance=1.1cm] (output) {};
    \node [block, below of=plant, node distance=2.5cm, align=center] (observer) {\small LSTM\\ \small based\\ \small observer};
    \draw [->] (plant) -- node [name=y_phi] {$y_\phi$}(output);
    \node [normalization, below of=y_phi, node distance=1.2cm] (n_output) {N};
    \node [normalization, below of=u_phi, node distance=1.2cm] (n_input2) {N};

    \draw [draw,->] (input) -- node [pos=0.25] {$y^0_\phi$} (n_setpoint);
    \draw [->] (n_setpoint) -- node [name=y0, pos=0.3] {$y^0$}  (ref_calc);
    \node [coordinate, right of=n_setpoint, node distance=0.7cm] (y0_avanti) {};
    \node [coordinate, above of=y0_avanti, node distance = 1cm] (angolo) {};
    \draw [-] (y0_avanti) -- (angolo);
    \draw [->] (angolo) -| node {$y^0$} (mpc);
    \draw [->] (u_phi) -- (n_input2);
    \draw [->] (n_input2) -| node [pos=0.8] {$u$} (observer.120);
    \draw [->] (y_phi) -- (n_output);
    \draw [->] (n_output) -| node [pos=0.8] {$y$} (observer.60);
    \draw [->] (observer.160) -| node [pos=0.75] {$\hat{x}$} (mpc);
    \draw [->] (observer.200) -| node [pos=0.75] {$\hat{d}$} (ref_calc);
    \draw [->] (ref_calc) -- node {$\bar{x}, \bar{u}$} (mpc);
\end{tikzpicture}
\caption{Block diagram of the control scheme.} 
\label{fig:block_diagram}
\end{figure}
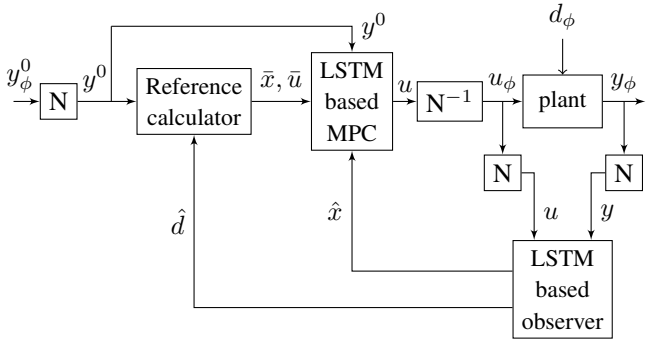

The objective of the control is to achieve null error at steady state for an asymptotically constant reference $y^0_\phi$
%such that $y^0_k \to y^0_\infty$ for $k \to \infty$,
also in presence of model-plant mismatch and of bounded asymptotically constant plant disturbances $d_\phi$, and to respect input and output constraints.
The proposed control scheme, whose block diagram is reported in Fig. \ref{fig:block_diagram}, is based on an LSTM nominal model of the plant.
A key element to obtain a well tuned LSTM model is the normalization of input and output signals that is represented in the scheme with the blocks ``N" and that consists in an affine transformation that scales the variables in a predefined range (typically [-1,1]). The normalized input is denoted with $u$ and the normalized output with $y$. For the normalized variables constraints \eqref{eq:input_saturation_phi}-\eqref{eq:output_constraint_phi} become
\begin{equation}    \label{eq:input_saturation}
    u \in \mathcal{U} = \{ u \in \mathbb{R}^m : \norm{u}_\infty \leq u_{max} \}
\end{equation}
%with $u_{lb}, u_{ub} \in \mathbb{R}^m$, 
where $u_{max} \in \mathbb{R}$ and is typically equal to 1,
and
\begin{equation}    \label{eq:output_constraint}
    y \in \mathcal{Y} = \{ y \in \mathbb{R}^p : y_{min} \leq y \leq y_{max} \}
\end{equation}
where $y_{min}, y_{max} \in \mathbb{R}^p$.

To achieve null error at steady state the LSTM model is augmented with an asymptotically constant disturbance term $d$.
Then an observer based on the equations of the LSTM model provides an estimation $\hat{x}$ of the state $x$ of the LSTM and an estimation $\hat{d}$ of the disturbance $d$.
At each sampling instant the current values of the normalized set-point $y^0$ and of the disturbance estimation $\hat{d}$ are used by a Reference calculator to compute the current values of input and state set-points $\Bar{u}$ and $\Bar{x}$ for the MPC. 
The last block is the MPC, that solves a %Finite Horizon Optimal Control Problem 
FHOCP and gives in output the first element of the achieved optimal control sequence. Since in the control problem formulation output constraints are considered, to ensure recursive feasibility in presence of the observer estimation error and time variant set-points, it is necessary to rely on a robust MPC.

\begin{figure}
\centering
\begin{tikzpicture}[auto, node distance=1.8cm,>=latex']
    % We start by placing the blocks
    \node [input, name=input] {};
    \node [block, right of=input, align=center] (ref_calc) {\small Reference\\ \small calculator};
    \node [block, right of=ref_calc, node distance=2.5cm, align=center] (mpc) {\small LSTM\\ \small based\\ \small MPC};
    %\node [normalization, right of=mpc, node distance=1.8cm] (n_input) {N};
    \node [block, right of=mpc,
           node distance=1.8cm, align=center] (plant) {\small LSTM\\ \small model};
    \node [sum, right of=plant, node distance=1.4cm] (summation) {};
    \node [block, above of=summation, node distance=1.3cm, pin={[pinstyle, pin distance=0.3cm]above:$w$}, minimum height=0.8cm, minimum width=1cm] (integrator) {\small $\sum$};
    \draw [draw, ->] (mpc) -- node[name=u] {$u$} (plant);
    %\draw [->] (n_input) -- node[name=u_phi] {$u_\phi$} (plant);
    \node [output, right of=summation, node distance=1cm] (output) {};
    \node [block, below of=plant, node distance=2.1cm, align=center] (observer) {\small LSTM\\ \small based\\ \small observer};
    \draw [->] (summation) -- node [name=y_phi, pos=0.75] {$y$}(output);
    \node [coordinate, below of=y_phi, node distance=1cm] (n_output) {};
    \node [coordinate, below of=u, node distance=1cm] (n_input2) {N};

    \draw [->] (input) -- node [name=y0, pos=0.3] {$y^0$}  (ref_calc);
    \node [coordinate, right of=input, node distance=0.6cm] (y0_avanti) {};
    \node [coordinate, above of=y0_avanti, node distance = 1cm] (angolo) {};
    \draw [-] (y0_avanti) -- (angolo);
    \draw [->] (angolo) -| node {$y^0$} (mpc);
    \draw [-] (u) -- (n_input2);
    \draw [->] (n_input2) -| (observer.120);
    \draw [-] (y_phi) -- (n_output);
    \draw [->] (n_output) -|  (observer.60);
    \draw [->] (observer.160) -| node [pos=0.75] {$\hat{x}$} (mpc);
    \draw [->] (observer.200) -| node [pos=0.75] {$\hat{d}$} (ref_calc);
    \draw [->] (ref_calc) -- node {$\bar{x}, \bar{u}$} (mpc);
    \draw [->] (plant) -- node [pos=0.3] {$\xi$} node [pos=0.8] {$+$} (summation);
    \draw [->] (integrator) -- node [pos=0.3] {$d$} node [pos=0.85] {$+$} (summation);
\end{tikzpicture}
\caption{Block diagram of the nominal closed-loop control scheme, where the real plant is replaced by its LSTM model augmented with a disturbance term.} 
\label{fig:nominal_block_diagram}
\end{figure}
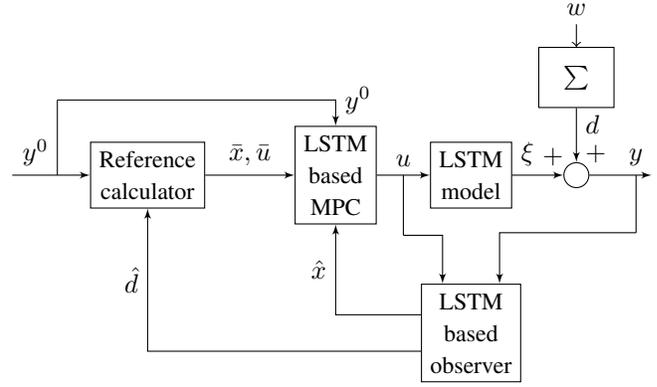

As usually done for the analysis of offset-free control schemes, closed-loop stability and constraint satisfaction will be proven in Section \ref{subsec:feasibility-stability} under the assumption that the plant behaves according to its LSTM model with an asymptotically constant additive disturbance $d$, as reported in Fig. \ref{fig:nominal_block_diagram}, while offset-free will be shown in Section \ref{subsec:offset-free} for the real closed-loop system (Fig. \ref{fig:block_diagram}) under the assumption that closed-loop convergence is not lost.

In the following subsections the different components of the control scheme are presented in detail.

\subsection{Long Short-Term Memory neural network model}
Long-Short Term Memory (LSTM) neural networks are a particular kind of RNN, capable of learning long-term dependencies between data.
%In particular its structure is built to address the vanishing gradient problem, creating paths where the gradient can flow for long duration. 
The LSTM module is composed by two states: the cell state $c \in \mathbb{R}^n$ and the hidden state $h \in \mathbb{R}^n$, where $n$ is also called number of neurons.
%The cell state encodes the memory of past computations, while the hidden state is used to compute the output of the network.
At each time-step $k$, the LSTM network receives an input $u \in \mathbb{R}^m$ and produces an output prediction $\xi \in \mathbb{R}^p$.
Cell state and hidden state are modified through structures called gates.
The equations that describe the LSTM network are the following
\begin{subequations} \label{eq:lstm}
    \begin{align}
    \begin{split}
        c_{k+1} &= \sigma(W_{\mathrm{f}} u_k + U_{\mathrm{f}} h_k + b_{\mathrm{f}}) \circ c_k \\
        & + \sigma(W_{\mathrm{i}} u_k + U_{\mathrm{i}} h_k + b_{\mathrm{i}}) \circ \tanh(W_{\mathrm{c}} u_k + U_{\mathrm{c}} h_k + b_{\mathrm{c}})
    \end{split}     \label{eq:c_lstm}\\
    h_{k+1} &= \sigma(W_{\mathrm{o}} u_k + U_{\mathrm{o}} h_k + b_{\mathrm{o}}) \circ \tanh(c_{k+1}) \label{eq:h_lstm}\\
    \xi_k &= W_{\mathrm{y}} h_k + b_{\mathrm{y}}    \label{eq:y_lstm}
    \end{align}
\end{subequations}
where $\sigma(z) = \frac{1}{1+e^{-z}}$ is the sigmoid activation function. $\tanh(\cdot)$ and $\sigma(\cdot)$ activation functions are applied to vectors element by element. Both $\tanh(\cdot)$ and $\sigma(\cdot)$ are Lipschitz continuous functions, with Lipschitz constants respectively $1$ and $\frac{1}{4}$. Matrices $W_{\mathrm{f}}, W_{\mathrm{i}}, W_{\mathrm{c}}, W_{\mathrm{o}} \in \mathbb{R}^{n \times m}$, $U_{\mathrm{f}}, U_{\mathrm{i}}, U_{\mathrm{c}}, U_{\mathrm{o}} \in \mathbb{R}^{n \times n}$, $W_{\mathrm{y}} \in \mathbb{R}^{p \times n}$ and vectors $b_{\mathrm{f}}, b_{\mathrm{i}}, b_{\mathrm{c}}, b_{\mathrm{o}} \in \mathbb{R}^{n}$, $b_{\mathrm{y}} \in \mathbb{R}^p$ contain the trainable weights of the network.

The LSTM network \eqref{eq:lstm} is a discrete-time time invariant dynamical system, that can be written in a more compact way as
\begin{subequations} \label{eq:model}
    \begin{align}
        x_{k+1} &= f(x_k, u_k)  \label{eq:state_model}\\
        \xi_k &= g(x_k)   \label{eq:output_model}
    \end{align}
\end{subequations}
where $x = [c^\top \; h^\top]^\top \in \mathbb{R}^{2n}$.

\subsubsection{LSTM stability properties}
In the following the $\delta$ISS stability condition and theorem for the LSTM network proposed in \cite{terzi2021mpc_lstm} are recalled.

\begin{assumption}  \label{ass:delta_iss_condition}
    The weights of the LSTM network \eqref{eq:lstm} respect the condition $\rho(A_\delta) < 1$, where
    \begin{equation} \label{eq:delta_iss_matrix}
    A_{\delta} =
        \begin{bmatrix}
            \bar{\sigma}^{\mathrm{f}} & \alpha \\
            \bar{\sigma}^{\mathrm{o}} \bar{\sigma}^{\mathrm{f}} & \alpha \bar{\sigma}^{\mathrm{o}} + \frac{1}{4} \bar{\sigma}^{\mathrm{x}} \|U_{\mathrm{o}}\| \\
        \end{bmatrix}
    \end{equation}
    and
    \begin{equation*} %\label{eq:sigma_f_def}
        %| \sigma(W_f u + U_f h_x + b_f)_{(j)} | \leq \\
        \bar{\sigma}^{\mathrm{f}} = \sigma ( \| [W_{\mathrm{f}} u_{max} \; U_{\mathrm{f}} \; b_{\mathrm{f}}]  \|_{\infty} )
    \end{equation*}
    \begin{equation*} %\label{eq:sigma_i_def}
        %| \sigma(W_i u + U_i h_x + b_i)_{(j)} | \leq \\
        \bar{\sigma}^{\mathrm{i}} = \sigma ( \| [W_{\mathrm{i}} u_{max} \; U_{\mathrm{i}} \; b_{\mathrm{i}}]  \|_{\infty} )
    \end{equation*}
    \begin{equation*} %\label{eq:sigma_o_def}
        %| \sigma(W_o u + U_o h_x + b_o)_{(j)} | \leq \\
        \bar{\sigma}^{\mathrm{o}} = \sigma ( \| [W_{\mathrm{o}} u_{max} \; U_{\mathrm{o}} \; b_{\mathrm{o}}]  \|_{\infty} )
    \end{equation*}
    \begin{equation*} %\label{eq:sigma_c_def}
        %| \tanh(W_c u + U_c h_x + b_c)_{(j)} | \leq \\
        \bar{\sigma}^{\mathrm{c}} = \tanh( \| [W_{\mathrm{c}} u_{max} \; U_{\mathrm{c}} \; b_{\mathrm{c}}]  \|_{\infty} )
    \end{equation*}
    \begin{equation*} %\label{eq:sigma_x_def}
        \bar{\sigma}^{\mathrm{x}} = \tanh\left( \frac{\bar{\sigma}^{\mathrm{i}} \bar{\sigma}^{\mathrm{c}}}{1 - \bar{\sigma}^{\mathrm{f}}} \right)
    \end{equation*}
    \begin{equation*}
        \alpha = \frac{1}{4} \|U_{\mathrm{f}}\| \frac{\bar{\sigma}^{\mathrm{i}} \bar{\sigma}^{\mathrm{c}}}{1 - \bar{\sigma}^{\mathrm{f}}} + \bar{\sigma}^{\mathrm{i}} \|U_{\mathrm{c}}\| + \frac{1}{4} \|U_{\mathrm{c}}\| \bar{\sigma}^{\mathrm{c}}
    \end{equation*}
    \hfill $\square$
\end{assumption}

\begin{theorem}[\cite{terzi2021mpc_lstm}]  \label{th:delta_iss_condition} 
    Let Assumption \ref{ass:delta_iss_condition} hold. Then it is possible to upper bound the difference between any couple of system trajectories $x_{\mathrm{a}} = [c_{\mathrm{a}}^\top \; h_{\mathrm{a}}^\top]^\top$ and $x_{\mathrm{b}} = [c_{\mathrm{b}}^\top \; h_{\mathrm{b}}^\top]^\top$ with the following inequality
    \begin{equation}    \label{eq:bound_lstm_trajectory}
    \begin{split}
        \begin{bmatrix}
            \norm{c_{\mathrm{a},k+1} - c_{\mathrm{b},k+1}} \\
            \norm{h_{\mathrm{a},k+1} - h_{\mathrm{b},k+1}}
        \end{bmatrix} &\leq A_\delta 
        \begin{bmatrix}
            \norm{c_{\mathrm{a},k} - c_{\mathrm{b},k}} \\
            \norm{h_{\mathrm{a},k} - h_{\mathrm{b},k}}
        \end{bmatrix} \\
        & + B_\delta \norm{u_{\mathrm{a},k} - u_{\mathrm{b},k}}
    \end{split}
    \end{equation}
    where
    \begin{equation*}
        B_{\delta} =
            \begin{bmatrix}
                \beta \\
                \beta \Bar{\sigma}^{\mathrm{o}} + \frac{1}{4} \Bar{\sigma}^{\mathrm{x}} \|W_{\mathrm{o}}\|
            \end{bmatrix}
    \end{equation*}
    \begin{equation*}
        \beta = \frac{1}{4} \|W_{\mathrm{f}}\| \frac{\Bar{\sigma}^{\mathrm{i}} \Bar{\sigma}^{\mathrm{c}}}{1 - \bar{\sigma}^{\mathrm{f}}} + \Bar{\sigma}^{\mathrm{i}} \|W_{\mathrm{c}}\| + \frac{1}{4} \|W_{\mathrm{i}}\| \Bar{\sigma}^{\mathrm{c}}
    \end{equation*}
    and the LSTM system \eqref{eq:lstm} is exponentially $\delta$ISS in the sets $\mathcal{X}$ and $\mathcal{U}$, where $\mathcal{X} = \mathcal{C} \times \mathcal{H}$, with
    \begin{equation*}
        \mathcal{C} = \left\{ c \in \mathbb{R}^n : \norm{c}_\infty \leq \frac{\bar{\sigma}^{\mathrm{i}} \bar{\sigma}^{\mathrm{c}}}{1 - \bar{\sigma}^{\mathrm{f}}} \right\}
    \end{equation*}
    %and
    \begin{equation*}
        \mathcal{H} = \{ h \in \mathbb{R}^n : \norm{h}_\infty \leq 1 \}
    \end{equation*}
    %\hfill $\square$
\end{theorem}

\begin{remark}  \label{rmk:delta-iss-inequality}
    Using the Jury criterion in \cite{terzi2021mpc_lstm} it is shown that it is possible to turn the condition of Assumption \ref{ass:delta_iss_condition} on eigenvalues of $A_{\delta}$ (i.e. $\rho(A_\delta) < 1$) into the following inequality on the trainable weights of the LSTM model
    \begin{equation*} %\label{eq:delta_iss_inequality}
        -1 + \bar{\sigma}^{\mathrm{f}} + \alpha \bar{\sigma}^{\mathrm{o}} + \frac{1}{4} \bar{\sigma}^{\mathrm{x}} \|U_{\mathrm{o}}\| < \frac{1}{4} \bar{\sigma}^{\mathrm{f}} \bar{\sigma}^{\mathrm{x}} \|U_{\mathrm{o}}\| < 1
    \end{equation*}
    The $\delta$ISS condition can be checked a posteriori or imposed as constraint during the training procedure.
    %\hfill $\square$
\end{remark}

\subsubsection{Incremental Lyapunov function for the LSTM}
Under Assumption \ref{ass:delta_iss_condition} it is possible to compute an incremental Lyapunov function for the $\delta$ISS LSTM model. Such Lyapunov function will be used to define the terminal cost, the terminal constraint and the constraint tightening for the MPC. 
\begin{lemma} \label{lem:incremental_lyap}
    Let Assumption \ref{ass:delta_iss_condition} hold, and denote $x_{\mathrm{a}} = [c_{\mathrm{a}}^\top \; h_{\mathrm{a}}^\top]^\top$, $x_{\mathrm{b}} = [c_{\mathrm{b}}^\top \; h_{\mathrm{b}}^\top]^\top$ with $x_{\mathrm{a}}, x_{\mathrm{b}} \in \mathcal{X}$, and $P_{\mathrm{s}}$ the solution of the Lyapunov equation $A_\delta^\top P_{\mathrm{s}} A_\delta - P_{\mathrm{s}} = -Q_{\mathrm{s}}$ for some symmetric positive definite matrix $Q_{\mathrm{s}}$. Then the function
    \begin{equation*}
        V_{\mathrm{s}} (x_{\mathrm{a}}, x_{\mathrm{b}}) = \norm{
    \begin{bmatrix}
        \norm{c_{\mathrm{a}} - c_{\mathrm{b}}} \\
        \norm{h_{\mathrm{a}} - h_{\mathrm{b}}}
    \end{bmatrix}}_{P_{\mathrm{s}}}
    \end{equation*}
    is an incremental Lyapunov function for the system \eqref{eq:lstm}, such that
    %If Assumption \ref{ass:delta_iss_condition} is satisfied, then for the system \eqref{eq:lstm} there exists an incremental Lyapunov function $V_s(x_a, x_b)$, such that, denoting $x_a = [c_a^\top \; h_a^\top]^\top$ and $x_b = [c_b^\top \; h_b^\top]^\top$ with $x_a, x_b \in \mathcal{X}$, one has
    \begin{subequations}   \label{eq:incremental_lyap_assumptions} 
        \begin{align}
            & c_{\mathrm{s,l}} \norm{x_{\mathrm{a}} - x_{\mathrm{b}}} \leq V_{\mathrm{s}}(x_{\mathrm{a}}, x_{\mathrm{b}}) \leq c_{\mathrm{s,u}} \norm{x_{\mathrm{a}} - x_{\mathrm{b}}} \label{eq:incremental_lyap_pos_def} \\
            & V_{\mathrm{s}}(x_{\mathrm{a}}^+, x_{\mathrm{b}}^+) \leq \rho_{\mathrm{s}} V_{\mathrm{s}} (x_{\mathrm{a}}, x_{\mathrm{b}}) \label{eq:incremental_lyap_neg_def} \\
            & |W_{\mathrm{y}} (h_{\mathrm{a}} - h_{\mathrm{b}})| \leq c_{\mathrm{s}} V_{\mathrm{s}} (x_{\mathrm{a}}, x_{\mathrm{b}})  \label{eq:incremental_lyap_constraints}
        \end{align}   
    \end{subequations}
    %where $c_{s,l}$, $c_{s,u} > 0$, $c_s \in \mathbb{R}^p$ with $c_{s(j)} > 0$ for $j=1,...,p$, $\rho_s \in (0,1)$ $x_a^+ = f(x_a, u)$ and $x_b^+ = f(x_b, u)$.
    where $x_{\mathrm{a}}^+ = f(x_{\mathrm{a}}, u)$, $x_{\mathrm{b}}^+ = f(x_{\mathrm{b}}, u)$, $c_{\mathrm{s,l}} = \sqrt{\lambda_{min}(P_{\mathrm{s}})}$, $c_{\mathrm{s,u}} = \sqrt{\lambda_{max}(P_{\mathrm{s}})}$, $\rho_{\mathrm{s}} = \sqrt{1 - \frac{\lambda_{min} (Q_{\mathrm{s}})}{\lambda_{max} (P_{\mathrm{s}})}}$ and $c_{\mathrm{s}} \in \mathbb{R}^p$ with $c_{\mathrm{s}(j)} = \norm{\left[0 \;  \norm{W_{\mathrm{y}(j*)}}\right] P_{\mathrm{s}}^{-1/2}}$.
    %for $j=1,...,m$, 
    %$x_a^+ = f(x_a, u)$ and $x_b^+ = f(x_b, u)$.
    %\hfill $\square$
\end{lemma}

\textit{Proof:} The proof is reported in Appendix \ref{sec:proof-incremental-lyap}.

The focus on $\delta$ISS LSTM networks limits the applicability of this work to $\delta$ISS plants. This is standard when black-box identification is used. In fact if small changes of the inputs can modify significantly the outputs (non $\delta$ISS systems) model identification is very difficult. In these cases an inner closed-loop can be introduced in order to achieve a $\delta$ISS system to be modelled with the LSTM and controlled with the MPC.

\subsection{State and disturbance observer}
In this subsection first a disturbance model is introduced in order to take into account the inaccuracy of the model and the effect of the possible disturbance $d_\phi$. Then an observer is proposed to provide estimations of the states of the LSTM model and of the disturbance model. The disturbance is modelled as the integral of a bounded unknown input $w$, that represents the variation of the disturbance. This representation is motivated by the fact that one of the goals of the paper is to guarantee zero error at steady state when $y^0$ and $d_\phi$ are constant, and therefore $w$ is null.

\subsubsection{Disturbance model}
The state of the disturbance model is denoted by $d \in \mathbb{R}^p$, while its variation is indicated with $w \in \mathbb{R}^p$. 
The equations of the LSTM model augmented with the disturbance are then the following
\begin{subequations} \label{eq:augmented_lstm}
    \begin{align}
    \begin{split}
        c_{k+1} &= \sigma(W_{\mathrm{f}} u_k + U_{\mathrm{f}} h_k + b_{\mathrm{f}}) \circ c_k \\
        &+ \sigma(W_{\mathrm{i}} u_k + U_{\mathrm{i}} h_k + b_{\mathrm{i}}) \circ \tanh(W_{\mathrm{c}} u_k + U_{\mathrm{c}} h_k + b_{\mathrm{c}}) 
    \end{split}    \label{eq:c_augmented_lstm}\\
    h_{k+1} &= \sigma(W_{\mathrm{o}} u_k + U_{\mathrm{o}} h_k + b_{\mathrm{o}}) \circ \tanh(c_{k+1}) \label{eq:h_augmented_lstm}\\
    d_{k+1} &= d_k + w_k \label{eq:d_augmented_lstm}\\
    y_k &= W_{\mathrm{y}} h_k + b_{\mathrm{y}} + d_k \label{eq:y_augmented_lstm}
    \end{align}
\end{subequations}
and will also be denoted by 
\begin{subequations}    \label{eq:augmented_model}
    \begin{align}
        \chi_{k+1} &= f_{aug}(\chi_k, u_k) + E w_k \label{eq:state_augmented_model}\\
        y_k &= g_{aug}(\chi_k) \label{eq:output_augmented_model}
    \end{align}
\end{subequations}
where $\chi = [x^\top \; d^\top]^\top$ and $E$ is matrix that can be obtained from \eqref{eq:augmented_lstm}.
Observer convergence, robust constraint satisfaction and offset-free can be achieved only for bounded disturbances. For this reason in the following $d$ and $w$ are assumed to be bounded, i.e. $d \in \mathcal{D} = \{ d \in \mathbb{R}^p : \norm{d}_\infty \leq d_{max} \}$ and $w \in \mathcal{W} = \{ w \in \mathbb{R}^p : \norm{w} \leq w_{max} \}$.

Moreover, to achieve offset-free it is necessary that for the model augmented with the constant disturbance there exists an equilibrium associated to any possible pair $(u, y)$, and that this equilibrium is unique (see \cite{morari2012nonlinear_offset_free}). In the following lemma and corollary it is shown that in view of the $\delta$ISS property of the LSTM model it is sufficient to add the disturbance only on the output transformation to satisfy this property. Hence the disturbance was not included also in the state equations.

\begin{lemma} \label{lem:observability}
    If the dynamical system $x_{k+1} = f(x_k, u_k)$ is exponentially $\delta$ISS in the sets $\mathcal{X}$ and $\mathcal{U}$, then given a constant input $u \in \mathcal{U}$ there exists a unique corresponding equilibrium state $x^* \in \mathcal{X}$, such that
    \begin{equation}
        x^* = f(x^*, u) 
    \end{equation}
\end{lemma}
\begin{corollary}   \label{cor:observability}
    For the augmented system \eqref{eq:augmented_lstm}, given $(u, y)$ with $u \in \mathcal{U}$, there exist unique values $(x^*, d^*)$ with $x^* \in \mathcal{X}$ such that
    \begin{subequations}
        \begin{align}
            x^* &= f(x^*, u)    \label{eq:equilibrium_lemma_state} \\
            y &= g (x^*) + d^*  \label{eq:equilibrium_lemma_output}
        \end{align}
    \end{subequations}
\end{corollary}

\textit{Proofs:} The proofs of Lemma \ref{lem:observability} and of Corollary \ref{cor:observability} are reported in Appendix \ref{sec:proof-observability}.

\subsubsection{Observer design}
An observer is designed to obtain an estimation $\hat{\chi} = [\hat{x}^\top \; \hat{d}^\top]^\top$, with $\hat{x} = [\hat{c}^\top \; \hat{h}^\top]^\top$, of the state $\chi$ of the augmented LSTM model \eqref{eq:augmented_lstm}.
The structure of the observer is based on the LSTM observer proposed in \cite{terzi2021mpc_lstm}, but it has an additional equation for the estimation of the disturbance $d$. The equations of the observer are the following
\begin{subequations}    \label{eq:lstm_observer}
    \begin{align}
        \begin{split}
            \hat{c}_{k+1} &= \sigma(W_{\mathrm{f}} u_k + U_{\mathrm{f}} \hat{h}_k + b_{\mathrm{f}} + L_{\mathrm{f}} (y_k - \hat{y}_k)) \circ \hat{c}_k \\
            & + \sigma(W_{\mathrm{i}} u_k + U_{\mathrm{i}} \hat{h}_k + b_{\mathrm{i}} + L_{\mathrm{i}} (y_k - \hat{y}_k)) \\
            & \circ \tanh(W_{\mathrm{c}} u_k + U_{\mathrm{c}} \hat{h}_k + b_{\mathrm{c}})
        \end{split} \label{eq:c_lstm_observer} \\
        \begin{split}
            \hat{h}_{k+1} &= \sigma(W_{\mathrm{o}} u_k + U_{\mathrm{o}} \hat{h}_k + b_{\mathrm{o}} + L_{\mathrm{o}} (y_k - \hat{y}_k)) \\
            & \circ \tanh(\hat{c}_{k+1})
        \end{split} \label{eq:h_lstm_observer} \\
        \hat{d}_{k+1} &= sat(\hat{d}_k +  L_{\mathrm{d}} (y_k - \hat{y}_k), d_{max}) \label{eq:d_lstm_observer}\\
        \hat{y}_k &= W_{\mathrm{y}} \hat{h}_k + b_{\mathrm{y}} + \hat{d}_k  \label{eq:y_lstm_observer} 
    \end{align}
\end{subequations}
where $L_{\mathrm{f}}, L_{\mathrm{i}}, L_{\mathrm{o}} \in \mathbb{R}^{n \times p}$ and $L_{\mathrm{d}} \in \mathbb{R}^{p \times p}$ are the gains of the observer, that are selected according to the following assumption.

\begin{assumption}  \label{ass:observer_gains}
    Denoting  
    \begin{equation*} %\label{eq:sigma_hat_f_def}
        \hat{\bar{\sigma}}^{\mathrm{f}} = \sigma ( \| [W_{\mathrm{f}} u_{max}, \,  U_{\mathrm{f}}-L_{\mathrm{f}} W_{\mathrm{y}}, \,  b_{\mathrm{f}}, \, L_{\mathrm{f}} W_{\mathrm{y}}, \, 2 L_{\mathrm{f}} d_{max}]  \|_{\infty} )
    \end{equation*}
    \begin{equation*} %\label{eq:sigma_hat_i_def}
        \hat{\bar{\sigma}}^{\mathrm{i}} = \sigma ( \| [W_{\mathrm{i}} u_{max}, \,  U_{\mathrm{i}}-L_{\mathrm{i}} W_y, \,  b_{\mathrm{i}}, \, L_{\mathrm{i}} W_{\mathrm{y}}, \, 2 L_{\mathrm{i}} d_{max}]  \|_{\infty} )
    \end{equation*}
    \begin{equation*} %\label{eq:sigma_hat_o_def}
        \hat{\bar{\sigma}}^{\mathrm{o}}= \sigma ( \| [W_{\mathrm{o}} u_{max}, \, U_{\mathrm{o}}-L_{\mathrm{o}} W_{\mathrm{y}}, \, b_{\mathrm{o}}, \, L_{\mathrm{o}} W_{\mathrm{y}}, \, 2 L_{\mathrm{o}} d_{max}]  \|_{\infty} )
    \end{equation*}
    \begin{equation*}
        \hat{\alpha} = \frac{1}{4} \frac{\bar{\sigma}^{\mathrm{i}} \bar{\sigma}^{\mathrm{c}}}{1 - \bar{\sigma}^{\mathrm{f}}} \| U_{\mathrm{f}} - L_{\mathrm{f}} W_{\mathrm{y}} \| + \bar{\sigma}^{\mathrm{i}} \| U_{\mathrm{c}} \| + \frac{1}{4} \bar{\sigma}^{\mathrm{c}} \| U_{\mathrm{i}} - L_{\mathrm{i}} W_{\mathrm{y}} \|
    \end{equation*}
    \begin{equation*}
        \hat{\beta} = \frac{1}{4} \frac{\bar{\sigma}^{\mathrm{i}} \bar{\sigma}^{\mathrm{c}}}{1 - \bar{\sigma}^{\mathrm{f}}} \| L_{\mathrm{f}} \| + \frac{1}{4} \bar{\sigma}^{\mathrm{c}} \| L_{\mathrm{i}} \|
    \end{equation*}
    \begin{equation*}
        \hat{\gamma} = \hat{\bar{\sigma}}^{\mathrm{o}} \hat{\alpha} + \frac{1}{4} \bar{\sigma}^{\mathrm{x}} \|U_{\mathrm{o}} - L_{\mathrm{o}} W_{\mathrm{y}} \|
    \end{equation*}
    %and
    \begin{equation} \label{eq:Ad_observer}
        A_{\mathrm{d}} = 
        \begin{bmatrix}
            \hat{\bar{\sigma}}^{\mathrm{f}} & \hat{\alpha} & \hat{\beta} \\
            \hat{\bar{\sigma}}^{\mathrm{o}} \hat{\bar{\sigma}}^{\mathrm{f}} & \hat{\gamma} & \hat{\bar{\sigma}}^{\mathrm{o}} \hat{\beta} + \frac{1}{4} \bar{\sigma}^{\mathrm{x}} \|L_{\mathrm{o}} \| \\
            0 & \| L_{\mathrm{d}} W_{\mathrm{y}} \| & \| I_{p} - L_{\mathrm{d}} \|
        \end{bmatrix}
    \end{equation}
    the observer gains $L_{\mathrm{f}}, L_{\mathrm{i}}, L_{\mathrm{o}}, L_{\mathrm{d}}$ are selected so that $\rho(A_{\mathrm{d}}) < 1$. 
\end{assumption}

\begin{remark}
    In Theorem \ref{th:observer} it will be shown that $A_{\mathrm{d}}$ is the matrix dynamic for an upper bound of the estimation error dynamic.
\end{remark}

\begin{remark}
    A possible suboptimal choice for the gains of the observer that satisfies Assumption \ref{ass:observer_gains} is $L_{\mathrm{f}} = L_{\mathrm{i}} = L_{\mathrm{o}} = \mathbf{0}_{n,p}$ and $L_{\mathrm{d}} = l_{\mathrm{d}} I_{p}$ with $0 < l_{\mathrm{d}} < 2$. In fact with this choice the matrix $A_{\mathrm{d}}$ becomes
    \begin{equation*}
        A_{\mathrm{d}} = 
        \left[\begin{array}{@{}c|c@{}}
      A_{\delta} & \mathbf{0}_{2,1} \\
    \hline
      \begin{matrix}
       0 & \| L_{\mathrm{d}} W_{\mathrm{y}}\|
      \end{matrix} &
      |l_{\mathrm{d}} - 1|
    \end{array}\right]
    \end{equation*}
    and therefore the eigenvalues of $A_d$ are the eigenvalues of $A_{\delta}$ and an eigenvalue in $|l_{\mathrm{d}} - 1| \in [0,1)$. Hence, because the LSTM is tuned with the constraint that $\rho(A_{\delta}) < 1$, $\rho(A_{\mathrm{d}}) < 1$.
    %\hfill $\square$
\end{remark}

\subsubsection{Observer convergence}
The following lemma defines a positive invariant set for the state $\hat{\chi}$ of the observer, in which it is possible to prove the convergence of the observer estimation. 

\begin{lemma}   \label{lem:invariant_set_observer}
    If $h_k \in \mathcal{H}$ and $d_k \in \mathcal{D}$, $\forall k \in \mathbb{Z}_{\geq 0}$, then the set $\hat{\mathcal{I}} = \hat{\mathcal{C}} \times \mathcal{H} \times \mathcal{D}$, with
    \begin{equation*}
        \hat{\mathcal{C}} = \left\{ \hat{c} \in \mathbb{R}^n : \norm{\hat{c}}_\infty \leq \frac{\hat{\bar{\sigma}}^{\mathrm{i}} \bar{\sigma}^{\mathrm{c}}}{1 - \hat{\bar{\sigma}}^{\mathrm{f}}} \right\}
    \end{equation*}
    is a positive invariant set for the observer \eqref{eq:lstm_observer}.    
\end{lemma}

\textit{Proof:} The proof is reported in Appendix \ref{sec:proof-invariant-set-observer}.

The following theorem reports the main results related to the observer convergence.

\begin{theorem} \label{th:observer}
    If the plant behaves according to \eqref{eq:augmented_lstm} with $x \in \mathcal{X}$, $d \in \mathcal{D}$ and $w \in \mathcal{W}$, Assumption \ref{ass:delta_iss_condition} holds, the observer parameters are selected according to Assumption \ref{ass:observer_gains} and $\hat{\chi} \in \hat{\mathcal{I}}$,
    %then the observer \eqref{eq:lstm_observer} provides a converging state estimation, i.e. $\hat{\chi}_k \to \chi_k$ for $k \to \infty$,
    then the function
    \begin{equation}
        V_{\mathrm{o}}(\hat{\chi}, \chi) = \norm{\begin{bmatrix}
            \| \hat{c} - c\| \\
            \| \hat{h} - h\| \\
            \| \hat{d} - d\|
        \end{bmatrix}}_{P_{\mathrm{o}}}
    \end{equation}
    where $P_\mathrm{o}$ is the solution of the Lyapunov equation $A_{\mathrm{d}}^\top P_{\mathrm{o}} A_{\mathrm{d}} - P_{\mathrm{o}} = - Q_{\mathrm{o}}$ for a symmetric positive definite matrix $Q_{\mathrm{o}}$,
    is an incremental Lyapunov function for the observer estimation error, such that
    \begin{subequations}    \label{eq:lyap_observer_assumptions}
        \begin{align}
            &c_{\mathrm{o,l}} \norm{\hat{\chi} - \chi} \leq V_{\mathrm{o}}(\hat{\chi}, \chi) \leq c_{\mathrm{o,u}} \norm{\hat{\chi} - \chi} \label{eq:lyap_observer_pos_def} \\
            &V_{\mathrm{o}}(\hat{\chi}^+, \chi^+) \leq \rho_{\mathrm{o}} V_{\mathrm{o}}(\hat{\chi}, \chi) + \Bar{w} \label{eq:lyap_observer_neg_def} \\
            &|W_{\mathrm{y}} (h - \hat{h}) + (d - \hat{d})| \leq c_{\mathrm{o}} V_{\mathrm{d}}(\hat{\chi}, \chi)  \label{eq:lyap_observer_constraint} \\
            &\norm{\hat{\chi}^+ - f_{aug}(\hat{\chi}, u)} \leq L_{max} V_{\mathrm{o}}(\hat{\chi}, \chi) \label{eq:lyap_observer_max_gain}
        \end{align}
    \end{subequations}
    where $\chi^+ = f_{aug}(\chi, u) + Ew$, $\hat{\chi}^+$ is the next state computed by the observer \eqref{eq:lstm_observer}, $c_{\mathrm{o,l}} = \sqrt{\lambda_{min} (P_{\mathrm{o}})}$, $c_{\mathrm{o,u}} = \sqrt{\lambda_{max} (P_{\mathrm{o}})}$, $\rho_{\mathrm{o}} = \sqrt{1 - \frac{\lambda_{min} (Q_{\mathrm{o}})}{\lambda_{max}(P_{\mathrm{o}})}} \in (0,1)$, $\Bar{w} = \sqrt{P_{\mathrm{o}(3,3)}} w_{max}$, $L_{max} > 0$, and $c_{\mathrm{o}} \in \mathbb{R}^{p}$ with $c_{\mathrm{o}(j)} = \norm{\left[0 \;  \norm{W_{\mathrm{y}(j*)}} \; 1\right] P_{\mathrm{o}}^{-1/2}}$. 
    %where $\Bar{W}_y = [\mathbf{0}_{p,n} \; W_y \; I_p]$.
    
    Moreover, if $w_k \to 0$ for $k \to \infty$, then the observer provides a converging state estimation, i.e. $\norm{\chi_k - \hat{\chi}_k} \to 0$ for $k \to \infty$. 
    %\hfill $\square$
\end{theorem}

\textit{Proof:} The proof is reported in Appendix \ref{sec:proof-observer}.

\begin{remark}
    In view of the properties of the Lyapunov function $V_{\mathrm{o}}(\hat{\chi}, \chi)$, the observer estimation error is also ISS with respect to the disturbance $w$. 
\end{remark}

\subsection{Reference calculator}

One of the most important characteristics of the proposed algorithm is the possibility to be applied with time varinat set-point and disturbances unknown in advance. In order to do this the MPC assumes a constant set-point along the prediction horizon but it is designed in order to preserve, under suitable assumptions, recursive feasibility even if the exogenous signals change at any time instant.
To manage possible variations of the set-point $y^0$ and/or of the disturbance estimation $\hat{d}$, a Reference calculator is introduced in the control loop. The goal of the Reference calculator is to provide
%, at any time instant, 
the state and input references $\bar{x} = [\bar{c}^\top \; \bar{h}^\top]^\top$ and $\bar{u}$ for the MPC, that are computed by solving the following equations:
\begin{subequations} \label{eq:reference_calculation}
    \begin{align}
        \bar{x} &= f(\bar{x}, \bar{u})  \label{eq:reference_calculation_state} \\
        y^0 &= g(\bar{x})+  \hat{d}     \label{eq:reference_calculation_output}
    \end{align}
\end{subequations}

In order to guarantee offset-free of the closed-loop system, the following assumption on the set-point $y^0$ and on the LSTM model is introduced. An additional assumption on the set-point $y^0$ will be introduced in the next section.

\begin{assumption}  \label{ass:setpoint_and_jacobian}
    The set-point $y^0$ and the LSTM model \eqref{eq:lstm} respect the following conditions:
    \begin{enumerate}
        \item there exists a bounded set $\mathcal{Y}^0 \subset \mathbb{R}^p$ such that $y^0_k \in \mathcal{Y}^0$ $\forall k \in \mathbb{Z}_{\geq 0}$;
        \item the set-point is asymptotically constant, i.e. $y^0_k \to y^0_\infty$ for $k \to \infty$;
        \item $\forall y^0 \in \mathcal{Y}^0$, $\forall \hat{d} \in \mathcal{D}$, there exist $(\Bar{x}, \Bar{u})$ with $\Bar{u} \in \mathcal{U}$ solving \eqref{eq:reference_calculation}, and the Jacobian matrix 
        \begin{equation}    \label{eq:jacobian}
            \begin{bmatrix}
                \frac{\partial}{\partial x} (f(\bar{x}, \bar{u}) - \bar{x}) & \frac{\partial}{\partial u} (f(\bar{x}, \bar{u}) - \bar{x}) \\
                \frac{\partial}{\partial x} (g(\bar{x}) +  \hat{d} - y^0 ) & \frac{\partial}{\partial u} (g(\bar{x})+  \hat{d} - y^0)
            \end{bmatrix}
        \end{equation}
        is invertible.
    \end{enumerate}
    %\hfill $\square$
\end{assumption}

\begin{remark}
    The assumption that the Jacobian matrix \eqref{eq:jacobian} is invertible implies the uniqueness of the couple $(\Bar{x}_k, \Bar{u}_k)$.
    %\hfill$\square$
    \raggedleft$\square$
\end{remark}

Moreover, since a large variation of the set-points $\Bar{x}, \Bar{u}$ is critical for the MPC recursive feasibility, in the following lemma an upper bound for the variation of the state set-point $\bar{x}$ between consecutive time-steps is derived. This upper bound depends on the variation of the set-point $y^0$ and on the observer parameters. 

\begin{lemma}   \label{lem:K_bar}
    Under Assumption \ref{ass:setpoint_and_jacobian}, given the maximum output estimation error $\Bar{e}_{y} = \max_{k \in \mathbb{Z}_{\geq 0}} \norm{y_k - \hat{y}_k}$, there exists a finite constant $\Bar{K}$ such that
    \begin{equation} \label{eq:max_setpoint_variation_y_variable}
        \norm{\Bar{x}_{k+1} - \Bar{x}_{k}} \leq \Bar{K} \norm{L_{\mathrm{d}}} \Bar{e}_y + \Bar{K} \norm{y^0_{k+1} - y^0_{k}}
    \end{equation}
    %\hfill $\square$
\end{lemma}

\textit{Proof:} The proof is reported in Appendix \ref{sec:proof-K-bar}.

\begin{remark}
    In general $\Bar{K}$ cannot be computed explicitly, but it can be estimated numerically by gridding.
\end{remark}

\subsection{Robust MPC formulation}
The robust MPC solves at every time-step a Finite Horizon Optimal Control Problem (FHOCP), where the evolution of the states of the system is predicted with the LSTM model and is initialized with the observer state estimation. The cost function penalizes the deviation from the state and the input set-points computed at the current time instant by the Reference calculator, that are assumed constant along the prediction horizon. The terminal cost and the terminal set are designed to stabilize the closed-loop system. To ensure satisfaction of the output constraints despite the observer estimation error, the disturbance and the set-point variation, a constraint tightening approach similar to the one proposed in \cite{kohler2019simple_robust_mpc} and a time variant terminal set are employed. 
For the constraint tightening, the MPC uses also a time variant term $\hat{e}_\mathrm{o} \in \mathbb{R}$ related to the uncertainty of the observer. This term has the same time evolution of the incremental Lyapunov function for the observer estimation error \eqref{eq:lyap_observer_neg_def}, that is described by the following equation
\begin{equation}    \label{eq:eo_evolution}
     \hat{e}_{\mathrm{o},k+1} = \rho_\mathrm{o} \hat{e}_{\mathrm{o},k} + \Bar{w}
\end{equation}
Note that depending on the values of $\hat{e}_{\mathrm{o},0}$, of $\rho_\mathrm{o}$ and of $\Bar{w}$, $\hat{e}_\mathrm{o}$ can increase or decrease, but its behavior is always monotonic with
\begin{equation}    \label{eq:eo-limit}
    \lim_{k \to \infty} \hat{e}_{\mathrm{o},k} = \Bar{e}_\infty = \frac{\Bar{w}}{1 - \rho_\mathrm{o}}
\end{equation}

\textit{Definition (FHOCP):} Given the prediction horizon $N$,
the FHOCP for the robust MPC is the following
\begin{subequations} \label{eq:optimization}
    \begin{align}
        \begin{split}
            \min_{u_{\cdot|k}} & \sum_{i=0}^{N-1} \left( \norm{x_{i|k} - \Bar{x}_k}_Q^2 + \norm{u_{i|k} - \Bar{u}_k}_R^2 \right) \\
            & + \norm{ \begin{bmatrix}
                \| c_{N|k} - \bar{c}_k \| \\
                \| h_{N|k} - \bar{h}_k \|
            \end{bmatrix}}_{P_\mathrm{f}}^2
        \end{split}  \label{eq:optimization_cost} \\
        \textrm{s.t.} \quad & x_{0|k} = \hat{x}_k \label{eq:optimization_initialization} \\
        & x_{i+1|k} = f(x_{i|k}, u_{i|k}) \\
        & W_\mathrm{y} h_{i|k} + b_\mathrm{y} + d_{max} \mathbf{1}_{p} \leq y_{max} - a_{i} \hat{e}_{\mathrm{o},k} - b_i  \label{eq:optimization_constraint_yub} \\
        & W_\mathrm{y} h_{i|k} + b_\mathrm{y} - d_{max} \mathbf{1}_{p} \geq y_{min} + a_{i} \hat{e}_{\mathrm{o},k} + b_i \label{eq:optimization_constraint_ylb} \\
        & u_{i|k} \in \mathcal{U} \label{eq:optimization_input_constraint} \\
        & \text{for } i = 0, ..., N-1 \\
        & x_{N|k} \in \mathcal{X}_f (k) \label{eq:optimization_terminal_constraint} 
    \end{align}
\end{subequations}
where $Q$ and $R$ are positive definite matrices and are design choices, while $P_\mathrm{f}$ is a positive definite matrix satisfying the Lyapunov condition $A_{\delta}^\top P_\mathrm{f} A_{\delta} - P_\mathrm{f} < - q I_2$,
where $q = \lambda_{max}(Q)$. 
Coefficients $a_i \in \mathbb{R}^{p}$ and $b_i \in \mathbb{R}^p$ for the constraint tightening are defined as 
\begin{subequations}    \label{eq:constraint_tightening_parameters}
    \begin{align}
        a_{0} & = c_\mathrm{o}, \quad b_0 = \mathbf{0}_{p,1} \label{eq:initialization_a_ji} \\
        a_{i+1} & = \rho_\mathrm{o} a_{i} +  \rho_\mathrm{s}^i c_{\mathrm{s,u}} L_{max} c_\mathrm{o,s} \label{eq:recursion_a_ji} \\
        %b_0 & = 0  \label{eq:initialization_b_i} \\
        b_{i+1} &= b_i + a_i \Bar{w}    \label{eq:recursion_b_i}
    \end{align}
\end{subequations}
$\mathcal{X}_f (k)$ is a time variant terminal set, chosen as a sublevel set of the terminal cost:
\begin{equation}    \label{eq:terminal_constraint}
    \mathcal{X}_f (k) = \left\{ \begin{bmatrix}
        c \\
        h
    \end{bmatrix} \in \mathbb{R}^{2n} : \norm{\begin{bmatrix}
        \norm{c - \Bar{c}_k} \\
        \norm{h - \Bar{h}_k}
    \end{bmatrix} }_{P_f} \leq \alpha_k \right\}
\end{equation}
with
\begin{subequations}
    \begin{equation}    \label{eq:def_alpha}
        \begin{split}
            &\alpha_k = \min_{j=1,...,p} \min \left\{ \alpha^{max}_{j,k}, \alpha^{min}_{j,k} \right\} 
        \end{split}    
    \end{equation}
    where
    \begin{equation}    \label{eq:alpha_ub}
        \begin{split}
            & \alpha^{max}_{j,k} = \norm{[0 \; \norm{W_{\mathrm{y}(j*)}}] P_\mathrm{f}^{-1/2}}^{-1} \\
            & \cdot (y_{max(j)} - y^0_{k(j)} - 2 d_{max} - a_{N(j)} \tilde{e}_{\mathrm{o},k} - b_{N(j)})
        \end{split}
    \end{equation}
    \begin{equation}    \label{eq:alpha_lb}
        \begin{split}
            &\alpha^{min}_{j,k} = \norm{[0 \; \norm{W_{\mathrm{y}(j*)}}] P_\mathrm{f}^{-1/2}}^{-1} \\
            & \cdot (y^0_{k(j)} - y_{min(j)} - 2 d_{max} - a_{N(j)} \tilde{e}_{\mathrm{o},k} - b_{N(j)})
        \end{split}
    \end{equation}
    \begin{equation}    \label{eq:e_ok_bar}
        \tilde{e}_{\mathrm{o},k} = \max \left\{ \hat{e}_{\mathrm{o},k}, \bar{e}_\infty  \right\}
    \end{equation}
\end{subequations}
\hfill $\square$

At each time-step $k$ the solution of the FHOCP is denoted by $u^*_{0|k},...,u^*_{N-1|k}$. According to the Receding Horizon principle, the MPC control law is obtained applying only the first element of the optimal input sequence
\begin{equation}    \label{eq:mpc}
    u_k = k^{MPC}(\hat{x}_k, \hat{e}_{\mathrm{o},k}, \Bar{x}_k, \Bar{u}_k, y^0_k) = u^*_{0|k}
\end{equation}

\section{Stability and offset-free results} \label{sec:stability_offset_free}
In this section the properties of the proposed control schema are analysed. In order to guarantee offset-free results a mild assumption on the convergence of the closed-loop of Fig. \ref{fig:block_diagram} will be introduced in Section \ref{subsec:offset-free}. However, first it is necessary to prove recursive feasibility and stability for the nominal closed-loop system reported in Fig. \ref{fig:nominal_block_diagram}, where the plant has been substituted by the LSTM with an additive disturbance on the output \eqref{eq:augmented_lstm}.

\subsection{Recursive feasibility and stability analysis}   \label{subsec:feasibility-stability}
In this subsection recursive feasibility and stability of the nominal closed-loop system reported in Fig. \ref{fig:nominal_block_diagram} are analysed. %The Plant is assumed to behave according to the LSTM model \eqref{eq:c_augmented_lstm}-\eqref{eq:h_augmented_lstm}, with the additive disturbance on the output described by Equation \eqref{eq:d_augmented_lstm}.
First note that in order to have a solution of the FHOCP, it is necessary that $\alpha_k > 0$ for all $k \in \mathbb{Z}_{\geq 0}$. To guarantee this condition, the following assumption on the set-point is introduced.
\begin{assumption}  \label{ass:setpoint}
    The set-point $y^0$ is such that
    \begin{subequations}
        \begin{align}
            y^0_{k(j)} &< y_{max(j)} - 2 d_{max} - a_{N(j)} \tilde{e}_{\mathrm{o},k} - b_{N(j)} \\
            y^0_{k(j)} &> y_{min(j)} + 2 d_{max} + a_{N(j)} \tilde{e}_{\mathrm{o},k} + b_{N(j)}
        \end{align}
    \end{subequations}
    for all $j = 1,...,p$, for all $k \in \mathbb{Z}_{\geq 0}$.
    \hfill $\square$
\end{assumption}

Consider the closed-loop system composed by the augmented LSTM \eqref{eq:augmented_model}, the observer \eqref{eq:lstm_observer}, the Reference calculator \eqref{eq:reference_calculation} and the MPC \eqref{eq:mpc}. This system has state
\begin{equation*}
    \psi = [c^\top \; h^\top \; \hat{c}^\top \; \hat{h}^\top \; \hat{d}^\top \; \hat{e}_\mathrm{o}]^\top
\end{equation*}
and inputs $d$ and $y^0$. Let's now define the feasible set of states and inputs
\begin{equation*}
    \begin{split}
        \mathcal{X}^{MPC} = &\{ (\psi, d, y^0) : x \in \mathcal{X}, \hat{\chi} \in \hat{\mathcal{I}}, d \in \mathcal{D}, y^0 \in \mathcal{Y}^0,\\
        & \hat{e}_{\mathrm{o}} \text{ is such that } V_\mathrm{o}(\hat{\chi}, \chi) \leq \hat{e}_{\mathrm{o}} \\
        & \text{and } \exists \text{ a solution of the FHOCP} \}
    \end{split}
\end{equation*}

\begin{remark}  \label{rmk:meaning_eo}
    In the feasible set $\mathcal{X}^{MPC}$ it is required that $\hat{e}_\mathrm{o}$ is an upper bound for $V_\mathrm{o}(\hat{\chi}, \chi)$. This condition is similar to require that $\hat{e}_\mathrm{o}$ is an upper bound of the observer estimation error. In fact, using Equation \eqref{eq:lyap_observer_pos_def} and $V_\mathrm{o}(\hat{\chi}, \chi) \leq \hat{e}_{\mathrm{o}}$, the following inequality relating $\hat{e}_\mathrm{o}$ and $\norm{\chi - \hat{\chi}}$ can be derived:
    \begin{equation*}
        \norm{\chi - \hat{\chi}} \leq \frac{\hat{e}_\mathrm{o}}{c_{\mathrm{o,l}}}
    \end{equation*}
    \hfill $\square$
\end{remark}

In the following theorem the main result for the nominal closed-loop schema described in Fig. \ref{fig:nominal_block_diagram} is derived.

\begin{theorem} \label{th:feasibility_stability}
    Let Assumptions \ref{ass:input_equal_output}, \ref{ass:delta_iss_condition}, \ref{ass:observer_gains}, \ref{ass:setpoint_and_jacobian}, \ref{ass:setpoint} hold. Then there exist $\Bar{L}_{max} > 0$, $\Bar{L_\mathrm{d}} > 0$ and $\Delta y^0_{max} > 0$ such that for $L_{max} \leq \Bar{L}_{max}$, $\norm{L_\mathrm{d}} \leq \Bar{L_\mathrm{d}}$ and $y^0$ such that $\norm{y^0_{k+1} - y^0_k} \leq \Delta y^0_{max}$ for all $k \in \mathbb{Z}_{\geq 0}$, for the closed-loop system composed by the augmented LSTM \eqref{eq:augmented_model}, the observer \eqref{eq:lstm_observer}, the Reference calculator \eqref{eq:reference_calculation} and the MPC \eqref{eq:mpc} the following properties hold:
   \begin{itemize}
        \item constraints \eqref{eq:input_saturation} and \eqref{eq:output_constraint} are satisfied $\forall (\psi, d, y^0) \in \mathcal{X}^{MPC}$;
        \item the FHOCP is recursively feasible, i.e. $(\psi_k, d_k, y^0_k) \in \mathcal{X}^{MPC} \implies (\psi_{k+1}, d_{k+1}, y^0_{k+1}) \in \mathcal{X}^{MPC}$;
        \item the closed-loop system \eqref{eq:augmented_model}-\eqref{eq:reference_calculation}-\eqref{eq:mpc} is ISpS with respect to the observer estimation error $\chi - \hat{\chi}$ in $\mathcal{X}^{MPC}$;
        \item if $d \to \Bar{d}_\infty$ for $k \to \infty$, then
        \begin{equation*}
            \lim_{k \to \infty} \norm{\psi - \psi_\infty} = 0
        \end{equation*}
        where
        \begin{equation*}
            \psi_\infty = [\Bar{c}_\infty^\top \; \Bar{h}_\infty^\top \; \Bar{c}_\infty^\top \; \Bar{h}_\infty^\top \; \bar{d}_\infty^\top \; \Bar{e}_\infty]^\top
        \end{equation*}
        and $\Bar{x}_\infty = [\Bar{c}_\infty^\top \; \Bar{h}_\infty^\top]^\top $ and $\Bar{u}_\infty$ are the solution of \eqref{eq:reference_calculation} when $y^0 = y^0_\infty$ and $\hat{d} = \bar{d}_\infty$.
        
    \end{itemize}
    %\hfill$\square$
\end{theorem}

\textit{Proof:} The proof is reported in Appendix \ref{sec:proof-feasibility-stability}.

\begin{remark}
    In order to satisfy the condition on the rate of change of $y^0$ it is sufficient to pass the set-point through a rate limiter.
\end{remark}

\subsection{Offset-free result} \label{subsec:offset-free}
It is now possible to follow the results in \cite{morari2012nonlinear_offset_free} to show that the proposed scheme based on the model augmented with the disturbance and on the Reference calculator guarantees offset-free at steady state also when applied to the real plant (see Fig. \ref{fig:block_diagram}), provided that the uncertainty on the model is sufficiently small to preserve convergence and constraints satisfaction, as assumed in the following assumption.
\begin{assumption}  \label{ass:converging_plant}
    The plant disturbance $d_{\phi}$ is bounded and asymptotically constant, and the closed-loop system composed by the plant, the observer \eqref{eq:lstm_observer}, the Reference calculator \eqref{eq:reference_calculation} and the MPC \eqref{eq:mpc} respects the constraints and converges to constant values strictly in the interior of the feasible set.
\end{assumption}
\begin{theorem} \label{th:offset-free}
    If Assumptions \ref{ass:input_equal_output}, \ref{ass:delta_iss_condition}, \ref{ass:observer_gains}, \ref{ass:setpoint_and_jacobian}, \ref{ass:setpoint}, \ref{ass:converging_plant} are satisfied, then the closed-loop system composed by the plant, the observer \eqref{eq:lstm_observer}, the Reference calculator \eqref{eq:reference_calculation} and the MPC \eqref{eq:mpc} is offset-free at steady state, i.e. $y_{\phi,k} \to y^0_{\phi,\infty}$ for $k \to \infty$, where $y^0_{\phi,\infty} = \lim_{k \to \infty} y^0_\phi$.
    %\hfill$\square$
\end{theorem}

\textit{Proof:} The proof is reported in Appendix \ref{sec:proof-offset-free}.

\section{Numerical example} \label{sec:numerical_example}

\begin{figure}
    \centering
    \includegraphics[width=5.0cm]{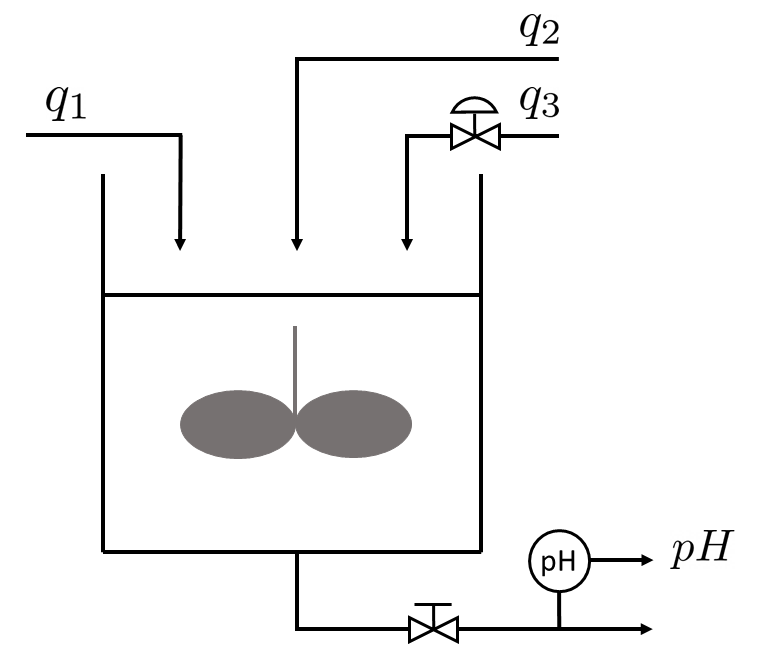}
    \caption{Schematic layout of the pH neutralization process.}
    \label{fig:ph_process}
\end{figure}
As benchmark example to test the proposed control algorithm a pH neutralization process described in \cite{henson1994ph} and used also in \cite{terzi2021mpc_lstm} was considered. The schematic layout of the plant is reported in Fig. \ref{fig:ph_process}. The process consists in a tank where three flows of substances are mixed: an acid flow $q_1$, a buffer flow $q_2$ and an alkaline flow $q_3$.
The measured variable is the pH on the output flow.  
Describing states, disturbance, input and output as follow 
\begin{equation*}
    \begin{split}
        &x = [W_{a4} \; W_{b4} \; h_1]^\top, \quad d_\phi = q_2, \\
        &u_\phi = q_3, \quad  y_\phi = pH
    \end{split}
\end{equation*}
where $W_{a4}$ is a charge related quantity in the output flow, $W_{b4}$ is the concentration of CO$_3^{2-}$ ions in the output flow, $h$ is the tank level and the acid flow $q_1$ is assumed to be a constant parameter, the process model can be written in the form
\begin{equation*}
    \dot{x} = f(x) + g(x)u_\phi + p(x)d_\phi
\end{equation*}
\begin{equation*}    \label{eq:ph_output}
    c(x,y_\phi) = 0
\end{equation*}
where
\begin{equation*}
    f(x) = \begin{bmatrix}
        \frac{q_1}{A_1 x_{(3)}} (W_{a1} - x_{(1)}) \\
        \frac{q_1}{A_1 x_{(3)}} (W_{b1} - x_{(2)}) \\
        \frac{1}{A_1}(q_1 - C_{v4} (x_{(3)} + z)^n) 
    \end{bmatrix}
\end{equation*}
\begin{equation*}
    g(x) = \begin{bmatrix}
        \frac{1}{A_1 x_{(3)}} (W_{a3} - x_{(1)}) \\
        \frac{1}{A_1 x_{(3)}} (W_{b3} - x_{(2)}) \\
        \frac{1}{A_1} 
    \end{bmatrix}
\end{equation*}
\begin{equation*}
    p(x) = \begin{bmatrix}
        \frac{1}{A_1 x_{(3)}} (W_{a2} - x_{(1)}) \\
        \frac{1}{A_1 x_{(3)}} (W_{b2} - x_{(2)}) \\
        \frac{1}{A_1} 
    \end{bmatrix}
\end{equation*}
\begin{equation*} 
\begin{split}
    c(x,y_\phi) &= x_{(1)} + 10^{y_\phi-14} - 10^{-y_\phi} \\
    & + x_{(2)} \frac{1 + 2 \times 10^{y_\phi - pK_2}}{1 + 10^{pK_1 - y_\phi} + 10^{y_\phi - pK_2}}
\end{split}
\end{equation*}

\begin{table}
    \caption{Nominal operating conditions of the pH system.}
    \label{tab:ph_param}
    \centering
    \begin{tabular}{c c c}
    \hline
    $z = 11.5 cm$ & $W_{a1} = 3.00 \cdot 10^{-3} M$ & $q_1 = 16.6 mL/s$ \\
    $C_{v4} = 4.59$ & $W_{b1} = 0.00 M$ & $q_2 = 0.55 mL/s$ \\
    $n = 0.607$ & $W_{a2} = -0.03 M$ & $q_3 = 15.6 mL/s$ \\
    $pK_1 = 6.35$ & $W_{b2} = 0.03 M$ & $q_4 = 32.8 mL/s$ \\
    $pK_2 = 10.25$ & $W_{a3} = 3.05 \cdot 10^{-3} M$ & $A_1 = 207 cm^2$ \\
    $h_1 = 14 cm$ & $W_{b3} = 5.00 \cdot 10^{-5} M$ & $W_{a4} = -4.32 \cdot 10^{-4} M$ \\
    $pH = 7.0$ & $W_{b4} = 5.28 \cdot 10^{-4} M$ & \\
    \hline
    \end{tabular}
\end{table}

In Table \ref{tab:ph_param} the nominal values of the model parameters are reported.
The objective is to control the pH by acting on the alkaline flow $q_3$, while the buffer flow $q_2$ is considered as a disturbance. The alkaline flow is considered saturated, so that 
\begin{equation*}
    u_\phi \in \mathcal{U}_\phi = [12.5, 17] \; mL/s
    %12.5 mL/s \leq u_\phi \leq 17mL/s
\end{equation*}

\subsection{LSTM model identification}
To extract the dataset for the training of the LSTM model, a simulator of the plant was forced with multilevel pseudo-random signals, and the input-output response was sampled with a sampling period of $T_s = 10s$, as done in \cite{terzi2021mpc_lstm}. In particular the input signals for the plant simulator are piecewise constant signals with random values in $\mathcal{U}_\phi$, where each constant value is applied for a random time between $10T_s$ and $100T_s$. 
15 sequences of $N_t = 1500$ time-steps each have been extracted, and have been split in 10 sequences for training, 3 for validation and 2 for testing. All the sequences have been generated considering the disturbance constant at its nominal value $q_2 = 0.55mL/s$. Before the training, inputs and outputs have been normalized using their maximum and minimum values present in the training dataset, so that $-1 \leq u \leq 1$ and $-1 \leq y \leq 1$. 

The LSTM neural network was developed in Python 3.9, using Tensorflow 2.8. To obtain a final network respecting the $\delta$ISS condition of Remark \ref{rmk:delta-iss-inequality}, the training loss and the stopping rule for the training have been modified as suggested in \cite{terzi2021mpc_lstm} and \cite{bonassi2022rnn}.
In particular, the training loss $L$ was selected as the sum of the Mean Squared Error (MSE) and of some terms penalizing weights that do not respect the $\delta$ISS condition. The loss for the single sequence is
\begin{equation*}    %\label{eq:loss_with_constraint}
\begin{split}
    L &= \frac{1}{N_t} \sum_{k = 1}^{N_t} \|y_k - y_{real,k}\|^2 + \lambda_1 \max\{r_1, 0\} \\
    & + \lambda_1 \max\{r_2, 0\} + \lambda_2 \min\{r_1, 0\} + \lambda_2 \min\{r_2, 0\}
    \end{split}
\end{equation*}
where $y_{real}$ is the output in the dataset and $y$ is the output predicted by the LSTM model. $r_1 = -1 + \bar{\sigma}^\mathrm{f} + \alpha \bar{\sigma}^\mathrm{o} + \frac{1}{4} \bar{\sigma}^\mathrm{x} \|U_\mathrm{o}\| -  \frac{1}{4} \bar{\sigma}^\mathrm{f} \bar{\sigma}^\mathrm{x} \|U_\mathrm{o}\|$ and $r_2 = \frac{1}{4} \bar{\sigma}^\mathrm{f} \bar{\sigma}^\mathrm{x} \|U_\mathrm{o}\| - 1$ are terms that need to be negative to satisfy the condition of Remark \ref{rmk:delta-iss-inequality}. $\lambda_1$ and $\lambda_2$ are two positive hyperparameters of the training, that weight the different terms in the loss. In particular $\lambda_1$ determines how much to penalize sets of weights that do not respect the $\delta$ISS condition, while $\lambda_2$ can be tuned to obtain a model that satisfies the $\delta$ISS condition by a larger margin, providing a model with smaller values of $\rho_\mathrm{s}$. In fact a small value of $\rho_\mathrm{s}$, together with a small value of $\rho_\mathrm{o}$, implies that coefficients $a_i$ defined in \eqref{eq:constraint_tightening_parameters} are smaller, leading to a less conservative constraint tightening for the MPC.
Even if the $\delta$ISS condition was included in the training optimization as a soft constraint, it was possible to ensure its satisfaction by introducing it also as a condition for the termination of the training. In particular the training was terminated after a predefined number of epochs only if the $\delta$ISS condition was satisfied, otherwise it was left running for a larger number of epochs.

The performances of the final model were assessed with the FIT index, defined as
\begin{equation*}
    FIT = 100 \left( 1 - \frac{\| y_{real} - y \|}{\| y_{real} - y_{avg} \|} \right) \%
\end{equation*}
where $y$ and $y_{real}$ are the vectors containing the predicted and the real evolution of the system, and $y_{avg}$ is the average of $y_{real}$. 
A trained network with $n = 5$ neurons and with a FIT on the test set of 94.0\% was used for the MPC simulations. The network was trained using Adam optimizer, with a learning rate of 0.001 and hyperparameters $\lambda_1 = 0.03$ and $\lambda_2 = 0.02$. The training took 1000 epochs. 
The final model respects the $\delta$ISS condition of Assumption \ref{ass:delta_iss_condition}, and has $\rho_\mathrm{s} = 0.92$ when the incremental Lyapunov function $V_s$ is computed using $Q_\mathrm{s} = 1000I_2$.

\subsection{Control implementation}
In order to tune the controller it was assumed $d_{max} = 0.1$, that corresponds to the 5\% of the range of variation of the output in the training dataset, since the output was normalized so that $-1 \leq y \leq 1$, and $\Bar{w} = 0.01$. 
Under this assumption, the following parameters have been used. The cost matrices were set to $Q = I_{2n}$ and $R = 1$, and the prediction horizon to $N = 5$. 
The output constraint set was selected as $\mathcal{Y}_\phi = [6.0, 9.0]$.  The observer gains $L_\mathrm{f}, L_\mathrm{i}, L_\mathrm{o}$ were chosen by solving the optimization proposed in \cite{terzi2021mpc_lstm}, while $L_\mathrm{d}$ was set equal to 0.01. The matrix $P_\mathrm{o}$ was obtained by solving the Lyapunov equation with $Q_\mathrm{o} = 1000I_3$, leading to $\rho_\mathrm{o} = 0.99$, $L_{max} = 8.4 \times 10^{-4}$, $w_{max} = 4.3 \times 10^{-5}$ and $\Bar{e}_\infty = 1.04$. 
Concerning the parameters related to the recursive feasibility condition, $\Bar{K} = 2.67$ has been estimated by gridding, while $\Bar{e}_y$ was set to 0.03. This value is clearly affected by the quality of the initialization of the observer. In the considered simulations this bound is always largely respected.
The initial value for the observer state was set to the state equilibrium of the LSTM model associated with the initial output of the system under control and $\hat{d}_0 = 0$. 
With the considered parameters, the time variant interval for $y^0$ needed to satisfy Assumption \ref{ass:setpoint} converges to [6.49, 8.51] after an initial transient.

\subsection{Simulation results}
\begin{figure}
    \centering
    \includegraphics[width=8.6cm]{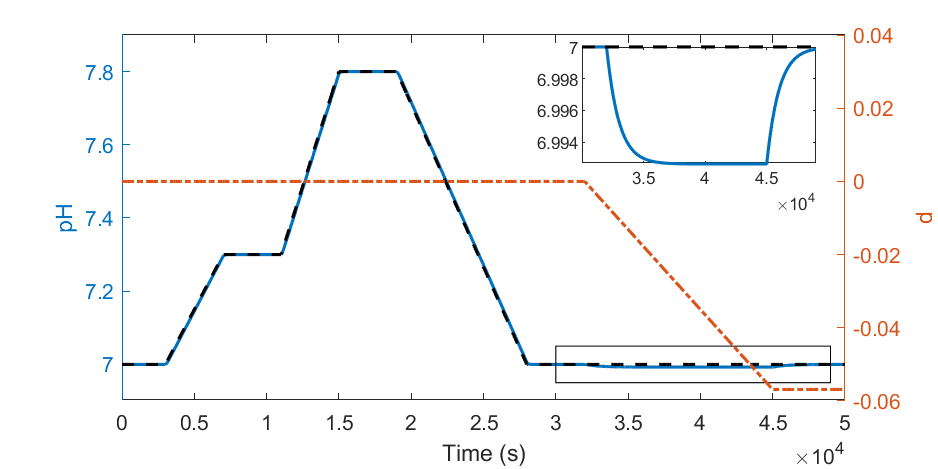}
    \caption{Evolution of the output of the LSTM model (blue line) compared with reference $y^0$ (black dashed line), and evolution of the disturbance $d$ (orange dash-dotted line). $y^0$ and $d$ are selected to respect the sufficient conditions for recursive feasibility of Theorem \ref{th:feasibility_stability}.}
    \label{fig:simulation-rec-feas-lstm}
\end{figure}
\begin{figure}
    \centering
    \includegraphics[width=8.6cm]{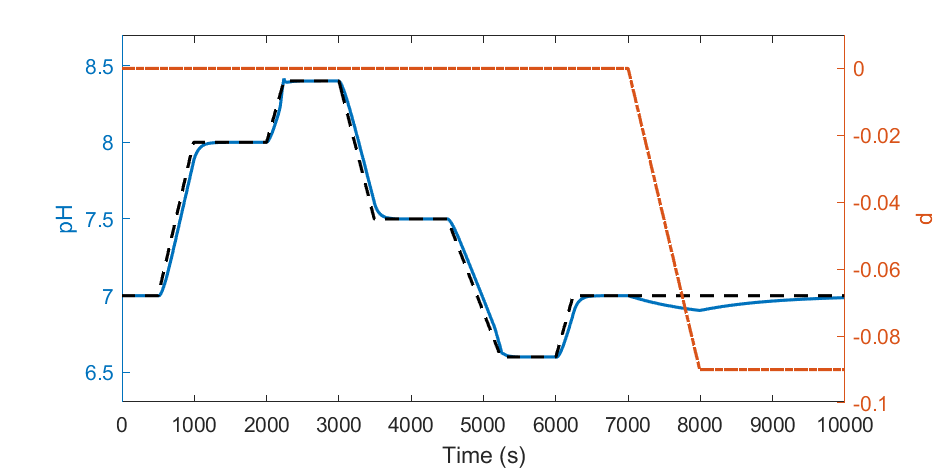}
    \caption{Evolution of the output of the LSTM model (blue line) compared with reference $y^0$ (black dashed line), and evolution of the disturbance $d$ (orange dash-dotted line). The sufficient conditions for recursive feasibility of Theorem \ref{th:feasibility_stability} are not respected, but recursive feasibility is maintained in practice.}
    \label{fig:simulation-performance-lstm}
\end{figure}
\begin{figure}
    \centering
    \includegraphics[width=8.6cm]{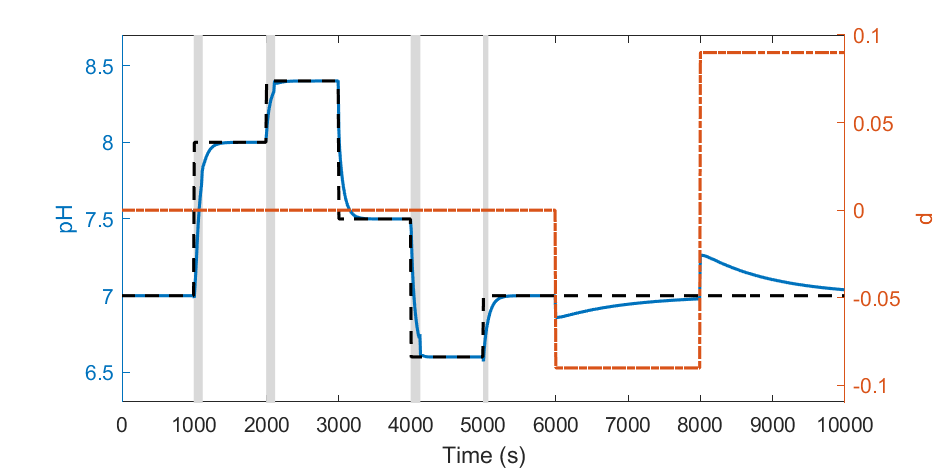}
    \caption{Evolution of the output of the LSTM model (blue line) compared with reference $y^0$ (black dashed line), and evolution of the disturbance $d$ (orange dash-dotted line). In the time instants highlighted in grey the optimization problem \eqref{eq:optimization} is not feasible and the control law is selected by solving \eqref{eq:optimization} without constraints \eqref{eq:optimization_constraint_yub}-\eqref{eq:optimization_constraint_ylb}-\eqref{eq:optimization_terminal_constraint}.}
    \label{fig:simulation-no-rec-feas-lstm}
\end{figure}

First, the conservatism of the feasibility/stability conditions has been tested by performing some simulations where the controller was applied on a perturbed version of the LSTM model. In the first simulation, reported in Fig. \ref{fig:simulation-rec-feas-lstm}, slow variations of the set-point $y^0$ and of the disturbance $d$ are applied, in order to respect all the sufficient conditions required by Theorem \ref{th:feasibility_stability}. As predicted by the theory, recursive feasibility, stability and zero error are fulfilled.
Then, considering that all the bounds have been derived using a sequence of possibly conservative inequalities, an analysis has been carried out to verify the applicability of the control in presence of larger variations of the set-point $y^0$ and of the disturbance $d$, that do not satisfy the sufficient conditions required by Theorem \ref{th:feasibility_stability}. In particular in Fig. \ref{fig:simulation-performance-lstm} and \ref{fig:simulation-no-rec-feas-lstm} simulations with increasingly larger set-point and disturbance variations are reported. 
In the simulation reported in Fig. \ref{fig:simulation-performance-lstm} feasibility is never lost, while in Fig. \ref{fig:simulation-no-rec-feas-lstm} in the time instants highlighted in grey the FHOCP is not feasible because of the large variations of $y^0$. Notably, feasibility is maintained also with large step variations of the disturbance $d$.

\begin{figure}
    \centering
    \includegraphics[width=8.6cm]{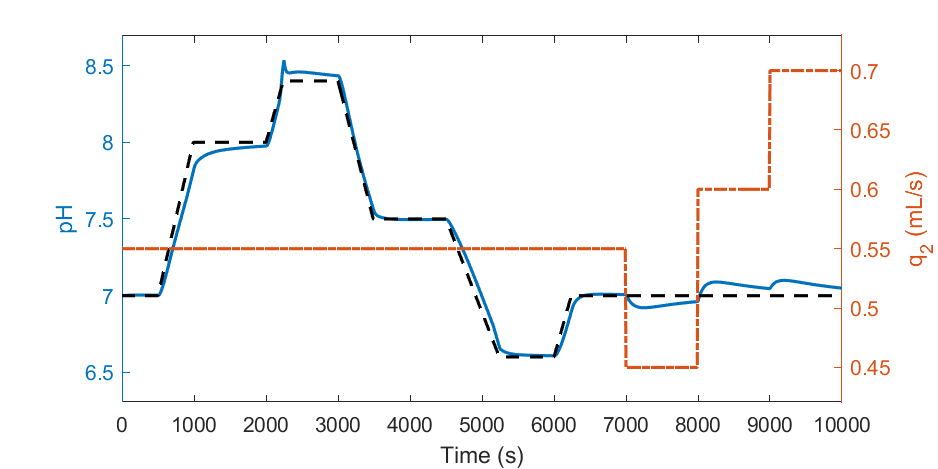}
    \caption{Evolution of the output of the real plant (blue line) compared with reference $y^0_\phi$ (black dashed line), and evolution of the plant disturbance $d_\phi$ (orange dash-dotted line), when $L_\mathrm{d} = 0.01$.}
    \label{fig:output-smallLd}
\end{figure}
\begin{figure}
    \centering
    \includegraphics[width=8.6cm]{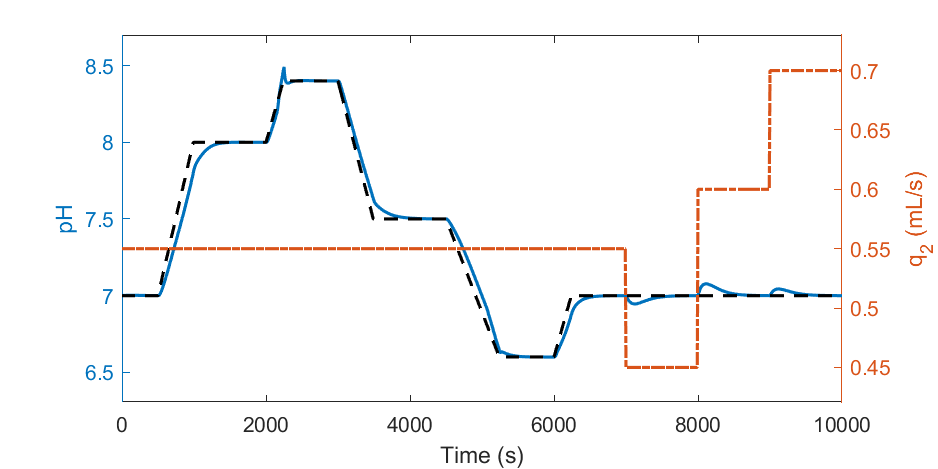}
    \caption{Evolution of the output of the real plant (blue line) compared with reference $y^0_\phi$ (black dashed line), and evolution of the plant disturbance $d_\phi$ (orange dash-dotted line), when $L_\mathrm{d} = 0.1$.}
    \label{fig:output}
\end{figure}

Then, the controller has been applied to the real plant (represented by the physical equations describing the system). Since model uncertainty has not been estimated the bounds $d_{max}$ and $\Bar{w}$ are considered as tuning parameters of the control. Also the parameter $\hat{e}_\mathrm{o}$ has been considered as an additional tuning parameter of the control, since the states of the plant do not coincide with the states of the LSTM.  The first simulation has been carried out using the same parameters of the simulations performed using the LSTM model, and is reported in Fig. \ref{fig:output-smallLd}. It can be noticed that the response of the plant to the disturbances is very slow, and null error is reached only after a very long transient. Hence, in a second simulation, $L_\mathrm{d}$ has been increased to 0.1 to make the disturbance estimation in the observer faster. With this tuning of the observer $\rho_\mathrm{o} = 0.97$, $L_{max} = 8.5 \times 10^{-3}$ and $\Bar{e}_\infty = 0.38$.
In this case, the asymptotic interval for $y^0$ needed to satisfy Assumption \ref{ass:setpoint} is $[6.57, 8.43]$.
The resulting trajectory is reported in Fig. \ref{fig:output}, and it can be seen that the control is able to correctly manage the plant, maintain the recursive feasibility and achieve null tracking error also in presence of disturbances.

\section{Conclusion}    \label{sec:conclusion}
In this paper an MPC control scheme based on an LSTM model of the plant under control has been proposed. The MPC algorithm is able to take into account input and output constraints and to guarantee null error at steady state also in presence of modelling errors and of asymptotically constant bounded disturbances.
Offset-free is obtained by introducing the estimation of a disturbance term in the observer, that is used to update at every time-step the reference values for the MPC cost.
The satisfaction of output constraints in presence of the observer estimation error and time variant set-point is obtained using a constraint tightening technique, based on an upper bound of the norm of the observer estimation error.
The key element for the design of the constraint tightening is the derivation of incremental Lyapunov functions for the LSTM model and for the observer, whose parameters are employed in the definition of the coefficients for the constraint tightening. 
In the FHOCP formulation for the MPC a time variant terminal constraint is also introduced, that guarantees recursive feasibility in presence of variations of the reference values. 
Then, ISpS and convergence are proven by means of a Lyapunov function, different from the optimal cost because considers the deviation from the asymptotic values of state and input set-points that are a-priori  unknown.

The main limitation of the proposed approach is the conservatism of the robust constrained MPC algorithm that must cope with both the disturbances and the set-point variations. In order to partially reduce this limitation an artificial reference approach could be adopted, avoiding the restriction on the tightened constraints introduced by the set-point variation. Further improvement could be achieved by deriving less conservative bounds along all the proofs. However most of the conservatism is inherent in the worst case deterministic approach. As shown in the simulation example it is possible to find a compromise between a-priori feasibility guarantee and performance by adapting some of the algorithm parameters.

\appendices


%%%%%%%%%%%%%%%%%%%%%%%%%%%%%%%%%%%%%%%%%%%%%%%%%%%%%%%%%%%%%%%%%%%%%%%%%%%%%%%%

%%%%%%%%%%%%%%%%%%%%%%%%%%%%%%%%%%%%%%%%%%%%%%%%%%%%%%%%%%%%%%%%%%%%%%%%%%%%%%%%

%%%%%%%%%%%%%%%%%%%%%%%%%%%%%%%%%%%%%%%%%%%%%%%%%%%%%%%%%%%%%%%%%%%%%%%%%%%%%%%%
\section{Proofs} \label{sec:appendix_proof}
The following lemmas will be used for the proofs.
\begin{lemma} [\cite{terzi2021mpc_lstm}] \label{lem:norm_inequality}
   Given two vectors $a, b \in \mathbb{R}^n$ and a positive definite matrix $M$, for any $\varphi \neq 0$ it holds that
   \begin{equation} \label{eq:norm_inequality}
        \|a+b\|_M^2 \leq (1 + \varphi^2)  \|a\|_M^2 + \left(1 + \frac{1}{\varphi^2}\right)  \|b\|_M^2
    \end{equation}
    %\hfill $\square$
\end{lemma}

\begin{lemma}
    Given two sets of $n$ numbers each, $\{ a_1, ..., a_n \}$ and $\{ b_1,  ..., b_n \}$, one has that
    \begin{equation}    \label{eq:inequality_minimums}
       \begin{split}
             & \min_{i = 1, ..., n} a_i - \min_{i = 1, ..., n} b_i \geq \min_{i = 1, ..., n} (a_i - b_i)
        \end{split}
    \end{equation}
    %\hfill $\square$
\end{lemma}

\textit{Proof: } Equation \eqref{eq:inequality_minimums} can be rearranged as 
\begin{equation*}
    \min_{i = 1, ..., n} a_i \geq \min_{i = 1, ..., n} b_i + \min_{i = 1, ..., n} (a_i - b_i)
\end{equation*}
Let's now denote $c_i = a_i - b_i$ for $i = 1, ..., n$. Then
\begin{equation*}
    \begin{split}
        &\min_{i = 1, ..., n} a_i = \min_{i = 1, ..., n} (b_i + c_i) \geq \min_{i = 1,...,n,\: j = 1,...,n} (b_i + c_j) \\
        & = \min_{i = 1, ..., n} b_i + \min_{j = 1, ..., n} c_j = \min_{i = 1, ..., n} b_i + \min_{i = 1, ..., n} c_i \\
        & = \min_{i = 1, ..., n} b_i + \min_{i = 1, ..., n} (a_i - b_i)
    \end{split}
\end{equation*}
%\hfill $\square$

\subsection{Proof of Lemma \ref{lem:incremental_lyap}} \label{sec:proof-incremental-lyap}
Condition \eqref{eq:incremental_lyap_pos_def} is easily verified for the definition of $V_\mathrm{s}$.

Condition \eqref{eq:incremental_lyap_neg_def} can be verified as follows:
\begin{equation*}
\begin{split}
    &V_{\mathrm{s}}(x_\mathrm{a}^+, x_\mathrm{b}^+) = \sqrt{ \begin{bmatrix}
        \| c_\mathrm{a}^+ - c_\mathrm{b}^+ \| \\
        \| h_\mathrm{a}^+ - h_\mathrm{b}^+ \|
    \end{bmatrix}^\top P_\mathrm{s} \begin{bmatrix}
        \| c_\mathrm{a}^+ - c_\mathrm{b}^+ \| \\
        \| h_\mathrm{a}^+ - h_\mathrm{b}^+ \|
    \end{bmatrix}} \\
    & \stackrel{\eqref{eq:bound_lstm_trajectory}}{\leq} 
    \sqrt{ \begin{bmatrix}
        \norm{c_\mathrm{a} - c_\mathrm{b}} \\
        \norm{h_\mathrm{a} - h_\mathrm{b}}
    \end{bmatrix}^\top A_\delta^\top P_\mathrm{s} A_\delta \begin{bmatrix}
        \norm{c_\mathrm{a} - c_\mathrm{b}} \\
        \norm{h_\mathrm{a} - h_\mathrm{b}}
    \end{bmatrix}} \\
    \end{split}
\end{equation*}
\begin{equation*}
    \begin{split}
    & = \sqrt{\begin{bmatrix}
        \norm{c_\mathrm{a} - c_\mathrm{b}} \\
        \norm{h_\mathrm{a} - h_\mathrm{b}}
    \end{bmatrix}^\top  (P_\mathrm{s} - Q_\mathrm{s})  \begin{bmatrix}
        \norm{c_\mathrm{a} - c_\mathrm{b}} \\
        \norm{h_\mathrm{a} - h_\mathrm{b}}
    \end{bmatrix}} \\ 
    & \leq \sqrt{\norm{\begin{bmatrix}
        \norm{c_\mathrm{a} - c_\mathrm{b}} \\
        \norm{h_\mathrm{a} - h_\mathrm{b}}
    \end{bmatrix}}^2_{P_\mathrm{s}} - \lambda_{min}(Q_\mathrm{s}) \norm{
    \begin{bmatrix}
        \norm{c_\mathrm{a} - c_\mathrm{b}} \\
        \norm{h_\mathrm{a} - h_\mathrm{b}}
    \end{bmatrix}}^2} \\
    & \leq \sqrt{\left( 1 - \frac{\lambda_{min}(Q_\mathrm{s})}{\lambda_{max}(P_\mathrm{s})} \right) \norm{
    \begin{bmatrix}
        \norm{c_\mathrm{a} - c_\mathrm{b}} \\
        \norm{h_\mathrm{a} - h_\mathrm{b}}
    \end{bmatrix}}^2_{P_\mathrm{s}} } = \rho_\mathrm{s}  V_\mathrm{s}(x_\mathrm{a}, x_\mathrm{b})
    %& = \sqrt{1 - \frac{\lambda_{min}(Q_s)}{\lambda_{max}(P_s)}} V_s(x_a, x_b)
\end{split}  
\end{equation*}

Finally, to verify condition \eqref{eq:incremental_lyap_constraints}, consider the $j$-th row of $| W_\mathrm{y} (h_\mathrm{a} - h_\mathrm{b}) |$ for $j = 1, ..., p$:
\begin{equation*}
    \begin{split}
        &|W_{\mathrm{y}(j*)} (h_\mathrm{a} - h_\mathrm{b})| = \abs{[\mathbf{0}_{1,n} \;  W_{\mathrm{y}(j*)}] \begin{bmatrix}
            c_\mathrm{a} - c_\mathrm{b} \\
            h_\mathrm{a} - h_\mathrm{b}
        \end{bmatrix}} \\
        & \leq \abs{\left[\norm{\mathbf{0}_{n,1}} \;  \norm{W_{\mathrm{y}(j*)}}\right] \begin{bmatrix}
            \| c_\mathrm{a} - c_\mathrm{b} \| \\
            \| h_\mathrm{a} - h_\mathrm{b} \|
        \end{bmatrix}} \\
        & = \abs{\left[0 \;  \norm{W_{\mathrm{y}(j*)}}\right] P_\mathrm{s}^{-1/2} P_\mathrm{s}^{1/2} \begin{bmatrix}
            \| c_\mathrm{a} - c_\mathrm{b} \| \\
            \| h_\mathrm{a} - h_\mathrm{b} \|
        \end{bmatrix}} \\
        & \leq \norm{\left[0 \;  \norm{W_{\mathrm{y}(j*)}}\right] P_\mathrm{s}^{-1/2}} \cdot \norm{P_\mathrm{s}^{1/2} \begin{bmatrix}
            \| c_\mathrm{a} - c_\mathrm{b} \| \\
            \| h_\mathrm{a} - h_\mathrm{b} \|
        \end{bmatrix}} \\
        & = c_{\mathrm{s}(j)} V_\mathrm{s} (x_\mathrm{a}, x_\mathrm{b})
    \end{split}
\end{equation*}

\subsection{Proofs of Lemma \ref{lem:observability} and of Corollary \ref{cor:observability}} \label{sec:proof-observability}
%The proof is divided in the proof of existence of $(x^*, d^*)$ and proof of uniqueness.
\textit{Proof of Lemma \ref{lem:observability}:}

\textbf{Existence of $x^*$:} It is first proven that if a constant input $u$ is applied to a $\delta$ISS system, then the resulting state trajectory is asymptotically constant. Considering any possible initial state $x_0$, and applying the definition of $\delta$ISS  \eqref{eq:def_delta_iss} with $x_{\mathrm{b},0} = x_{0}$, $x_{\mathrm{a},0} = x_1 = f(x_{0}, u)$ and $u_{\mathrm{a},h} = u_{\mathrm{b},h} = u$ for all $h = 0, ..., k-1$, it follows that:
\begin{equation*}
    \| x_{k+1} - x_{k} \| \leq \beta(\|x_{1} - x_{0}\|, k)
\end{equation*}
Hence the difference between $x_{k+1}$ and $x_k$ becomes small when $k$ increases. %i.e. the trajectory is asymptotically constant.
By summing up the previous inequality one has that $\lim_{T \to \infty} \norm{x_T - x_0} \leq \sum_{k = 0}^\infty \beta(\|x_{1} - x_{0}\|, k)$, that is finite in view of the exponential nature of the function $\beta$. Then the resulting trajectory is asymptotically constant.
Moreover, given any $u$ there exists $x^*$ solving \eqref{eq:equilibrium_lemma_state}, that corresponds to the asymptotic value of the trajectories associated to the constant input $u$. % Once found $x^*$, for any $y$ there exist $d^* = y - g(x^*)$ solving \eqref{eq:equilibrium_lemma_output}.
Finally, $x^* \in \mathcal{X}$ in view of the positive invariance of $\mathcal{X}$.

\textbf{Uniqueness of $x^*$:} Consider the definition of $\delta$ISS (\ref{eq:def_delta_iss}), and
assume that there exist two different equilibrium states $x^*_\mathrm{a}$ and $x^*_\mathrm{b}$ corresponding to the same input $u$. Under this assumption it is possible to study the evolution of the system with $x_{\mathrm{a},0} = x^*_\mathrm{a}$, $x_{\mathrm{b},0} = x^*_\mathrm{b}$ and $u_{\mathrm{a},h} = u_{\mathrm{b},h} = u$ for all $h = 0, ..., k-1$. Then, for the $\delta$ISS assumption, it follows that $\| x_{\mathrm{a},k} - x_{\mathrm{b},k} \| \to 0$ for $k \to \infty$. This is a contradiction with the fact that $x^*_\mathrm{a}$ and $x^*_\mathrm{b}$ are two different equilibrium states. Then the equilibrium state $x^*$ is unique.

\textit{Proof of Corollary \ref{cor:observability}:}
The existence and uniqueness of $x^*$ follows from Lemma \ref{lem:observability}. Then
%\textbf{Existence and uniqueness of $d^*$:} Once found $x^*$, 
there exists a unique value $d^*$ satisfying \eqref{eq:equilibrium_lemma_output}, that is $d^* = y - g(x^*)$.
%\hfill$\square$

\subsection{Proof of Lemma \ref{lem:invariant_set_observer}} \label{sec:proof-invariant-set-observer}
First of all, recall from \cite{terzi2021mpc_lstm} that $\Bar{\sigma}^\mathrm{f}$, $\Bar{\sigma}^\mathrm{i}$, $\Bar{\sigma}^\mathrm{o}$ and $\Bar{\sigma}^\mathrm{c}$ are bounds on the values of the gates of the LSTM. In particular, if $h \in \mathcal{H}$ and $u \in \mathcal{U}$ one has that for $j = 1,...,n$
\begin{subequations}
    \begin{equation}    \label{eq:sigma_f_bound}
         | \sigma(W_\mathrm{f} u + U_\mathrm{f} h + b_\mathrm{f})_{(j)} | \leq \bar{\sigma}^\mathrm{f}
    \end{equation}
    \begin{equation}    \label{eq:sigma_i_bound}
         | \sigma(W_\mathrm{i} u + U_\mathrm{i} h + b_\mathrm{i})_{(j)} | \leq \bar{\sigma}^\mathrm{i}
    \end{equation}
    \begin{equation}    \label{eq:sigma_o_bound}
         | \sigma(W_\mathrm{o} u + U_\mathrm{o} h + b_\mathrm{o})_{(j)} | \leq \bar{\sigma}^\mathrm{o}
    \end{equation}
    \begin{equation}    \label{eq:sigma_c_bound}
         | \tanh(W_\mathrm{c} u + U_\mathrm{c} h + b_\mathrm{c})_{(j)} | \leq \bar{\sigma}^\mathrm{c}
    \end{equation}
\end{subequations}

Consider Equation \eqref{eq:h_lstm_observer} for $\hat{h}$. $\mathcal{H}$ is an invariant set for $\hat{h}$ because each component of $\hat{h}_{k+1}$ is computed as the product between a sigmoid function, whose output lies in $(0,1)$, and an hyperbolic tangent, whose output lies in $(-1,1)$.

Moreover, $\mathcal{D}$ is an invariant set for $\hat{d}$ in view of the saturation in Equation \eqref{eq:d_lstm_observer}.

It is now proven that $\hat{\mathcal{C}}$ is a positive invariant set for $\hat{c}$.
First note that if $u \in \mathcal{U}$, $h, \hat{h} \in \mathcal{H}$, $d, \hat{d} \in \mathcal{D}$, in view of \eqref{eq:y_augmented_lstm}, one has that
\begin{equation}    \label{eq:sigma_hat_f_bound}
    \begin{split}
        &| \sigma(W_\mathrm{f} u + U_\mathrm{f} \hat{h} + b_\mathrm{f} + L_\mathrm{f} (y - \hat{y}))_{(j)} | \\
        & = \sigma \left( (W_\mathrm{f} u + U_\mathrm{f} \hat{h} + b_\mathrm{f} + L_\mathrm{f} (y - \hat{y}))_{(j)} \right) \\
        & \leq \sigma \left( \abs{(W_\mathrm{f} u + U_\mathrm{f} \hat{h} + b_\mathrm{f} + L_\mathrm{f} (y - \hat{y}))_{(j)} } \right) \\
        & \leq \sigma \left( \max_{u, h, \hat{h}, d, \hat{d}} \max_j \abs{(W_\mathrm{f} u + U_\mathrm{f} \hat{h} + b_\mathrm{f} + L_\mathrm{f} (y - \hat{y}))_{(j)} } \right) \\
        & = \sigma \left( \max_{u, h, \hat{h}, d, \hat{d}} \| W_\mathrm{f} u + U_\mathrm{f} \hat{h} + b_\mathrm{f} + L_\mathrm{f} (y - \hat{y}) \|_{\infty} \right)  \\
        &\leq \sigma ( \| [W_\mathrm{f} u_{max}, \ U_\mathrm{f}-L_\mathrm{f} W_\mathrm{y}, \ b_\mathrm{f}, \ L_\mathrm{f} W_\mathrm{y}, \ 2 L_\mathrm{f} d_{max}]  \|_{\infty} ) \\
        & = \hat{\bar{\sigma}}^\mathrm{f}
    \end{split}
\end{equation}
and that
\begin{equation}    \label{eq:sigma_hat_i_bound}
    | \sigma(W_\mathrm{i} u + U_\mathrm{i} \hat{h} + b_\mathrm{i} + L_\mathrm{i} (y - \hat{y}))_{(j)} | \leq \hat{\bar{\sigma}}^\mathrm{i}
\end{equation}

Consider now the $j$-th component of $\hat{c}$, and assume $\hat{c}_k \in \hat{\mathcal{C}}$, $h_k, \hat{h}_k \in \mathcal{H}$ and $d_k, \hat{d}_k \in \mathcal{D}$. By taking the absolute value of the $j$-th component of \eqref{eq:c_lstm_observer} one gets
\begin{equation*}
    \begin{split}   
        & |\hat{c}_{k+1(j)}| \leq | \sigma(W_\mathrm{f} u_k + U_\mathrm{f} \hat{h}_k + b_\mathrm{f} + L_\mathrm{f} (y - \hat{y}))_{(j)} | \cdot |\hat{c}_{k(j)}| \\
        & + | \sigma(W_\mathrm{i} u_k + U_\mathrm{i} \hat{h}_k + b_\mathrm{i} + L_\mathrm{i} (y - \hat{y}))_{(j)} | \\
        & \cdot |\tanh(W_\mathrm{c} u_k + U_\mathrm{c} \hat{h}_k + b_\mathrm{c})_{(j)}| \\
        & \stackrel{\eqref{eq:sigma_c_bound}\eqref{eq:sigma_hat_f_bound}\eqref{eq:sigma_hat_i_bound}}{\leq} \hat{\bar{\sigma}}^\mathrm{f} |\hat{c}_{k(j)}| + \hat{\bar{\sigma}}^\mathrm{i} \bar{\sigma}^\mathrm{c} \leq \frac{\hat{\bar{\sigma}}^\mathrm{i} \bar{\sigma}^\mathrm{c}}{1 - \hat{\bar{\sigma}}^\mathrm{f}}
    \end{split}
\end{equation*}
i.e. $\hat{\mathcal{C}}$ is a positive invariant set for state $\hat{c}$.
%\hfill $\square$

\subsection{Proof of Theorem \ref{th:observer}} \label{sec:proof-observer}
The proof is divided in three parts:
\begin{enumerate}
    \item Derivation of an upper bound for the evolution of the observer estimation error;
    \item Proof of the properties \eqref{eq:lyap_observer_assumptions} of the observer Lyapunov function $V_\mathrm{o}$;
    \item Proof of convergence, under the assumption that $w_k \to 0$ for $k \to \infty$.
\end{enumerate}

\textbf{Part 1: } First note that in a similar way to \eqref{eq:sigma_hat_f_bound}, it is possible to show that
\begin{equation}    \label{eq:sigma_hat_o_bound}
    | \sigma(W_\mathrm{o} u + U_\mathrm{o} \hat{h} + b_\mathrm{o} + L_\mathrm{o} (y - \hat{y}))_{(j)} | \leq \hat{\bar{\sigma}}^\mathrm{o}
\end{equation}

Let's now define the error variables $e_{\mathrm{c}} = c - \hat{c}$, $e_{\mathrm{h}} = h - \hat{h}$ and $e_{\mathrm{d}} = d - \hat{d}$, and let's study their evolution in time. 

Consider the time evolution of $e_c$: 
\begin{equation*}
    \begin{split}
        & e_{\mathrm{c}}^+ = c^+ - \hat{c}^+ \\
        & = \sigma(W_\mathrm{f} u + U_\mathrm{f} h + b_\mathrm{f}) \circ c \\
        & + \sigma(W_\mathrm{i} u + U_\mathrm{i} h + b_\mathrm{i}) \circ \tanh(W_\mathrm{c} u + U_\mathrm{c} h + b_\mathrm{c}) \\
        & - \left( \sigma(W_\mathrm{f} u + U_\mathrm{f} \hat{h} + b_\mathrm{f} + L_\mathrm{f} (y - \hat{y})) \circ \hat{c} \right. \\
        & + \sigma(W_\mathrm{i} u + U_\mathrm{i} \hat{h} + b_\mathrm{i} + L_\mathrm{i} (y - \hat{y})) \\
        & \circ  \left. \tanh(W_\mathrm{c} u + U_\mathrm{c} \hat{h} + b_\mathrm{c}) \right) \\
        & = \sigma(W_\mathrm{f} u + U_\mathrm{f} \hat{h} + b_\mathrm{f} + L_\mathrm{f} (y - \hat{y})) \circ (c - \hat{c}) \\
        & + c \circ \Big( \sigma(W_\mathrm{f} u + U_\mathrm{f} h + b_\mathrm{f})  \\
        & - \left. \sigma(W_\mathrm{f} u + U_\mathrm{f} \hat{h} + b_\mathrm{f} + L_\mathrm{f} (y - \hat{y})) \right) \\
        & + \sigma(W_\mathrm{i} u + U_\mathrm{i} h + b_\mathrm{i}) \circ \\
        & \left( \tanh(W_\mathrm{c} u + U_\mathrm{c} h + b_\mathrm{c}) - \tanh(W_\mathrm{c} u + U_\mathrm{c} \hat{h} + b_\mathrm{c}) \right) \\
        & + \tanh(W_\mathrm{c} u + U_\mathrm{c} \hat{h} + b_\mathrm{c}) \circ \\
        & \left( \sigma(W_\mathrm{i} u + U_\mathrm{i} h + b_\mathrm{i}) - \sigma(W_\mathrm{i} u + U_\mathrm{i} \hat{h} + b_\mathrm{i} + L_\mathrm{i} (y - \hat{y})) \right)
    \end{split}
\end{equation*}
Taking the norm of both sides, exploiting $c \in \mathcal{C}$, \eqref{eq:y_lstm}, (\ref{eq:sigma_f_bound}), (\ref{eq:sigma_i_bound}), (\ref{eq:sigma_o_bound}), (\ref{eq:sigma_c_bound}), %(\ref{eq:sigma_x_def}), 
(\ref{eq:sigma_hat_f_bound}), (\ref{eq:sigma_hat_o_bound}), the definition of $\Bar{\sigma}^\mathrm{x}$, and the fact that $\sigma(\cdot)$ and $\tanh(\cdot)$ are Lipschitz continuous functions with Lipschitz constants respectively of $\frac{1}{4}$ and $1$ (see \eqref{eq:lstm}), one has that
\begin{equation*}
    \begin{split}
        & \| e_{\mathrm{c}}^+ \| \leq \hat{\bar{\sigma}}^\mathrm{f} \| e_{\mathrm{c}} \| \\
        & + \frac{\bar{\sigma}^\mathrm{i} \bar{\sigma}^\mathrm{c}}{1 - \bar{\sigma}^\mathrm{f}} \frac{1}{4} \| U_\mathrm{f} h - U_\mathrm{f} \hat{h} - L_\mathrm{f} W_\mathrm{y} (h - \hat{h}) - L_\mathrm{f} (d - \hat{d}) \| \\
        & + \bar{\sigma}^\mathrm{i} \| U_\mathrm{c} (h - \hat{h}) \| \\
        & + \bar{\sigma}^\mathrm{c} \frac{1}{4} \| U_\mathrm{i} h - U_\mathrm{i} \hat{h} - L_\mathrm{i} W_\mathrm{y} (h - \hat{h}) - L_\mathrm{i} (d - \hat{d}) \| \\
        & \leq \hat{\bar{\sigma}}^\mathrm{f} \| e_{\mathrm{c}} \| + \frac{\bar{\sigma}^\mathrm{i} \bar{\sigma}^\mathrm{c}}{1 - \bar{\sigma}^\mathrm{f}} \frac{1}{4} (\| U_\mathrm{f} - L_\mathrm{f} W_\mathrm{y} \| \|e_{\mathrm{h}}\| + \| L_\mathrm{f} \| \|e_{\mathrm{d}}\|) \\
        & + \bar{\sigma}^\mathrm{i} \| U_\mathrm{c}\| \| e_{\mathrm{h}} \| + \frac{1}{4} \bar{\sigma}^\mathrm{c} (\| U_\mathrm{i} - L_\mathrm{i} W_\mathrm{y} \| \|e_{\mathrm{h}}\| + \| L_\mathrm{i} \| \|e_{\mathrm{d}}\|)
    \end{split}
\end{equation*}
Ordering all terms one has
\begin{equation}    \label{eq:bound_ec_observer}
    \begin{split}
        & \| e_{\mathrm{c}}^+ \| \leq \hat{\bar{\sigma}}^\mathrm{f} \| e_{\mathrm{c}} \| \\
        & + \bigg( \frac{1}{4} \frac{\bar{\sigma}^\mathrm{i} \bar{\sigma}^\mathrm{c}}{1 - \bar{\sigma}^\mathrm{f}} \| U_\mathrm{f} - L_\mathrm{f} W_\mathrm{y} \| + \bar{\sigma}^\mathrm{i} \| U_\mathrm{c} \| \\
        & + \frac{1}{4} \bar{\sigma}^\mathrm{c} \| U_\mathrm{i} - L_\mathrm{i} W_\mathrm{y} \| \bigg) \| e_{\mathrm{h}} \| \\
        & + \left( \frac{1}{4} \frac{\bar{\sigma}^\mathrm{i} \bar{\sigma}^\mathrm{c}}{1 - \bar{\sigma}^\mathrm{f}} \| L_\mathrm{f} \| + \frac{1}{4} \bar{\sigma}^\mathrm{c} \| L_\mathrm{i} \| \right) \|e_{\mathrm{d}}\| \\ 
        & = \hat{\bar{\sigma}}^\mathrm{f} \| e_{\mathrm{c}} \| + \hat{\alpha} \| e_{\mathrm{h}} \| + \hat{\beta} \| e_{\mathrm{d}} \|
    \end{split}
\end{equation}
Consider the time evolution of $e_\mathrm{h}$: 
\begin{equation*}
    \begin{split}
         & e_{\mathrm{h}}^+ = h^+ - \hat{h}^+ \\
         & = \sigma(W_\mathrm{o} u + U_\mathrm{o} h + b_\mathrm{o}) \circ \tanh(c^+) \\
         & - \sigma(W_\mathrm{o} u + U_\mathrm{o} \hat{h} + b_\mathrm{o} + L_\mathrm{o} (y - \hat{y})) \circ \tanh(\hat{c}^+) \\
      \end{split}
\end{equation*}
\begin{equation*}
    \begin{split}
         & = \sigma(W_\mathrm{o} u + U_\mathrm{o} \hat{h} + b_\mathrm{o} + L_\mathrm{o} (y - \hat{y})) \\
         & \circ (\tanh(c^+) - \tanh(\hat{c}^+)) \\
         & + \tanh(c^+) \circ \Big(\sigma(W_\mathrm{o} u + U_\mathrm{o} h + b_\mathrm{o}) \\
         & - \sigma(W_\mathrm{o} u + U_\mathrm{o} \hat{h} + b_\mathrm{o} + L_\mathrm{o} (y - \hat{y}))\Big)
    \end{split}
\end{equation*}
Taking the norm of both sides, exploiting (\ref{eq:sigma_f_bound}), (\ref{eq:sigma_i_bound}), (\ref{eq:sigma_o_bound}), (\ref{eq:sigma_c_bound}), %(\ref{eq:sigma_x_def}), 
(\ref{eq:sigma_hat_f_bound}), (\ref{eq:sigma_hat_o_bound}), the definition of $\Bar{\sigma}^\mathrm{x}$, lipschitzianity of $\sigma(\cdot)$ and $\tanh(\cdot)$, one has that
\begin{equation*}
    \begin{split}
        &\| e_{\mathrm{h}}^+ \| \leq \hat{\bar{\sigma}}^\mathrm{o} \| c^+ - \hat{c}^+ \| \\
        & + \bar{\sigma}^\mathrm{x} \frac{1}{4} \| U_\mathrm{o} h - U_\mathrm{o} \hat{h} - L_\mathrm{o} W_\mathrm{y} (h - \hat{h}) - L_\mathrm{o} (d - \hat{d}) \| \\
        & \leq \hat{\bar{\sigma}}^\mathrm{o} \| e_{\mathrm{c}}^+ \| + \frac{1}{4} \bar{\sigma}^\mathrm{x} \| U_\mathrm{o} - L_\mathrm{o} W_\mathrm{y} \| \| e_{\mathrm{h}} \| + \frac{1}{4} \bar{\sigma}^\mathrm{x} \| L_\mathrm{o} \| \| e_{\mathrm{d}} \|
    \end{split}
\end{equation*}
By substituting $\| e_{\mathrm{c}}^+ \|$ with its upper bound (\ref{eq:bound_ec_observer}) and ordering all terms one has
\begin{equation} \label{eq:bound_eh_observer}
    \begin{split}
        \| e_{\mathrm{h}}^+ \| & \leq \hat{\bar{\sigma}}^\mathrm{o} \hat{\bar{\sigma}}^\mathrm{f} \| e_{\mathrm{c}} \| + \hat{\gamma} \| e_{\mathrm{h}} \| + (\hat{\bar{\sigma}}^\mathrm{o} \hat{\beta} + \frac{1}{4} \bar{\sigma}^\mathrm{x} \| L_\mathrm{o} \|) \| e_{d} \|
    \end{split}
\end{equation}

Consider the evolution in time of $e_\mathrm{d}$. First define $\hat{d}^+_{ns} = \hat{d} + L_\mathrm{d} (y - \hat{y})$, that is \eqref{eq:d_lstm_observer} without the saturation operator. Then
\begin{equation*}
    \begin{split}
        e_{\mathrm{d},ns}^+ &= d^+ - \hat{d}^+_{ns} \\
        % &= d + w - \hat{d} - L_d (y - \hat{y}) \\
        & = d + w - \hat{d} - L_\mathrm{d} W_\mathrm{y} (h - \hat{h}) - L_\mathrm{d} (d - \hat{d}) \\
        & = (I_{p} - L_\mathrm{d}) e_{\mathrm{d}} - L_\mathrm{d} W_\mathrm{y} e_{\mathrm{h}} + w
    \end{split}
\end{equation*}
By taking the norm of both sides one has that
\begin{equation}    \label{eq:bound_ed_observer}
    \begin{split}
        & \| e_{\mathrm{d},ns}^+ \| \leq \| L_\mathrm{d} W_\mathrm{y} \| \| e_{\mathrm{h}} \| + \|I_{p} - L_\mathrm{d}\| \| e_{\mathrm{d}} \| + \| w \|
    \end{split}
\end{equation}
Moreover, noting that $\| e_\mathrm{d}^+ \| \leq \| e_{\mathrm{d},ns}^+ \|$, 
it is possible to write in matrix form (\ref{eq:bound_ec_observer}), (\ref{eq:bound_eh_observer}) and (\ref{eq:bound_ed_observer}), obtaining
\begin{equation}    \label{eq:observer_convergence}
    \begin{bmatrix}
        \| e_{\mathrm{c}}^+ \| \\
        \| e_{\mathrm{h}}^+ \| \\
        \| e_{\mathrm{d}}^+ \|
    \end{bmatrix}
    \leq A_d
    \begin{bmatrix}
        \| e_{\mathrm{c}} \| \\
        \| e_{\mathrm{h}} \| \\
        \| e_{\mathrm{d}} \|
    \end{bmatrix} + 
    \begin{bmatrix}
        0 \\
        0 \\
        1
    \end{bmatrix} \| w \|
\end{equation}
where $A_\mathrm{d}$ is defined in (\ref{eq:Ad_observer}).

\textbf{Part 2: }
Condition \eqref{eq:lyap_observer_pos_def} is easily verified for the definition of $V_\mathrm{o}$.

Proof of \eqref{eq:lyap_observer_neg_def} is shown in the following:
\begin{equation*}
    \begin{split}
         &V_{\mathrm{o}}(\hat{\chi}^+, \chi^+) = \norm{\begin{bmatrix}
            \| \hat{c}^+ - c^+ \| \\
            \| \hat{h}^+ - h^+ \| \\
            \| \hat{d}^+ - d^+ \| 
         \end{bmatrix}}_{P_\mathrm{o}} \\
         & \stackrel{\eqref{eq:observer_convergence}}{\leq} \norm{ A_\mathrm{d} \begin{bmatrix}
            \| \hat{c} - c \| \\
            \| \hat{h} - h \| \\
            \| \hat{d} - d \| 
         \end{bmatrix} + \begin{bmatrix}
             0 \\
             0 \\
             1
         \end{bmatrix} \norm{w}}_{P_\mathrm{o}} \\
         & \leq \norm{ A_\mathrm{d} \begin{bmatrix}
            \| \hat{c} - c \| \\
            \| \hat{h} - h \| \\
            \| \hat{d} - d \| 
         \end{bmatrix}}_{P_\mathrm{o}} + \norm{\begin{bmatrix}
             0 \\
             0 \\
             1
         \end{bmatrix} \norm{w}}_{P_\mathrm{o}}
    \end{split}
\end{equation*}

The two terms of this sum are now considered separately.

For the first term one has that 
\begin{equation*}
    \begin{split}
    &\norm{ A_\mathrm{d} \begin{bmatrix}
            \| \hat{c} - c \| \\
            \| \hat{h} - h \| \\
            \| \hat{d} - d \| 
         \end{bmatrix}}_{P_\mathrm{o}} = \sqrt{\begin{bmatrix}
        \| \hat{c} - c \| \\
        \| \hat{h} - h \| \\
        \| \hat{d} - d \| 
    \end{bmatrix}^\top A_\mathrm{d}^\top P_\mathrm{o} A_\mathrm{d} \begin{bmatrix}
        \| \hat{c} - c \| \\
        \| \hat{h} - h \| \\
        \| \hat{d} - d \| 
    \end{bmatrix}}  \\
    & = \sqrt{\begin{bmatrix}
        \| \hat{c} - c \| \\
        \| \hat{h} - h \| \\
        \| \hat{d} - d \| 
    \end{bmatrix}^\top (P_\mathrm{o} - Q_\mathrm{o}) \begin{bmatrix}
        \| \hat{c} - c \| \\
        \| \hat{h} - h \| \\
        \| \hat{d} - d \| 
    \end{bmatrix}} \\
    & \leq \sqrt{\norm{\begin{bmatrix}
        \| \hat{c} - c \| \\
        \| \hat{h} - h \| \\
        \| \hat{d} - d \| 
    \end{bmatrix}}^2_{P_\mathrm{o}} - \lambda_{min}(Q_\mathrm{o}) \norm{
    \begin{bmatrix}
        \| \hat{c} - c \| \\
        \| \hat{h} - h \| \\
        \| \hat{d} - d \| 
    \end{bmatrix}}^2} \\
    & \leq \sqrt{\left( 1 - \frac{\lambda_{min}(Q_\mathrm{o})}{\lambda_{max}(P_\mathrm{o})} \right) \norm{
    \begin{bmatrix}
        \| \hat{c} - c \| \\
        \| \hat{h} - h \| \\
        \| \hat{d} - d \| 
    \end{bmatrix}}^2_{P_\mathrm{o}} } \\
    & = \sqrt{1 - \frac{\lambda_{min}(Q_\mathrm{o})}{\lambda_{max}(P_\mathrm{o})}} V_\mathrm{o}(\hat{\chi}, \chi) = \rho_\mathrm{o} V_\mathrm{o}(\hat{\chi}, \chi)
    \end{split}
\end{equation*}

For the second term, in view of boundedness of set $\mathcal{W}$, one has that there exists a finite constant $\Bar{w}$
such that for all $w \in \mathcal{W}$
\begin{equation*}
\begin{split}
    \norm{\begin{bmatrix}
             0 \\
             0 \\
             1
         \end{bmatrix} \norm{w}}_{P_\mathrm{o}} \leq \sqrt{P_{\mathrm{o}(3,3)}} w_{max} = \Bar{w}
\end{split}
\end{equation*}

It follows that 
\begin{equation*}
    %V_{o}(\hat{\chi}^+, \chi^+) \leq \sqrt{1 - \frac{\lambda_{min}(Q_o)}{\lambda_{max}(P_o)}} V_o(\hat{\chi}, \chi) + \Bar{w}
    V_{\mathrm{o}}(\hat{\chi}^+, \chi^+) \leq \rho_\mathrm{o} V_\mathrm{o}(\hat{\chi}, \chi) + \Bar{w}
\end{equation*}

Condition \eqref{eq:lyap_observer_constraint} is now verified. Consider the $j$-th row of $|W_\mathrm{y} (h - \hat{h}) + (d - \hat{d})|$:
\begin{equation*}
    \begin{split}
        &|W_{\mathrm{y}(j*)} (h - \hat{h}) + (d_{(j)} - \hat{d}_{(j)})| \\
        &= \abs{[\mathbf{0}_{1,n} \;  W_{\mathrm{y}(j*)}  \; I_{p(j*)}] \begin{bmatrix}
            c - \hat{c} \\
            h - \hat{h} \\
            d - \hat{d}
        \end{bmatrix}} \\
        & \leq \abs{\left[0 \;  \norm{W_{\mathrm{y}(j*)}} \; 1\right] \begin{bmatrix}
            \| c - \hat{c} \| \\
            \| h - \hat{h} \| \\
            \| d - \hat{d} \|
        \end{bmatrix}} \\
        & = \abs{\left[0 \;  \norm{W_{\mathrm{y}(j*)}} \; 1\right] P_\mathrm{o}^{-1/2} P_\mathrm{o}^{1/2} \begin{bmatrix}
            \| c - \hat{c} \| \\
            \| h - \hat{h} \| \\
            \| d - \hat{d} \|
        \end{bmatrix}} \\
        & \leq \norm{\left[0 \;  \norm{W_{\mathrm{y}(j*)}} \; 1\right] P_\mathrm{o}^{-1/2}} \cdot \norm{P_\mathrm{o}^{1/2} \begin{bmatrix}
            \| c - \hat{c} \| \\
            \| h - \hat{h} \| \\
            \| d - \hat{d} \|
        \end{bmatrix}} \\
        & = c_{\mathrm{o}(j)} V_\mathrm{o} (\hat{\chi}, \chi)
    \end{split}
\end{equation*}

Finally, to study condition \eqref{eq:lyap_observer_max_gain},
first note that 
\begin{equation*}
    \norm{\hat{\chi}^+ - f_{aug}(\hat{\chi}, u)} = \norm{\begin{bmatrix}
        \| \hat{c}^+ - f_\mathrm{c}(\hat{\chi}, u)\| \\
        \| \hat{h}^+ - f_\mathrm{h}(\hat{\chi}, u)\| \\
        \| \hat{d}^+ - f_\mathrm{d}(\hat{\chi}, u) \|
    \end{bmatrix} }
\end{equation*}
where $f_\mathrm{c}$ and  $f_\mathrm{h}$ represent Equations \eqref{eq:c_augmented_lstm} and \eqref{eq:h_augmented_lstm} respectively, and $f_\mathrm{d}$ represents Equation \eqref{eq:d_augmented_lstm} without the term $w_k$.

Consider $\hat{c}^+ - f_\mathrm{c}(\hat{\chi}, u)$:
\begin{equation*}
    \begin{split}
       & \hat{c}^+ - f_\mathrm{c}(\hat{\chi}, u) = \\
       & = \sigma(W_\mathrm{f} u + U_\mathrm{f} \hat{h} + b_\mathrm{f} + L_\mathrm{f} (y - \hat{y})) \circ \hat{c} \\
       &+ \sigma(W_\mathrm{i} u + U_\mathrm{i} \hat{h} + b_\mathrm{i} + L_\mathrm{i} (y - \hat{y})) \\
       & \circ \tanh(W_\mathrm{c} u + U_\mathrm{c} \hat{h} + b_\mathrm{c}) \\
       &- \left( \sigma(W_\mathrm{f} u + U_\mathrm{f} \hat{h} + b_\mathrm{f}) \circ \hat{c} \right. \\
       & + \left. \sigma(W_\mathrm{i} u + U_\mathrm{i} \hat{h} + b_\mathrm{i}) \circ \tanh(W_\mathrm{c} u + U_\mathrm{c} \hat{h} + b_\mathrm{c}) \right) \\
       & = \left( \sigma(W_\mathrm{f} u + U_\mathrm{f} \hat{h} + b_\mathrm{f} + L_\mathrm{f} (y - \hat{y})) \right. \\
       & - \left. \sigma(W_\mathrm{f} u + U_\mathrm{f} \hat{h} + b_\mathrm{f})  \right) \circ \hat{c} \\
       & + \left( \sigma(W_\mathrm{i} u + U_\mathrm{i} \hat{h} + b_\mathrm{i} + L_\mathrm{i} (y - \hat{y})) \right. \\
       & - \left. \sigma(W_\mathrm{i} u + U_\mathrm{i} \hat{h} + b_\mathrm{i})  \right) \\
       & \circ \tanh(W_\mathrm{c} u + U_\mathrm{c} \hat{h} + b_\mathrm{c})
    \end{split}  
\end{equation*}

Taking the norm of both sides, exploiting the positive invariance of the set $\hat{\mathcal{I}}$, the upper bound \eqref{eq:sigma_c_bound}, and lipscitzianity of $\sigma(\cdot)$ and $\tanh(\cdot)$ one has
\begin{equation}    \label{eq:observer_gain_c}
    \begin{split}
        & \norm{\hat{c}^+ - f_\mathrm{c}(\hat{\chi}, u)}  \leq \frac{1}{4} \norm{L_\mathrm{f} (y - \hat{y})} \norm{\hat{c}} \\
        & + \frac{1}{4} \norm{L_\mathrm{i} (y - \hat{y})} \norm{\tanh(W_\mathrm{c} u + U_\mathrm{c} \hat{h} + b_\mathrm{c})} \\
        & \leq \frac{1}{4} \frac{\hat{\bar{\sigma}}^\mathrm{i} \bar{\sigma}^\mathrm{c}}{1 - \hat{\bar{\sigma}}^\mathrm{f}} \norm{L_\mathrm{f} W_\mathrm{y}} \norm{h - \hat{h}} + \frac{1}{4} \norm{L_\mathrm{f}} \frac{\hat{\bar{\sigma}}^\mathrm{i} \bar{\sigma}^\mathrm{c}}{1 - \hat{\bar{\sigma}}^\mathrm{f}} \norm{d - \hat{d}} \\
        & + \frac{1}{4} \bar{\sigma}^\mathrm{c} \norm{L_\mathrm{i} W_\mathrm{y}} \norm{h - \hat{h}} + \frac{1}{4} \bar{\sigma}^\mathrm{c} \norm{L_\mathrm{i}} \norm{d - \hat{d}} \\
        & = \left( \frac{1}{4} \frac{\hat{\bar{\sigma}}^\mathrm{i} \bar{\sigma}^\mathrm{c}}{1 - \hat{\bar{\sigma}}^\mathrm{f}} \norm{L_\mathrm{f} W_\mathrm{y}} +  \frac{1}{4} \bar{\sigma}^\mathrm{c} \norm{L_\mathrm{i} W_\mathrm{y}}\right) \norm{h - \hat{h}} \\
        & + \left( \frac{1}{4} \norm{L_\mathrm{f}} \frac{\hat{\bar{\sigma}}^\mathrm{i} \bar{\sigma}^\mathrm{c}}{1 - \hat{\bar{\sigma}}^\mathrm{f}} + \frac{1}{4} \bar{\sigma}^\mathrm{c} \norm{L_\mathrm{i}} \right) \norm{d - \hat{d}}
    \end{split}
\end{equation}

Consider $\hat{h}^+ - f_h(\hat{\chi}, u)$: 
\begin{equation*}
    \begin{split}
        & \hat{h}^+ - f_\mathrm{h}(\hat{\chi}, u) = \\
        & = \sigma(W_\mathrm{o} u + U_\mathrm{o} \hat{h} + b_\mathrm{o} + L_\mathrm{o} (y - \hat{y})) \circ \tanh(\hat{c}^+) \\
        & - \left( \sigma(W_\mathrm{o} u + U_\mathrm{o} \hat{h} + b_\mathrm{o}) \circ \tanh(f_\mathrm{c}(\hat{\chi}, u)) \right) \\
        & = \Big( \sigma(W_\mathrm{o} u + U_\mathrm{o} \hat{h} + b_\mathrm{o} + L_\mathrm{o} (y - \hat{y})) \\
        & - \sigma(W_\mathrm{o} u + U_\mathrm{o} \hat{h} + b_\mathrm{o}) \Big) \circ \tanh(\hat{c}^+) \\
        & + \left( \tanh(\hat{c}^+) - \tanh(f_\mathrm{c}(\hat{\chi}, u)) \right) \circ \sigma(W_\mathrm{o} u + U_\mathrm{o} \hat{h} + b_\mathrm{o})
    \end{split}
\end{equation*}
Taking the norm of both sizes, exploiting the positive invariance of the set $\hat{\mathcal{I}}$, the upper bound \eqref{eq:sigma_o_bound}, lipschitzianity of $\sigma(\cdot)$ and $\tanh(\cdot)$, and by substituting \eqref{eq:observer_gain_c} one has
\begin{equation*}   %\label{eq:observer_gain_h}
    \begin{split}
        & \norm{\hat{h}^+ - f_\mathrm{h}(\hat{\chi}, u)} \leq \\
        & \leq \frac{1}{4} \norm{L_\mathrm{o} (y - \hat{y})} \tanh \left( \frac{\hat{\bar{\sigma}}^\mathrm{i} \bar{\sigma}^\mathrm{c}}{1 - \hat{\bar{\sigma}}^\mathrm{f}} \right) + \norm{\hat{c}^+ - f_\mathrm{c}(\hat{\chi}, u)} \Bar{\sigma}^\mathrm{o} \\
     \end{split}
\end{equation*}
\begin{equation} \label{eq:observer_gain_h}
    \begin{split}
        & \leq \frac{1}{4} \tanh \left( \frac{\hat{\bar{\sigma}}^\mathrm{i} \bar{\sigma}^\mathrm{c}}{1 - \hat{\bar{\sigma}}^\mathrm{f}} \right) \norm{L_\mathrm{o} W_\mathrm{y}} \norm{h - \hat{h}} \\
        & + \frac{1}{4} \tanh \left( \frac{\hat{\bar{\sigma}}^\mathrm{i} \bar{\sigma}^\mathrm{c}}{1 - \hat{\bar{\sigma}}^\mathrm{f}} \right) \norm{L_\mathrm{o}} \norm{d - \hat{d}} \\
        & + \Bar{\sigma}^\mathrm{o} \left( \frac{1}{4} \frac{\hat{\bar{\sigma}}^\mathrm{i} \bar{\sigma}^\mathrm{c}}{1 - \hat{\bar{\sigma}}^\mathrm{f}} \norm{L_\mathrm{f} W_\mathrm{y}} +  \frac{1}{4} \bar{\sigma}^\mathrm{c} \norm{L_\mathrm{i} W_\mathrm{y}}\right) \norm{h - \hat{h}} \\
        & + \Bar{\sigma}^\mathrm{o} \left( \frac{1}{4} \norm{L_\mathrm{f}} \frac{\hat{\bar{\sigma}}^\mathrm{f} \bar{\sigma}^\mathrm{c}}{1 - \hat{\bar{\sigma}}^\mathrm{f}} + \frac{1}{4} \bar{\sigma}^\mathrm{c} \norm{L_\mathrm{i}} \right) \norm{d - \hat{d}}
    \end{split}
\end{equation}

Consider $\hat{d}^+ - f_\mathrm{d}(\hat{\chi}, u)$:
\begin{equation}   \label{eq:observer_gain_d}
\begin{split}
    & \norm{\hat{d}^+ - f_\mathrm{d}(\hat{\chi}, u)} = \norm{sat(\hat{d} + L_\mathrm{d} (y - \hat{y}), d_{max}) - \hat{d}}  \\
    & \leq \norm{\hat{d} + L_\mathrm{d} (y - \hat{y}) - \hat{d}}\\
    & = \norm{L_\mathrm{d} (y - \hat{y})} \leq \norm{L_\mathrm{d} W_\mathrm{y}} \norm{h - \hat{h}} + \norm{L_\mathrm{d}} \norm{d - \hat{d}}
\end{split}
\end{equation}

Combining \eqref{eq:observer_gain_c}, \eqref{eq:observer_gain_h} and \eqref{eq:observer_gain_d}, one has that
\begin{equation}    \label{eq:inequality_L_bar}
    \begin{bmatrix}
        \| \hat{c}^+ - f_\mathrm{c}(\hat{\chi}, u)\| \\
        \| \hat{h}^+ - f_\mathrm{h}(\hat{\chi}, u)\| \\
        \| \hat{d}^+ - f_\mathrm{d}(\hat{\chi}, u) \|
    \end{bmatrix} 
    \leq L
    \begin{bmatrix}
        \| c - \hat{c}\| \\
        \| h - \hat{h}\| \\
        \| d - \hat{d}\|
    \end{bmatrix}
\end{equation}
where $L$ is a matrix that can be obtained from \eqref{eq:observer_gain_c}-\eqref{eq:observer_gain_h}-\eqref{eq:observer_gain_d}.
In particular
\begin{equation*}
    L = \begin{bmatrix}
        0 & \Bar{\alpha} & \Bar{\beta} \\
        0 & \Bar{\gamma} \norm{L_\mathrm{o} W_\mathrm{y}} + \Bar{\sigma}^\mathrm{o} \Bar{\alpha} & \Bar{\gamma} \norm{L_\mathrm{o}} + \Bar{\sigma}^\mathrm{o} \Bar{\beta} \\
        0 & \norm{L_\mathrm{d} W_\mathrm{y}} & \norm{L_\mathrm{d}}
    \end{bmatrix}
\end{equation*}
with
\begin{equation*}
    \Bar{\alpha} = \frac{1}{4} \frac{\hat{\bar{\sigma}}^\mathrm{i} \bar{\sigma}^\mathrm{c}}{1 - \hat{\bar{\sigma}}^\mathrm{f}} \norm{L_\mathrm{f} W_\mathrm{y}} +  \frac{1}{4} \bar{\sigma}^\mathrm{c} \norm{L_\mathrm{i} W_\mathrm{y}}
\end{equation*}
\begin{equation*}
    \Bar{\beta} = \frac{1}{4} \norm{L_\mathrm{f}} \frac{\hat{\bar{\sigma}}^\mathrm{i} \bar{\sigma}^\mathrm{c}}{1 - \hat{\bar{\sigma}}^\mathrm{f}} + \frac{1}{4} \bar{\sigma}^\mathrm{c} \norm{L_\mathrm{i}}
\end{equation*}
\begin{equation*}
    \Bar{\gamma} = \frac{1}{4} \tanh \left( \frac{\hat{\bar{\sigma}}^\mathrm{i} \bar{\sigma}^\mathrm{c}}{1 - \hat{\bar{\sigma}}^\mathrm{f}} \right)
\end{equation*}
%\end {comment}
Taking the norm of both sides of \eqref{eq:inequality_L_bar}, one has
\begin{equation*}
    \begin{split}
        & \norm{\hat{\chi}^+ - f_{aug}(\hat{\chi}, u)} \leq \norm{L
    \begin{bmatrix}
        \| c - \hat{c}\| \\
        \| h - \hat{h}\| \\
        \| d - \hat{d}\|
    \end{bmatrix}} \\
    & = \norm{L P_\mathrm{o}^{-1/2} P_\mathrm{o}^{1/2}
    \begin{bmatrix}
        \| c - \hat{c}\| \\
        \| h - \hat{h}\| \\
        \| d - \hat{d}\|
    \end{bmatrix}} \leq \norm{L P_\mathrm{o}^{-1/2}} V_\mathrm{o}(\hat{\chi}, \chi)
    \end{split}
\end{equation*}
so that Equation \eqref{eq:lyap_observer_max_gain} is verified with $L_{max} = \norm{L P_\mathrm{o}^{-1/2}}$.

\textbf{Part 3: } If $w_k \to 0$ for $k \to \infty$, then observer convergence immediately follows from \eqref{eq:observer_convergence}.
%\hfill$\square$

\subsection{Proof of Lemma \ref{lem:K_bar}} \label{sec:proof-K-bar}
Let's define $\hat{y}^0 = \hat{d} - y^0$ and $\zeta = [\Bar{x}^\top \; \Bar{u}^\top]^\top$.
Then the equilibrium problem \eqref{eq:reference_calculation} can be rewritten as the problem of finding the solution of $F(\hat{y}^0, \zeta) = 0$, with
\begin{equation*}
    F(\hat{y}^0, \zeta) = \begin{bmatrix}
        f(\bar{x}, \bar{u}) - \Bar{x} \\
        g(\bar{x})+  \hat{y}^0
    \end{bmatrix}
\end{equation*}

In this proof the function $\zeta = \gamma(\hat{y}^0)$ such that $F(\hat{y}^0, \gamma(\hat{y}^0)) = 0$ is studied.
Let's define the following Jacobian matrices of $F$ in the point $(\hat{y}^0, \gamma(\hat{y}^0))$:
\begin{equation*}
    J_{\zeta}(\hat{y}^0, \gamma(\hat{y}^0)) %= \frac{\partial F}{\partial \xi} (\hat{y}^0, \gamma(\hat{y}^0)) 
    = \begin{bmatrix}
        \frac{\partial}{\partial x} (f(\bar{x}, \bar{u}) - \bar{x}) & \frac{\partial}{\partial u} (f(\bar{x}, \bar{u}) - \bar{x}) \\
        \frac{\partial}{\partial x} (g(\bar{x}) +  \hat{y}^0 ) & \frac{\partial}{\partial u} (g(\bar{x})+  \hat{y}^0)
    \end{bmatrix}
\end{equation*}
\begin{equation*}
    J_{\hat{y}^0} (\hat{y}^0, \gamma(\hat{y}^0)) %= \frac{\partial F}{\partial \hat{y}^0} (\hat{y}^0, \gamma(\hat{y}^0)) 
    =  \begin{bmatrix}
        \frac{\partial}{\partial \hat{y}^0} (f(\bar{x}, \bar{u}) - \bar{x}) \\
        \frac{\partial}{\partial \hat{y}^0} (g(\bar{x})+  \hat{y}^0)
    \end{bmatrix} = \begin{bmatrix}
            \mathbf{0}_{2n,p} \\
            I_p
        \end{bmatrix}
\end{equation*}

Under Assumption \ref{ass:setpoint_and_jacobian},
the implicit function theorem states that 
\begin{equation*}
    \frac{\partial \gamma}{\partial \hat{y}^0} (\hat{y}^0) = - [J_{\zeta}(\hat{y}^0, \gamma(\hat{y}^0))]^{-1} J_{\hat{y}^0}(\hat{y}^0, \gamma(\hat{y}^0)) = K(\hat{y}^0)
\end{equation*}

Let's denote $K(\hat{y}^0) = [K_{\Bar{x}}(\hat{y}^0)^\top \; K_{\Bar{u}}(\hat{y}^0)^\top]^\top$, where $K_{\Bar{x}}(\hat{y}^0)$ are the first $2n$ rows of $K(\hat{y}^0)$ and $K_{\Bar{u}}(\hat{y}^0)$ are the remaining $m$ rows.
Then, denoting 
\begin{equation*}
    \Bar{K} = \max_{\hat{y}^0 : y^0 \in \mathcal{Y}^0, \hat{d} \in \mathcal{D}} \norm{K_{\Bar{x}}(\hat{y}^0)}
\end{equation*}
it is possible to obtain a relationship between the maximum variation of $\Bar{x}$ and the variation of $\hat{y}^0$:
\begin{equation*}
    \norm{\Bar{x}_{k+1} - \bar{x}_k} \leq \Bar{K} \norm{\hat{y}^0_{k+1} - \hat{y}^0_k}
\end{equation*}
Note that the maximum that defines $\Bar{K}$ is always finite in view of the boundedness of the sets $\mathcal{Y}^0$ and $\mathcal{D}$.

The variation of $\hat{y}^0 = \hat{d}-y^0$ can be given by a variation of the disturbance estimation $\hat{d}$ and/or by a variation of the set-point $y^0$. This two contributions can be separated, obtaining
\begin{equation}   \label{eq:max_variation_xbar}
    \norm{\Bar{x}_{k+1} - \bar{x}_k} \leq \Bar{K} \norm{\hat{d}_{k+1} - \hat{d}_k} + \Bar{K} \norm{y^0_{k+1} - y^0_k}
\end{equation}

The maximum variation of the disturbance estimation $\hat{d}$ depends on the maximum estimation error $\Bar{e}_{y}$ and on the observer gain $L_d$. In particular it holds that
\begin{equation}   \label{eq:max_variation_d_hat}
    \norm{\hat{d}_{k+1} - \hat{d}_{k}} \stackrel{\eqref{eq:d_lstm_observer}}{\leq} \norm{L_\mathrm{d} (y_k - \hat{y}_k)} \leq \norm{L_\mathrm{d}} \Bar{e}_y
\end{equation}
Then, combining \eqref{eq:max_variation_xbar} and \eqref{eq:max_variation_d_hat}, \eqref{eq:max_setpoint_variation_y_variable} is proven.
%\hfill $\square$

\subsection{Proof of Theorem \ref{th:feasibility_stability}} \label{sec:proof-feasibility-stability}
The proof is divided in 3 parts:  
\begin{enumerate}
    \item Proof of satisfaction of constraints \eqref{eq:input_saturation} and \eqref{eq:output_constraint}, given that $(\psi,d, y^0) \in \mathcal{X}^{MPC}$.
    \item Proof of recursive feasibility, that is divided in:
    \begin{enumerate}
        \item Show that if $[c^\top \; h^\top]^\top \in \mathcal{X}_f(k)$ then $ W_\mathrm{y} h + b_\mathrm{y} + d_{max} \mathbf{1}_{p} \leq y_{max} - a_{N} \hat{e}_{\mathrm{o},k} - b_N$ and $W_\mathrm{y} h + b_\mathrm{y} - d_{max} \mathbf{1}_{p} \geq y_{min} + a_{N} \hat{e}_{\mathrm{o},k} + b_N$;
        \item Show that the candidate solution satisfies \eqref{eq:optimization_constraint_yub} and \eqref{eq:optimization_constraint_ylb};
        \item Show that the candidate solution satisfies the terminal constraint \eqref{eq:optimization_terminal_constraint}.
    \end{enumerate}
    \item Proof of ISpS and convergence, that is divided in:
    \begin{enumerate}
        \item Proof of ISpS;
        \item Proof of convergence.
    \end{enumerate}
\end{enumerate}

\textbf{Part 1: } If $\psi \in \mathcal{X}^{MPC}$, the satisfaction of the input saturation \eqref{eq:input_saturation} follows from constraint \eqref{eq:optimization_input_constraint} of the FHOCP formulation. Also the satisfaction of the output constraint \eqref{eq:output_constraint} is guaranteed by the control design. In particular $y_k \leq y_{max}$ follows from
\begin{equation*}
    \begin{split}
        &y_k = W_\mathrm{y} h_k + b_\mathrm{y} + d_k \stackrel{\eqref{eq:lyap_observer_constraint}}{\leq} W_\mathrm{y} \hat{h}_k + b_\mathrm{y} + \hat{d}_k \\
        & +  c_\mathrm{o} V_\mathrm{o}(\hat{\chi}_k, \chi_k) \leq W_\mathrm{y} h_{0|k} + b_\mathrm{y} + d_{max} \mathbf{1}_p + c_\mathrm{o} \hat{e}_{\mathrm{o},k} \\
        & \stackrel{\eqref{eq:initialization_a_ji} \eqref{eq:optimization_constraint_yub}}{\leq} y_{max} - a_0  \hat{e}_{\mathrm{o},k} + a_0 \hat{e}_{\mathrm{o},k} = y_{max}
    \end{split}
\end{equation*}
The fact that $y_k \geq y_{min}$ can be proven in a similar way:
\begin{equation*}
    \begin{split}
        &y_k = W_\mathrm{y} h_k + b_\mathrm{y} + d_k \stackrel{\eqref{eq:lyap_observer_constraint}}{\geq} W_\mathrm{y} \hat{h}_k + b_\mathrm{y} + \hat{d}_k \\
        & - c_\mathrm{o} V_\mathrm{o}(\hat{\chi}_k, \chi_k)  \geq W_\mathrm{y} h_{0|k} + b_\mathrm{y} - d_{max} \mathbf{1}_p - c_\mathrm{o} \hat{e}_{\mathrm{o},k} \\
        & \stackrel{\eqref{eq:initialization_a_ji} \eqref{eq:optimization_constraint_ylb}}{\geq} y_{min} + a_0 \hat{e}_{\mathrm{o},k} - a_0 \hat{e}_{\mathrm{o},k} = y_{min}
    \end{split}
\end{equation*}

\textbf{Part 2a: } 
To verify that if $x \in \mathcal{X}_f(k)$ then the tightened output constraints are satisfied,
first note that if $x \in \mathcal{X}_f(k)$ one has that
\begin{equation}    \label{eq:bound_h_alpha}
\begin{split}
    & \abs{W_{\mathrm{y}(j*)} (h - \Bar{h}_k)} = \abs{[\mathbf{0}_{1,n} \; W_{\mathrm{y}(j*)}] \begin{bmatrix}
        c - \Bar{c}_k \\
        h - \Bar{h}_k
    \end{bmatrix}} \\
    & \leq \abs{[0 \; \norm{W_{\mathrm{y}(j*)}}]\begin{bmatrix}
        \norm{c - \Bar{c}_k} \\
        \norm{h - \Bar{h}_k}
    \end{bmatrix}} \\
    & = \abs{[0 \; \norm{W_{\mathrm{y}(j*)}}] P_\mathrm{f}^{-1/2} P_\mathrm{f}^{1/2} \begin{bmatrix}
        \norm{c - \Bar{c}_k} \\
        \norm{h - \Bar{h}_k}
    \end{bmatrix}} \\
    & \leq \norm{[0 \; \norm{W_{\mathrm{y}(j*)}}] P_\mathrm{f}^{-1/2}} \cdot \norm{\begin{bmatrix}
        \norm{c - \Bar{c}_k} \\
        \norm{h - \Bar{h}_k}
    \end{bmatrix}}_{P_\mathrm{f}} \\
    & \stackrel{\eqref{eq:terminal_constraint}}{\leq} \norm{[0 \; \norm{W_{\mathrm{y}(j*)}}] P_\mathrm{f}^{-1/2}} \alpha_k
\end{split} 
\end{equation}

Considering now the $j$-th row of $W_\mathrm{y} h + b_\mathrm{y} + d_{max} \mathbf{1}_{p}$ one has
\begin{equation*}
    \begin{split}
        & W_{\mathrm{y}(j*)} h + b_{\mathrm{y}(j)} + d_{max} \\
        & = W_{\mathrm{y}(j*)} (h - \Bar{h}_k) + W_{\mathrm{y}(j*)} \Bar{h}_k + b_{\mathrm{y}(j)} + \hat{d}_{k(j)} - \hat{d}_{k(j)} \\
        & + d_{max} \stackrel{\eqref{eq:reference_calculation_output}}{=} W_{\mathrm{y}(j*)} (h - \Bar{h}_k) + y^0_{k(j)} - \hat{d}_{k(j)} + d_{max} \\
        & \leq \abs{W_{\mathrm{y}(j*)} (h - \Bar{h}_k)} + y^0_{k(j)} + 2 d_{max} \\
        & \stackrel{\eqref{eq:bound_h_alpha}}{\leq} \norm{[0 \; \norm{W_{\mathrm{y}(j*)}}] P_\mathrm{f}^{-1/2}} \alpha_k + y^0_{k(j)} + 2 d_{max} \\
        & \stackrel{\eqref{eq:def_alpha}}{\leq} y_{max(j)} - a_{N(j)} \tilde{e}_{\mathrm{o},k} - b_{N(j)} \\
        & \stackrel{\eqref{eq:e_ok_bar}}{\leq} y_{max(j)} - a_{N(j)} \hat{e}_{\mathrm{o},k} - b_{N(j)}
    \end{split}
\end{equation*}

In a similar way it is possible to prove that $W_\mathrm{y} h + b_\mathrm{y} - d_{max} \mathbf{1}_{p} \geq  y_{min} + a_{N} \hat{e}_{\mathrm{o},k} + b_{N}$. In fact, considering the $j$-th row of $- W_\mathrm{y} h - b_\mathrm{y} + d_{max} \mathbf{1}_{p}$ one obtains that
\begin{equation*}
    \begin{split}
        & - W_{\mathrm{y}(j*)} h - b_{\mathrm{y}(j)} + d_{max} \\
        & = - W_{\mathrm{y}(j*)} (h - \Bar{h}_k) - W_{\mathrm{y}(j*)} \Bar{h}_k - b_{\mathrm{y}(j)} - \hat{d}_{k(j)} + \hat{d}_{k(j)} \\
        & + d_{max} \stackrel{\eqref{eq:reference_calculation_output}}{=} - W_{\mathrm{y}(j*)} (h - \Bar{h}_k) - y^0_{(j)} - \hat{d}_{k(j)} + d_{max} \\
        & \leq \abs{W_{\mathrm{y}(j*)} (h - \Bar{h}_k)} - y^0_{(j)} + 2 d_{max} \\
        & \stackrel{\eqref{eq:bound_h_alpha}}{\leq} \norm{[0 \; \norm{W_{\mathrm{y}(j*)}}] P_\mathrm{f}^{-1/2}} \alpha_k - y^0_{(j)} + 2 d_{max} \\
        & \stackrel{\eqref{eq:def_alpha}}{\leq} - y_{min(j)} - a_{N(j)} \tilde{e}_{\mathrm{o},k} - b_{N(j)} \\
        & \stackrel{\eqref{eq:e_ok_bar}}{\leq} - y_{min(j)} - a_{N(j)} \hat{e}_{\mathrm{o},k} - b_{N(j)}
    \end{split}
\end{equation*}

\textbf{Part 2b: } Given the optimal solution of the optimization problem at time-step $k$, that is $u^*_{0|k}, ..., u^*_{N-1|k}$, let's denote with $x^*_{0|k}, ..., x^*_{N-1|k}$ the associate state trajectory defined by $x^*_{i+1|k} = f(x^*_{i|k}, u^*_{i|k})$ with $x^*_{0|k} = \hat{x}_k$. Let's also define $x^*_{N+1|k} = f(x^*_{N|k}, \Bar{u}_{k+1})$.

Let's define $\Tilde{u}_{i|k+1}$ with $i = 0,...,N-1$ the candidate solution at time-step $k+1$, where $\tilde{u}_{i|k+1} = u^*_{i+1|k}$ for $i = 0,...,N-2$ and $\Tilde{u}_{N-1|k+1} = \Bar{u}_{k+1}$. Note that $\Bar{u}_{k+1} \in \mathcal{U}$ for Assumption \ref{ass:setpoint_and_jacobian}.
Consider also the associate trajectory $\Tilde{x}_{0|k+1}, ..., \Tilde{x}_{N|k+1}$ defined by $\Tilde{x}_{i+1|k+1} = f(\Tilde{x}_{i|k+1}, \Tilde{u}_{i|k+1})$ with $\Tilde{x}_{0|k+1} = \hat{x}_{k+1}$.

Preliminarily, note that thanks to Lemma \ref{lem:incremental_lyap} and Theorem \ref{th:observer} one has that
\begin{equation*}
    \begin{split}
        & V_\mathrm{s} (\Tilde{x}_{0|k+1}, x^*_{1|k}) \stackrel{\eqref{eq:incremental_lyap_pos_def}}{\leq} c_{\mathrm{s,u}} \norm{\Tilde{x}_{0|k+1} - x^*_{1|k}} \\
        & \leq c_{\mathrm{s,u}} \norm{\begin{bmatrix}
           \Tilde{x}_{0|k+1} \\
           \hat{d}_{k+1}
        \end{bmatrix} - \begin{bmatrix}
            x^*_{1|k} \\
            \hat{d}_k
        \end{bmatrix}} \\
        & \stackrel{\eqref{eq:lyap_observer_max_gain}}{\leq} c_{\mathrm{s,u}} L_{max} V_\mathrm{o} (\hat{\chi}_k, \chi_k)  \leq c_{\mathrm{s,u}} L_{max} \hat{e}_{\mathrm{o},k}
    \end{split}
\end{equation*}
and that using Lemma \ref{lem:incremental_lyap} and inequality \eqref{eq:incremental_lyap_neg_def} recursively it follows
\begin{equation}    \label{eq:bound_Vs}
    V_\mathrm{s} (\Tilde{x}_{i|k+1}, x^*_{i+1|k}) \leq c_{\mathrm{s,u}} \rho_\mathrm{s}^i L_{max} \hat{e}_{\mathrm{o},k}
\end{equation}
for $i = 0,...,N$.

Moreover, as shown in Part 2a of the proof, since $x^*_{N|k} \in \mathcal{X}_f(k)$
\begin{equation*}
    \begin{split}
        W_\mathrm{y} h^*_{N|k} + b_\mathrm{y} + d_{max} \mathbf{1}_p &\leq y_{max} - a_{N} \hat{e}_{\mathrm{o},k} - b_N\\
        W_\mathrm{y} h^*_{N|k} + b_\mathrm{y} - d_{max} \mathbf{1}_p &\geq y_{min} + a_{N} \hat{e}_{\mathrm{o},k} + b_N
    \end{split}
\end{equation*}

Hence, $\Tilde{x}_{i|k+1}$ satisfies \eqref{eq:optimization_constraint_yub} for $i = 0,...,N-1$ as shown in the following
\begin{equation*}
    \begin{split}
        & W_\mathrm{y} \Tilde{h}_{i|k+1} + b_\mathrm{y} + d_{max} \mathbf{1}_p \\
        & \stackrel{\eqref{eq:incremental_lyap_constraints}}{\leq} W_\mathrm{y} h^*_{i+1|k} +  c_\mathrm{s} V_\mathrm{s} (\Tilde{x}_{i|k+1}, x^*_{i+1|k}) + b_\mathrm{y} + d_{max} \mathbf{1}_p \\
        & \stackrel{\eqref{eq:optimization_constraint_yub} \eqref{eq:bound_Vs}}{\leq} y_{max} - a_{i+1} \hat{e}_{\mathrm{o},k} - b_{i+1} + c_{\mathrm{s,u}} \rho_\mathrm{s}^i L_{max} \hat{e}_{\mathrm{o},k} c_\mathrm{s} \\
        & \stackrel{\eqref{eq:recursion_a_ji} \eqref{eq:recursion_b_i}}{=} y_{max} - \rho_\mathrm{o} a_{i} \hat{e}_{\mathrm{o},k} - c_{\mathrm{s,u}} \rho_\mathrm{s}^i L_{max} \hat{e}_{\mathrm{o},k} c_\mathrm{s}  - b_i - a_i \Bar{w} \\
        & + c_{\mathrm{s,u}} \rho_\mathrm{s}^i L_{max} \hat{e}_{\mathrm{o},k} c_\mathrm{s}  \stackrel{\eqref{eq:eo_evolution}}{=} y_{max} - a_{i} \hat{e}_{\mathrm{o},k+1} - b_i
    \end{split}
\end{equation*}

In a similar way it follows that $\Tilde{x}_{i|k+1}$ also satisfies \eqref{eq:optimization_constraint_ylb} for $i = 0,...,N-1$:
\begin{equation*}
    \begin{split}
        & -W_\mathrm{y} \Tilde{h}_{i|k+1} - b_\mathrm{y} + d_{max} \mathbf{1}_p \\
        & \stackrel{\eqref{eq:incremental_lyap_constraints}}{\leq} -W_\mathrm{y} h^*_{i+1|k} + c_\mathrm{s} V_\mathrm{s} (\Tilde{x}_{i|k+1}, x^*_{i+1|k}) - b_\mathrm{y} + d_{max} \mathbf{1}_p \\
        & \stackrel{\eqref{eq:bound_Vs} \eqref{eq:optimization_constraint_ylb}}{\leq} -y_{min} - a_{i+1} \hat{e}_{\mathrm{o},k} - b_{i+1} + c_{\mathrm{s,u}} \rho_\mathrm{s}^i L_{max} \hat{e}_{\mathrm{o},k} c_\mathrm{s} \\
        & \stackrel{\eqref{eq:recursion_a_ji}}{=} -y_{min} - \rho_\mathrm{o} a_{i} \hat{e}_{\mathrm{o},k} - c_{\mathrm{s,u}} \rho_\mathrm{s}^i L_{max} \hat{e}_{\mathrm{o},k} c_\mathrm{s} - b_i - a_i \Bar{w} \\
        & + c_{\mathrm{s,u}} \rho_\mathrm{s}^i L_{max} \hat{e}_{\mathrm{o},k} c_\mathrm{s} \stackrel{\eqref{eq:eo_evolution}}{=} -y_{min} - a_{i} \hat{e}_{\mathrm{o},k+1} - b_i
    \end{split}
\end{equation*}

\textbf{Part 2c: } In this part it is proven that $\tilde{x}_{N|k+1} \in \mathcal{X}_f(k+1)$, i.e.
\begin{equation} \label{eq:x_tilde_in_Xf}
    \norm{\begin{bmatrix}
            \|  \tilde{c}_{N|k+1} - \Bar{c}_{k+1} \| \\
            \| \tilde{h}_{N|k+1} - \Bar{h}_{k+1} \|
        \end{bmatrix}}_{P_\mathrm{f}} \leq \alpha_{k+1}
\end{equation}
starting from the fact that $x^*_{N|k} \in \mathcal{X}_f(k)$, i.e.
\begin{equation}   \label{eq:x_star_in_Xf}
    \norm{\begin{bmatrix}
            \|  c^*_{N|k} - \Bar{c}_{k} \| \\
            \| h^*_{N|k} - \Bar{h}_{k} \|
        \end{bmatrix}}_{P_\mathrm{f}} \leq \alpha_{k}
\end{equation}
Comparing \eqref{eq:x_tilde_in_Xf} and \eqref{eq:x_star_in_Xf} it is apparent that both the sides of the inequalities change. Hence, an upper bound for the left hand side of \eqref{eq:x_tilde_in_Xf} and a lower bound for the variation of the right hand side are computed in the following.

Before deriving the upper bound for the left hand side, the following preliminary inequalities are derived:
\begin{equation}    \label{eq:preliminary1_part2c}
\begin{split}
    \norm{ \Tilde{x}_{N|k+1} - x^*_{N+1|k}} &\stackrel{\eqref{eq:incremental_lyap_pos_def}}{\leq} \frac{1}{c_{\mathrm{s,l}}} V_\mathrm{s}(\Tilde{x}_{N|k+1}, x^*_{N+1|k}) \\
    & \stackrel{\eqref{eq:bound_Vs}}{\leq} \frac{c_\mathrm{s,u}}{c_{\mathrm{s,l}}} \rho_\mathrm{s}^N L_{max} \hat{e}_{\mathrm{o},k}
\end{split}  
\end{equation}
and
\begin{equation}    \label{eq:preliminary2_part2c}
    \begin{split}
        \norm{\begin{bmatrix}
            \|  c^*_{N+1|k} - \Bar{c}_{k+1} \| \\
            \|  h^*_{N+1|k}- \Bar{h}_{k+1} \|
        \end{bmatrix}}_{P_f} \leq \rho_\mathrm{f} \norm{\begin{bmatrix}
            \|  c^*_{N|k} - \Bar{c}_{k+1} \| \\
            \|  h^*_{N|k}- \Bar{h}_{k+1} \|
        \end{bmatrix}}_{P_\mathrm{f}}
    \end{split}
\end{equation}
with $\rho_\mathrm{f} = \sqrt{1 - \frac{q}{\lambda_{max} (P_\mathrm{f})}}$. The proof of this inequality is similar to the proof of Equation \eqref{eq:incremental_lyap_neg_def} in Lemma \ref{lem:incremental_lyap}, using $Q_\mathrm{s} = q I_2$.

Then, the upper bound for the left hand side of \eqref{eq:x_tilde_in_Xf} is
\begin{equation}    \label{eq:upper_bound_part2c}
    \begin{split}
        & \norm{\begin{bmatrix}
            \|  \tilde{c}_{N|k+1} - \Bar{c}_{k+1} \| \\
            \| \tilde{h}_{N|k+1} - \Bar{h}_{k+1} \|
        \end{bmatrix}}_{P_\mathrm{f}} \leq \\
        &  \norm{\begin{bmatrix}
            \|  \tilde{c}_{N|k+1} - c^*_{N+1|k} \| \\
            \|  \tilde{h}_{N|k+1} - h^*_{N+1|k} \|
        \end{bmatrix}}_{P_\mathrm{f}} 
        + \norm{\begin{bmatrix}
            \|  c^*_{N+1|k} - \Bar{c}_{k+1} \| \\
            \|  h^*_{N+1|k}- \Bar{h}_{k+1} \|
        \end{bmatrix}}_{P_\mathrm{f}} \\
        & \stackrel{\eqref{eq:preliminary2_part2c}}{\leq} \sqrt{\lambda_{max}(P_\mathrm{f})} \norm{\begin{bmatrix}
            \|  \tilde{c}_{N|k+1} - c^*_{N+1|k} \| \\
            \|  \tilde{h}_{N|k+1} - h^*_{N+1|k} \|
        \end{bmatrix}} \\
        & + \rho_\mathrm{f} \norm{\begin{bmatrix}
            \|  c^*_{N|k} - \Bar{c}_{k+1} \| \\
            \|  h^*_{N|k}- \Bar{h}_{k+1} \|
        \end{bmatrix}}_{P_\mathrm{f}} \\
        & \leq \sqrt{\lambda_{max}(P_\mathrm{f})} \norm{ \Tilde{x}_{N|k+1} - x^*_{N+1|k} } \\
        & + \rho_\mathrm{f} \norm{\begin{bmatrix}
            \|  c^*_{N|k} - \Bar{c}_{k} \| \\
            \|  h^*_{N|k}- \Bar{h}_{k} \|
        \end{bmatrix}}_{P_\mathrm{f}} + \rho_f \norm{\begin{bmatrix}
            \|  \Bar{c}_k - \Bar{c}_{k+1} \| \\
            \|  \Bar{h}_k - \Bar{h}_{k+1} \|
        \end{bmatrix}}_{P_\mathrm{f}} \\
        & \stackrel{\eqref{eq:x_star_in_Xf} \eqref{eq:preliminary1_part2c}}{\leq} \sqrt{\lambda_{max}(P_\mathrm{f})} \bigg( \frac{c_{\mathrm{s,u}}}{c_{\mathrm{s,l}}} \rho_\mathrm{s}^N L_{max} \hat{e}_{\mathrm{o},k} \\
        & + \rho_\mathrm{f} \norm{\Bar{x}_{k+1} - \Bar{x}_k} \bigg)  + \rho_\mathrm{f} \alpha_k \\
        & \stackrel{\eqref{eq:max_setpoint_variation_y_variable}}{\leq} \sqrt{\lambda_{max}(P_\mathrm{f})} \left( \frac{c_{\mathrm{s,u}}}{c_{\mathrm{s,l}}} \rho_\mathrm{s}^N L_{max} \hat{e}_{\mathrm{o},k} + \rho_\mathrm{f} \Bar{K} \norm{L_\mathrm{d}} \Bar{e}_y  \right. \\
        & +  \rho_\mathrm{f} \Bar{K} \norm{y^0_{k+1} - y^0_k} \Big) + \rho_\mathrm{f} \alpha_k
    \end{split}
\end{equation}

Let's now compute a lower bound for the possible variation of $\alpha$ (right hand side of \eqref{eq:x_tilde_in_Xf} and \eqref{eq:x_star_in_Xf}): 
\begin{equation*}
    \begin{split}
        & \alpha_{k+1} - \alpha_k \\
        & \stackrel{\eqref{eq:def_alpha}}{=} \min_{j=1,...,p} \min \left\{ \alpha^{max}_{j,k+1}, \alpha^{min}_{j,k+1} \right\}  \\
        & - \min_{j=1,...,p} \min \left\{ \alpha^{max}_{j,k}, \alpha^{min}_{j,k} \right\} \\
        & \stackrel{\eqref{eq:inequality_minimums}}{\geq} \min_{j=1,...,p} \min \left\{ \alpha^{max}_{j,k+1} - \alpha^{max}_{j,k}, \alpha^{min}_{j,k+1} - \alpha^{min}_{j,k} \right\} \\
    \end{split}
\end{equation*}
Noting that
\begin{equation*}
    \begin{split}
        & \min \left\{ \alpha^{max}_{j,k+1} - \alpha^{max}_{j,k}, \alpha^{min}_{j,k+1} - \alpha^{min}_{j,k} \right\} \\
        & \stackrel{\eqref{eq:alpha_ub} \eqref{eq:alpha_lb}}{=} \min \left\{ - \norm{[0 \; \norm{W_{\mathrm{y}(j*)}}] P_\mathrm{f}^{-1/2}}^{-1} (y^0_{k+1(j)} - y^0_{k(j)}) \right. \\
        & - \norm{[0 \; \norm{W_{\mathrm{y}(j*)}}] P_\mathrm{f}^{-1/2}}^{-1} a_{N(j)} (\tilde{e}_{\mathrm{o},k+1} - \tilde{e}_{\mathrm{o},k}),   \\
        & \norm{[0 \; \norm{W_{\mathrm{y}(j*)}}] P_\mathrm{f}^{-1/2}}^{-1} (y^0_{k+1(j)} - y^0_{k(j)}) \\
        & - \left. \norm{[0 \; \norm{W_{\mathrm{y}(j*)}}] P_\mathrm{f}^{-1/2}}^{-1} a_{N(j)} (\tilde{e}_{\mathrm{o},k+1} - \tilde{e}_{\mathrm{o},k}) \right\} \\
        & = - \norm{[0 \; \norm{W_{\mathrm{y}(j*)}}] P_\mathrm{f}^{-1/2}}^{-1} \abs{y^0_{k+1(j)} - y^0_{k(j)}}  \\
        & + \norm{[0 \; \norm{W_{\mathrm{y}(j*)}}] P_\mathrm{f}^{-1/2}}^{-1} a_{N(j)} (\tilde{e}_{\mathrm{o},k} - \tilde{e}_{\mathrm{o},k+1})
    \end{split}
\end{equation*}
it is possible to derive that
\begin{equation}    \label{eq:lower_bound_part2c}
    \begin{split}
        & \alpha_k \leq \alpha_{k+1} \\
        & + \max_{j = 1,...,p} \left\{ \norm{[0 \; \norm{W_{\mathrm{y}(j*)}}] P_\mathrm{f}^{-1/2}}^{-1} \abs{y^0_{k+1(j)} - y^0_{k(j)}} \right\}\\
        & - \min_{j = 1,...,p} \left\{ \norm{[0 \; \norm{W_{\mathrm{y}(j*)}}] P_\mathrm{f}^{-1/2}}^{-1} a_{N(j)} (\tilde{e}_{\mathrm{o},k} - \tilde{e}_{\mathrm{o},k+1}) \right\} \\
        & \leq \alpha_{k+1} \\
        & + \max_{j = 1,...,p} \left\{ \norm{[0 \; \norm{W_{\mathrm{y}(j*)}}] P_\mathrm{f}^{-1/2}}^{-1} \abs{y^0_{k+1(j)} - y^0_{k(j)}} \right\} 
    \end{split}
\end{equation}

Moreover, from \eqref{eq:upper_bound_part2c} and \eqref{eq:lower_bound_part2c} it is possible to obtain
\begin{equation*}
    \begin{split}
        & \norm{\begin{bmatrix}
            \|  \tilde{c}_{N|k+1} - \Bar{c}_{k+1} \| \\
            \| \tilde{h}_{N|k+1} - \Bar{h}_{k+1} \|
        \end{bmatrix}}_{P_\mathrm{f}} \\
        & \stackrel{\eqref{eq:upper_bound_part2c}}{\leq} \sqrt{\lambda_{max}(P_\mathrm{f})} \bigg( \frac{c_{\mathrm{s,u}}}{c_{\mathrm{s,l}}} \rho_\mathrm{s}^N L_{max} \hat{e}_{\mathrm{o},k} + \rho_\mathrm{f} \Bar{K} \norm{L_\mathrm{d}} \Bar{e}_y \\
        & + \rho_\mathrm{f} \Bar{K} \norm{y^0_{k+1} - y^0_k} \bigg) + \rho_\mathrm{f} \alpha_k \\
        & \stackrel{\eqref{eq:lower_bound_part2c}}{\leq} \sqrt{\lambda_{max}(P_\mathrm{f})} \bigg( \frac{c_{\mathrm{s,u}}}{c_{\mathrm{s,l}}} \rho_\mathrm{s}^N L_{max} \hat{e}_{\mathrm{o},k} + \rho_\mathrm{f} \Bar{K} \norm{L_\mathrm{d}} \Bar{e}_y \\
        & + \rho_\mathrm{f} \Bar{K} \norm{y^0_{k+1} - y^0_k} \bigg) + (\rho_\mathrm{f} - 1) \alpha_{k+1} \\
        & + \rho_\mathrm{f} \max_{j = 1,...,p} \left\{ \norm{[0 \; \norm{W_{\mathrm{y}(j*)}}] P_\mathrm{f}^{-1/2}}^{-1} \abs{y^0_{k+1(j)} - y^0_{k(j)}} \right\}\\
        & + \alpha_{k+1}
    \end{split}
\end{equation*}
Finally, note that $\rho_\mathrm{f} < 1$ by definition and $\alpha_{k+1} > 0$ for all $k$ in view of Assumption \ref{ass:setpoint}. %and $\tilde{e}_{\mathrm{o},k} - \tilde{e}_{\mathrm{o},k+1} \geq 0$ because $\tilde{e}_{o}$ is always monotonically decreasing. 
Then there exist $\Bar{L}_{max} > 0$, $\Bar{L_\mathrm{d}} > 0$ and $\Delta y^0_{max} > 0$ such that for $L_{max} \leq \Bar{L}_{max}$, $\norm{L_\mathrm{d}} \leq \Bar{L_\mathrm{d}}$ and $y^0$ such that $\norm{y^0_{k+1} - y^0_k} \leq \Delta y^0_{max}$ 
\begin{equation*}
    \norm{\begin{bmatrix}
            \|  \tilde{c}_{N|k+1} - \Bar{c}_{k+1} \| \\
            \| \tilde{h}_{N|k+1} - \Bar{h}_{k+1} \|
        \end{bmatrix}}_{P_\mathrm{f}} \leq \alpha_{k+1}
\end{equation*}

\textbf{Part 3: }
To prove the ISpS and convergence properties of the closed-loop system a candidate Lyapunov function is introduced.
The candidate Lyapunov function has a structure similar to the optimal cost of the MPC, but it considers the asymptotic values of the state and input set-points instead of their values at time-step $k$. Note that these values cannot be used in the MPC cost function because are unknown. The expression of the candidate Lyapunov function is the following
\begin{equation*}
\begin{split}
    V_k & = \sum_{i=0}^{N-1} \left( \|x^*_{i|k} - \bar{x}_{\infty} \|_Q^2 + \|u^*_{i|k} - \bar{u}_{\infty} \|_R^2 \right) \\
    & + \norm{\begin{bmatrix}
        \| c^*_{N|k} - \bar{c}_{\infty} \| \\
        \| h^*_{N|k} - \bar{h}_{\infty} \|
    \end{bmatrix}}_{P_\mathrm{f}}^2 \\
\end{split}
\end{equation*}
where $u^*_{0|k}, ..., u^*_{N-1|k}$ and $x^*_{0|k}, ..., x^*_{N|k}$ are defined as in Part 2b of this proof, and $x^*_{N|k} = \left[(c^*_{N|k})^\top \; (h^*_{N|k})^\top \right]^\top$.
Note that $V_k$ is a function of $x = \hat{x}_k$, $\hat{e}_{\mathrm{o},k}$, $\bar{x}_k$ and $\Bar{u}_k$, %and $y^0_k$, 
because all these values are needed to compute the optimal input sequence $u^*_{0|k}, ..., u^*_{N-1|k}$. 

In Part 3a, a lower bound and an upper bound for $V_k$ and a bound on the variation of $V_k$ between subsequent time-steps are derived to prove ISpS. In Part 3b, convergence is shown by studying the asymptotic values of the bounds, under the assumption that $y^0_k \to y^0_\infty$ and $d_k \to \Bar{d}_\infty$ for $k \to \infty$.

\textbf{Part 3a: }
The lower bound for $V_k$ follows from
\begin{equation}   \label{eq:lower_bound_Vk}
    V_k \geq \|x^*_{0|k} - \bar{x}_{\infty} \|_Q^2 = \|\hat{x}_k - \bar{x}_{\infty} \|_Q^2 \geq \lambda_{min}(Q) \|\hat{x}_k - \bar{x}_{\infty} \|^2
\end{equation}

The upper bound for $V_k$ is now proven. 

Consider the possibly suboptimal control input $\tilde{u}_{i|k} = \bar{u}_k$ for all $i = 0, ..., N-1$, and denote with $\tilde{x}_{0|k}, ..., \tilde{x}_{N|k}$ the correspondent state trajectory with initial condition $\tilde{x}_{0|k} = \hat{x}_k$. 
Note that there exists $\mu > 0$ such that this control input is feasible at time-step $k$ for all $\hat{x}_k$ such that $\norm{\hat{x}_k - \Bar{x}_k} \leq \mu$.

Three different cases are considered:
\begin{enumerate}
    \item $\norm{\hat{x}_k - \Bar{x}_\infty} \leq \frac{\mu}{2}$ and $\norm{\Bar{x}_k - \Bar{x}_\infty} \leq \frac{\mu}{2}$;
    \item $\norm{\hat{x}_k - \Bar{x}_\infty} > \frac{\mu}{2}$;
    \item $\norm{\hat{x}_k - \Bar{x}_\infty} \leq \frac{\mu}{2}$ and $\norm{\Bar{x}_k - \Bar{x}_\infty} > \frac{\mu}{2}$.
\end{enumerate}

\textit{Case 1:} $\norm{\hat{x}_k - \Bar{x}_\infty} \leq \frac{\mu}{2}$ and $\norm{\Bar{x}_k - \Bar{x}_\infty} \leq \frac{\mu}{2}$.

In this case the sequence $\Tilde{u}_{i|k}$ is feasible. In fact
\begin{equation*}
    \norm{\hat{x}_k - \Bar{x}_k} \leq \norm{\hat{x}_k - \Bar{x}_\infty} + \norm{\bar{x}_k - \Bar{x}_\infty} \leq \frac{\mu}{2} + \frac{\mu}{2} = \mu
\end{equation*}

Let's introduce new variables for the difference between the set-point at step $k$ and the set-point for $k \to \infty$
\begin{equation*}
    \begin{split}
        \delta \bar{x}_k = \bar{x}_k - \bar{x}_{\infty}, \quad \delta \bar{u}_k = \bar{u}_k - \bar{u}_{\infty} \\
         \delta \bar{c}_k = \bar{c}_k - \bar{c}_{\infty}, \quad \delta \bar{h}_k = \bar{h}_k - \bar{h}_{\infty}
    \end{split}
\end{equation*}
Terms $\delta \bar{x}_k$, $\delta \bar{u}_k$, $\delta \bar{c}_k$, $\delta \bar{h}_k$ can be related to $\hat{d}_k - \Bar{d}_\infty$ and $y^0_k - y^0_\infty$ with a reasoning similar to the proof of Lemma \ref{lem:K_bar}. In particular, under Assumption \ref{ass:setpoint_and_jacobian}, there exist finite constants $\Bar{K}_{x}$ and $\Bar{K}_{u}$ such that
\begin{subequations}    \label{eq:bound-delta}
    \begin{align}
        \norm{\delta \bar{x}_k} &\leq \Bar{K}_{x} \norm{\hat{d}_k - \Bar{d}_\infty} + \Bar{K}_{x} \norm{y^0_k - y^0_\infty} \label{eq:bound-delta-x} \\
        \norm{\delta \bar{u}_k} &\leq \Bar{K}_{u} \norm{\hat{d}_k - \Bar{d}_\infty} + \Bar{K}_{u} \norm{y^0_k - y^0_\infty} \label{eq:bound-delta-u}
    \end{align}
\end{subequations}

Considering $V_k$, one has that
\begin{equation*}
    \begin{split}
        & V_k \leq \sum_{i=0}^{N-1} \left( \|x^*_{i|k} - \bar{x}_k \|_Q^2 + \| u^*_{i|k} - \bar{u}_k \|_R^2 \right) \\
        & + \norm{\begin{bmatrix}
            \| c^*_{N|k} - \bar{c}_k \| \\
            \| h^*_{N|k} - \bar{h}_k \|
        \end{bmatrix}}_{P_\mathrm{f}}^2 + \sum_{i=0}^{N-1} \Big( \| \delta \bar{x}_k \|_Q^2 + \| \delta \bar{u}_k \|_R^2  \\
        & +  2 (x^*_{i|k} - \bar{x}_k)^\top Q \delta \bar{x}_k  + 2 (u^*_{i|k} - \bar{u}_k)^\top R \delta \bar{u}_k \Big) \\
        & + \norm{\begin{bmatrix}
            \| \delta \bar{c}_k \| \\
            \| \delta \bar{h}_k \|
        \end{bmatrix}}_{P_\mathrm{f}}^2 + 2 \begin{bmatrix}
            \| c^*_{N|k} - \bar{c}_k \| \\
            \| h^*_{N|k} - \bar{h}_k \|
        \end{bmatrix}^\top P_\mathrm{f} \begin{bmatrix}
            \| \delta \bar{c}_k \| \\
            \| \delta \bar{h}_k \|
        \end{bmatrix} 
    \end{split}
\end{equation*}

This expression contains the optimal cost of the MPC optimization at step $k$, that is 
\begin{equation*}
\begin{split}
    J^*_k &= \sum_{i=0}^{N-1} \left( \|x^*_{i|k} - \bar{x}_k \|_Q^2 + \|u^*_{i|k} - \bar{u}_k \|_R^2 \right) \\
    & + \norm{\begin{bmatrix}
        \| c^*_{N|k} - \bar{c}_k \| \\
        \| h^*_{N|k} - \bar{h}_k \|
    \end{bmatrix}}_{P_\mathrm{f}}^2
\end{split}  
\end{equation*}

Consider now the cost associate to the suboptimal control input $\tilde{u}_{i|k}$, and note that it is greater or equal than the optimal cost, i.e.
\begin{equation*}
    J^*_k \leq \sum_{i=0}^{N-1} \| \tilde{x}_{i|k} - \bar{x}_k \|_Q^2 + \norm{\begin{bmatrix}
        \| \tilde{c}_{N|k} - \bar{c}_k \| \\
        \| \tilde{h}_{N|k} - \bar{h}_k \|
    \end{bmatrix}}_{P_\mathrm{f}}^2
\end{equation*}

It is now used that
\begin{equation*}
\begin{split}
    \norm{\begin{bmatrix}
        \| \tilde{c}_{N|k} - \bar{c}_k \| \\
        \| \tilde{h}_{N|k} - \bar{h}_k \|
    \end{bmatrix}}_{P_\mathrm{f}}^2 & \leq \lambda_{max}(P_\mathrm{f}) \norm{\begin{bmatrix}
        \| \tilde{c}_{N|k} - \bar{c}_k \| \\
        \| \tilde{h}_{N|k} - \bar{h}_k \|
    \end{bmatrix}}^2 \\
    & = \lambda_{max}(P_\mathrm{f}) \| \tilde{x}_{N|k} - \bar{x}_k \|^2
\end{split}
\end{equation*}
and that, in view of Assumption \ref{ass:delta_iss_condition} and Theorem \ref{th:delta_iss_condition}, there exist $\bar{\mu} \geq 0$ and $\lambda \in (0,1)$ such that $\forall i \geq 0$ 
\begin{equation*}
    \| \tilde{x}_{i|k} - \bar{x}_k \| \leq \bar{\mu} \lambda^i \| \tilde{x}_{0|k} - \bar{x}_k \| = \bar{\mu} \lambda^i \| \hat{x}_k - \bar{x}_k \|
\end{equation*}
to obtain that there exists a constant $\Tilde{b} \geq 0$ such that
\begin{equation*}
    \begin{split}
        J^*_k & \leq \Tilde{b} \| \hat{x}_k - \bar{x}_k \|^2 = \Tilde{b} \| \hat{x}_k - \bar{x}_{\infty} - \delta \bar{x}_k \|^2 \\
        & = \Tilde{b} \| \hat{x}_k - \bar{x}_{\infty} \|^2 + \Tilde{b} \| \delta \bar{x}_k \|^2 - 2\Tilde{b} (\hat{x}_k - \bar{x}_{\infty})^\top \delta \bar{x}_k
    \end{split}
\end{equation*}
Therefore %there exists a constant $\Tilde{b} \geq 0$ such that
\begin{equation}    \label{eq:upper_bound_Vk_caso3}
    \begin{split}
        V_k & \leq \Tilde{b} \| \hat{x}_k - \bar{x}_{\infty} \|^2 + \beta(k)
    \end{split}
\end{equation}
where
\begin{equation*}
    \begin{split}
        \beta(k) &= \sum_{i=0}^{N-1} \Big( \| \delta \bar{x}_k \|_Q^2 + \| \delta \bar{u}_k \|_R^2   \\
        & \left. + 2 (x^*_{i|k} - \bar{x}_k)^\top Q \delta \bar{x}_k + 2 (u^*_{i|k} - \bar{u}_k)^\top R \delta \bar{u}_k \right) \\
        & + \norm{\begin{bmatrix}
            \| \delta \bar{c}_k \| \\
            \| \delta \bar{h}_k \|
        \end{bmatrix}}_{P_\mathrm{f}}^2 + 2 \begin{bmatrix}
            \| c^*_{N|k} - \bar{c}_k \| \\
            \| h^*_{N|k} - \bar{h}_k \|
        \end{bmatrix}^\top P_\mathrm{f} \begin{bmatrix}
            \| \delta \bar{c}_k \| \\
            \| \delta \bar{h}_k \|
        \end{bmatrix} \\
        & + \Tilde{b} \| \delta \bar{x}_k \|^2 - 2\Tilde{b} (\hat{x}_k - \bar{x}_{\infty})^\top \delta \bar{x}_k
    \end{split}
\end{equation*}

In view of \eqref{eq:bound-delta} and boundedness of $\mathcal{D}$ and $\mathcal{Y}^0$, terms $\delta \bar{x}_k$, $\delta \bar{u}_k$, $\delta \bar{c}_k$ and $\delta \bar{h}_k$ are bounded for all $k \in \mathbb{Z}_{\geq 0}$. In addition, inputs are limited in $\mathcal{U}$ and states are limited in $\hat{\mathcal{C}} \times \mathcal{H}$. Then there exists a constant $\tilde{c}_1 \geq 0$ such that $\beta(k) \leq \tilde{c}_1$ for all $k \in \mathbb{Z}_{\geq 0}$.

\textit{Case 2:} $\norm{\hat{x}_k - \Bar{x}_\infty} > \frac{\mu}{2}$.

There exists $V_{max} > 0$ such that $V_k \leq V_{max}$ in $\mathcal{X}^{MPC}$. Then
\begin{equation}    \label{eq:upper_bound_Vk_caso1}
    V_k \leq \frac{4 V_{max}}{\mu^2} \norm{\hat{x}_k - \Bar{x}_\infty}^2
\end{equation}

\textit{Case 3:} $\norm{\hat{x}_k - \Bar{x}_\infty} \leq \frac{\mu}{2}$ and $\norm{\Bar{x}_k - \Bar{x}_\infty} > \frac{\mu}{2}$.

One has that
\begin{equation}    \label{eq:upper_bound_Vk_caso2}
    V_k \leq 0 \norm{\hat{x}_k - \Bar{x}_\infty}^2 + \tilde{\beta}(k)
\end{equation}
with $\tilde{\beta}(k) = V_k \leq V_{max}$.

Hence, in view of \eqref{eq:upper_bound_Vk_caso3}, \eqref{eq:upper_bound_Vk_caso1} and \eqref{eq:upper_bound_Vk_caso2},
\begin{equation} \label{eq:upper-bound-Vk}
    V_k \leq b \|\hat{x}_k - \bar{x}_{\infty} \|^2 + c_1
\end{equation}
with $ b = \max\left\{\Tilde{b}, \frac{4 V_{max}}{\mu^2}\right\}$ and $c_1 = \max \left\{ \Tilde{c}_1, V_{max} \right\}$.

Let's now study the variation of $V_k$ between subsequent time-steps.

Consider the Lyapunov function at time-step $k+1$:
\begin{equation*}
\begin{split}
    V_{k+1} &= \sum_{i=0}^{N-1} \left( \|x^*_{i|k+1} - \bar{x}_{\infty} \|_Q^2 + \|u^*_{i|k+1} - \bar{u}_{\infty} \|_R^2 \right) \\
    & + \norm{\begin{bmatrix}
            \| c^*_{N|k+1} - \Bar{c}_\infty \| \\
            \| h^*_{N|k+1} - \Bar{h}_\infty \|
        \end{bmatrix}}_{P_\mathrm{f}}^2
\end{split}
\end{equation*}
where $ u^*_{0|k+1},..., u^*_{N|k+1} $ is the optimal sequence given by the MPC at step $k+1$ and $x^*_{i+1|k+1} = f(x^*_{i|k+1}, u^*_{i|k+1})$ with $x^*_{0|k+1} = \hat{x}_{k+1} $. Note that in general $\hat{x}_{k+1}$ can be different from $x^*_{1|k}$.
It is possible to derive the following upper bound for $V_{k+1}$:
\begin{equation*}
    \begin{split}
        & V_{k+1} \leq \sum_{i=0}^{N-1} \left( \|x^*_{i|k+1} - \bar{x}_{k+1} \|_Q^2 + \| u^*_{i|k+1} - \bar{u}_{k+1} \|_R^2 \right) \\
        & + \norm{\begin{bmatrix}
            \| c^*_{N|k+1} - \bar{c}_{k+1} \| \\
            \| h^*_{N|k+1} - \bar{h}_{k+1} \|
        \end{bmatrix}}_{P_\mathrm{f}}^2 + \sum_{i=0}^{N-1} \Big( \| \delta \bar{x}_{k+1} \|_Q^2 \\
        & + \| \delta \bar{u}_{k+1} \|_R^2 + 2 (x^*_{i|k+1} - \bar{x}_{k+1})^\top Q \delta \bar{x}_{k+1} \\
        & + 2 (u^*_{i|k+1} - \bar{u}_{k+1})^\top R \delta \bar{u}_{k+1} \Big) + \norm{\begin{bmatrix}
            \| \delta \bar{c}_{k+1} \| \\
            \| \delta \bar{h}_{k+1} \|
        \end{bmatrix}}_{P_\mathrm{f}}^2 \\
        & + 2 \begin{bmatrix}
            \| c^*_{N|k+1} - \bar{c}_{k+1} \| \\
            \| h^*_{N|k+1} - \bar{h}_{k+1} \|
        \end{bmatrix}^\top P_\mathrm{f} \begin{bmatrix}
            \| \delta \bar{c}_{k+1} \| \\
            \| \delta \bar{h}_{k+1} \|
        \end{bmatrix}
    \end{split}
\end{equation*}
This expression contains the optimal cost of the MPC optimization at step $k+1$, that is 
\begin{equation*}
\begin{split}
    J^*_{k+1} &= \sum_{i=0}^{N-1} \left( \|x^*_{i|k+1} - \bar{x}_{k+1} \|_Q^2 + \| u^*_{i|k+1} - \bar{u}_{k+1} \|_R^2 \right) \\
    & + \norm{\begin{bmatrix}
            \| c^*_{N|k+1} - \bar{c}_{k+1} \| \\
            \| h^*_{N|k+1} - \bar{h}_{k+1} \|
        \end{bmatrix}}_{P_\mathrm{f}}^2
\end{split}
\end{equation*}

Consider now the feasible trajectory $\Tilde{u}_{i|k+1}$ for $i = 0, ..., N-1$ used in Part 2b of this proof.
Noting that the cost associated to the feasible trajectory is greater or equal than $J^*_{k+1}$, one has
\begin{equation*}
    \begin{split}
        & V_{k+1} \leq \sum_{i=0}^{N-2} \left( \| \tilde{x}_{i|k+1} - \bar{x}_{k+1} \|_Q^2 + \| u^*_{i+1|k} - \bar{u}_{k+1} \|_R^2 \right) \\
        & + \| \tilde{x}_{N-1|k+1} - \bar{x}_{k+1} \|_Q^2 + \| \bar{u}_{k+1} - \bar{u}_{k+1} \|_R^2 \\
        & + \norm{\begin{bmatrix}
            \| \tilde{c}_{N|k+1} - \bar{c}_{k+1} \| \\
            \| \tilde{h}_{N|k+1} - \bar{h}_{k+1} \|
        \end{bmatrix}}_{P_\mathrm{f}}^2 + N \| \delta \bar{x}_{k+1} \|_Q^2 + N \| \delta \bar{u}_{k+1} \|_R^2 \\
        & + \sum_{i=0}^{N-1} \left( 2 (x^*_{i|k+1} - \bar{x}_{k+1})^\top Q \delta \bar{x}_{k+1} \right. \\
        & + \left. 2 (u^*_{i|k+1} - \bar{u}_{k+1})^\top R \delta \bar{u}_{k+1} \right) + \norm{\begin{bmatrix}
            \| \delta \bar{c}_{k+1} \| \\
            \| \delta \bar{h}_{k+1} \|
        \end{bmatrix}}_{P_\mathrm{f}}^2 \\
        & + 2 \begin{bmatrix}
            \| c^*_{N|k+1} - \bar{c}_{k+1} \| \\
            \| h^*_{N|k+1} - \bar{h}_{k+1} \|
        \end{bmatrix}^\top P_\mathrm{f} \begin{bmatrix}
            \| \delta \bar{c}_{k+1} \| \\
            \| \delta \bar{h}_{k+1} \|
        \end{bmatrix}
    \end{split}
\end{equation*}
Therefore the variation of the Lyapunov function is bounded by
\begin{equation*}
    \begin{split}
        & V_{k+1} - V_k \leq \\
        & \sum_{i=0}^{N-2} \left( \| \tilde{x}_{i|k+1} - \bar{x}_{k+1} \|_Q^2  + \| u^*_{i+1|k} - \bar{u}_{k+1} \|_R^2 \right) \\
        & + \| \tilde{x}_{N-1|k} - \bar{x}_{k+1} \|_Q^2 + \norm{\begin{bmatrix}
            \| \tilde{c}_{N|k+1} - \bar{c}_{k+1} \| \\
            \| \tilde{h}_{N|k+1} - \bar{h}_{k+1} \|
        \end{bmatrix}}_{P_f}^2 \\
        & + N \| \delta \bar{x}_{k+1} \|_Q^2 + N \| \delta \bar{u}_{k+1} \|_R^2 \\
        & + \sum_{i=0}^{N-1} \bigg( 2 (x^*_{i|k+1} - \bar{x}_{k+1})^\top Q \delta \bar{x}_{k+1} \\
        & + 2 (u^*_{i|k+1} - \bar{u}_{k+1})^\top R \delta \bar{u}_{k+1} \bigg) + \norm{\begin{bmatrix}
            \| \delta \bar{c}_{k+1} \| \\
            \| \delta \bar{h}_{k+1} \|
        \end{bmatrix}}_{P_\mathrm{f}}^2 \\
        & + 2 \begin{bmatrix}
            \| c^*_{N|k+1} - \bar{c}_{k+1} \| \\
            \| h^*_{N|k+1} - \bar{h}_{k+1} \|
        \end{bmatrix}^\top P_\mathrm{f} \begin{bmatrix}
            \| \delta \bar{c}_{k+1} \| \\
            \| \delta \bar{h}_{k+1} \|
        \end{bmatrix} \\
        & - \sum_{i=0}^{N-1} \left( \|x^*_{i|k} - \bar{x}_{\infty} \|_Q^2 + \|u^*_{i|k} - \bar{u}_{\infty} \|_R^2 \right) \\
        & - \norm{\begin{bmatrix}
            \| c^*_{N|k} - \bar{c}_{\infty} \| \\
            \| h^*_{N|k} - \bar{h}_{\infty} \|
        \end{bmatrix}}_{P_f}^2 \\
    \end{split}
\end{equation*}
\begin{equation*}
    \begin{split}
        & = - \| x^*_{0|k} - \bar{x}_{\infty} \|_Q^2 - \| u^*_{0|k} - \bar{u}_{\infty} \|_R^2 \\
        & + \sum_{i=1}^{N-1} \left( \| \tilde{x}_{i-1|k+1} - \bar{x}_{k+1} \|_Q^2 - \| x^*_{i|k} - \bar{x}_{\infty} \|_Q^2 \right) \\
        & + \sum_{i=1}^{N-1} \left( \| u^*_{i|k} - \bar{u}_{k+1} \|_R^2 - \| u^*_{i|k} - \bar{u}_{\infty} \|_R^2 \right) \\
        & + \| \tilde{x}_{N-1|k+1} - \bar{x}_{k+1} \|_Q^2 \\
        & - \norm{\begin{bmatrix}
            \| c^*_{N|k} - \bar{c}_{\infty} \| \\
            \| h^*_{N|k} - \bar{h}_{\infty} \|
        \end{bmatrix}}_{P_\mathrm{f}}^2 + \norm{\begin{bmatrix}
            \| \tilde{c}_{N|k+1} - \bar{c}_{k+1} \| \\
            \| \tilde{h}_{N|k+1} - \bar{h}_{k+1} \|
        \end{bmatrix}}_{P_\mathrm{f}}^2 \\
        & +\sum_{i=0}^{N-1} \Big( 2 (x^*_{i|k+1} - \bar{x}_{k+1})^\top Q \delta \bar{x}_{k+1} \\
        & + 2 (u^*_{i|k+1} - \bar{u}_{k+1})^\top R \delta \bar{u}_{k+1} \Big) \\
        & + 2 \begin{bmatrix}
            \| c^*_{N|k+1} - \bar{c}_{k+1} \| \\
            \| h^*_{N|k+1} - \bar{h}_{k+1} \|
        \end{bmatrix}^\top P_\mathrm{f} \begin{bmatrix}
            \| \delta \bar{c}_{k+1} \| \\
            \| \delta \bar{h}_{k+1} \|
        \end{bmatrix} \\
        & + N \| \delta \bar{x}_{k+1} \|_Q^2 + N \| \delta \bar{u}_{k+1} \|_R^2 + \norm{\begin{bmatrix}
            \| \delta \bar{c}_{k+1} \| \\
            \| \delta \bar{h}_{k+1} \|
        \end{bmatrix}}_{P_\mathrm{f}}^2
    \end{split}
\end{equation*}

It is possible to introduce the error terms related to the fact that $\Tilde{x}_{0|k+1} = \hat{x}_{k+1} \neq x^*_{1|k}$ because of the presence of the observer in the control loop:
\begin{equation*}
    \varepsilon_{k+i+1} = \tilde{x}_{i|k+1} - x^*_{i+1|k}
\end{equation*}
\begin{equation*}
    \varepsilon_{c, k+N} = \tilde{c}_{N-1|k+1} - c^*_{N|k}
\end{equation*}
\begin{equation*}
    \varepsilon_{h, k+N} = \tilde{h}_{N-1|k+1} - h^*_{N|k}
\end{equation*}  
The different terms of the upper bound of $V_{k+1} - V_k$ are now considered separately.

Consider the state terms at time-steps between $k+1$ and $k+N-1$:
\begin{equation*}
    \begin{split}
        &\sum_{i=1}^{N-1} \left( \| \tilde{x}_{i-1|k+1} - \bar{x}_{k+1} \|_Q^2 - \| x^*_{i|k} - \bar{x}_{\infty} \|_Q^2 \right) \\
        & = \sum_{i=1}^{N-1} \left( \| x^*_{i|k} - \bar{x}_{\infty} + \varepsilon_{k+i} - \delta \bar{x}_{k+1} \|_Q^2  - \| x^*_{i|k} - \bar{x}_{\infty} \|_Q^2 \right) \\
        % & = \sum_{i=1}^{N-1} \left( \| x^*_{i|k} - \bar{x}_{\infty} \|_Q^2 - \| x^*_{i|k} - \bar{x}_{\infty} \|_Q^2 \right. \\
        % & + \left. 2(x^*_{i|k} - \bar{x}_{\infty})^\top Q (\varepsilon_{k+i} - \delta \bar{x}_{k+1}) + \| \varepsilon_{k+i} - \delta \bar{x}_{k+1} \|_Q^2 \right) \\
        & = \sum_{i=1}^{N-1} \Big( 2(x^*_{i|k} - \bar{x}_{\infty})^\top Q (\varepsilon_{k+i} - \delta \bar{x}_{k+1}) \\
        & + \| \varepsilon_{k+i} - \delta \bar{x}_{k+1} \|_Q^2\Big)
    \end{split}
\end{equation*}
Consider the input terms at time-steps between $k+1$ and $k+N-1$:
\begin{equation*}
    \begin{split}
        & \sum_{i=1}^{N-1} \left( \| u^*_{i|k} - \bar{u}_{k+1} \|_R^2 - \| u^*_{i|k} - \bar{u}_{\infty} \|_R^2 \right) \\
        & = \sum_{i=1}^{N-1} \left( \| u^*_{i|k} - \bar{u}_{\infty} - \delta \bar{u}_{k+1} \|_R^2 - \| u^*_{i|k} - \bar{u}_{\infty} \|_R^2 \right) \\
        % & = \sum_{i=1}^{N-1} \left( \| u^*_{i|k} - \bar{u}_{\infty} \|_R^2 - \| u^*_{i|k} - \bar{u}_{\infty} \|_R^2 \right. \\
        % & - \left. 2 (u^*_{i|k} - \bar{u}_{\infty})^\top R \delta \bar{u}_{k+1} + \| \delta \bar{u}_{k+1} \|_R^2 \right) \\
        & = \sum_{i=1}^{N-1} \left(- 2 (u^*_{i|k} - \bar{u}_{\infty})^\top R \delta \bar{u}_{k+1} \right) + (N-1)\| \delta \bar{u}_{k+1} \|_R^2
    \end{split}
\end{equation*}

Finally, consider the state terms at time-steps $k+N$ and $k+N+1$.
In view of Lemma \ref{lem:norm_inequality} one has that, for any $\varphi \neq 0$,
\begin{equation*}
    \begin{split}
        & \| \tilde{x}_{N-1|k+1} - \bar{x}_{k+1} \|_Q^2 \\
        %& =  \| \tilde{x}_{N-1|k+1} - x^*_{N|k} + x^*_{N|k} - \bar{x}_{k+1} + \bar{x}_{\infty} - \bar{x}_{\infty}\|_Q^2 \\
        & = \| x^*_{N|k} - \bar{x}_{\infty} + \varepsilon_{k+N} - \delta \bar{x}_{k+1} \|_Q^2 \\
        & \stackrel{\eqref{eq:norm_inequality}}{\leq} (1 + \varphi^2) \|x^*_{N|k} - \bar{x}_{\infty}\|_Q^2 \\
        & + \left(1 + \frac{1}{\varphi^2}\right) \|\varepsilon_{k+N} - \delta \bar{x}_{k+1} \|_Q^2 \\
        & \leq q(1 + \varphi^2) \|x^*_{N|k} - \bar{x}_{\infty}\|^2 \\
        & + \left(1 + \frac{1}{\varphi^2}\right) \|\varepsilon_{k+N} - \delta \bar{x}_{k+1} \|_Q^2 \\
        & = q(1 + \varphi^2) \norm{\begin{bmatrix}
        \| c^*_{N|k} - \bar{c}_{\infty} \| \\
        \| h^*_{N|k} - \bar{h}_{\infty} \|
    \end{bmatrix}}^2 \\
    & + \left(1 + \frac{1}{\varphi^2}\right) \|\varepsilon_{k+N} - \delta \bar{x}_{k+1} \|_Q^2 
    \end{split}
\end{equation*}
Moreover, for any $\varphi \neq 0$,
\begin{equation*}
\begin{split}
    & \norm{\begin{bmatrix}
        \| \tilde{c}_{N|k+1} - \bar{c}_{k+1} \| \\
        \| \tilde{h}_{N|k+1} - \bar{h}_{k+1} \|
    \end{bmatrix}}_{P_\mathrm{f}}^2 \\
    & \stackrel{\eqref{eq:bound_lstm_trajectory}}{\leq} \norm{ A_\delta \begin{bmatrix}
        \| \tilde{c}_{N-1|k+1} - \bar{c}_{k+1} \| \\
        \| \tilde{h}_{N-1|k+1} - \bar{h}_{k+1} \|
    \end{bmatrix}}_{P_\mathrm{f}}^2 \\
    & \leq \norm{\begin{bmatrix}
        \| c^*_{N|k} - \bar{c}_{\infty} + \varepsilon_{c,k+N} - \delta \Bar{c}_{k+1} \| \\
        \| h^*_{N|k} - \bar{h}_{\infty} + \varepsilon_{h,k+N} - \delta \Bar{h}_{k+1} \|
    \end{bmatrix}}^2_{A_\delta^\top P_\mathrm{f} A_\delta} \\
    &\stackrel{\eqref{eq:norm_inequality}}{\leq} (1 + \varphi^2) \norm{\begin{bmatrix}
        \| c^*_{N|k} - \bar{c}_{\infty} \| \\
        \| h^*_{N|k} - \bar{h}_{\infty} \|
    \end{bmatrix}}^2_{A_\delta^\top P_\mathrm{f} A_\delta} \\
    & + \left(1 + \frac{1}{\varphi^2}\right) \norm{\begin{bmatrix}
        \| \varepsilon_{c,k+N} - \delta \Bar{c}_{k+1} \| \\
        \| \varepsilon_{h,k+N} - \delta \Bar{h}_{k+1} \|
    \end{bmatrix}}^2_{A_\delta^\top P_\mathrm{f} A_\delta}
\end{split}
\end{equation*}

Then the following inequality holds:
\begin{equation*}
    \begin{split}
        & - \norm{\begin{bmatrix}
            \| c^*_{N|k} - \bar{c}_{\infty} \| \\
            \| h^*_{N|k} - \bar{h}_{\infty} \|
        \end{bmatrix}}_{P_\mathrm{f}}^2 + \| \tilde{x}_{N-1|k+1} - \bar{x}_{k+1} \|_Q^2 \\
        & + \norm{\begin{bmatrix}
            \| \tilde{c}_{N|k+1} - \bar{c}_{k+1} \| \\
            \| \tilde{h}_{N|k+1} - \bar{h}_{k+1} \|
        \end{bmatrix}}_{P_\mathrm{f}}^2 \\
        & \leq \norm{\begin{bmatrix}
            \| c^*_{N|k} - \bar{c}_{\infty} \| \\
            \| h^*_{N|k} - \bar{h}_{\infty} \|
        \end{bmatrix}}^2_{-P_\mathrm{f} + (1 + \varphi^2)q I_2 + (1 + \varphi^2) A_{\delta}^\top P_\mathrm{f} A_{\delta}} \\
        & + \left(1 + \frac{1}{\varphi^2}\right) \|\varepsilon_{k+N} - \delta \bar{x}_{k+1} \|_Q^2 \\
        & + \left(1 + \frac{1}{\varphi^2}\right) \norm{\begin{bmatrix}
            \| \varepsilon_{c,k+N} - \delta \Bar{c}_{k+1} \| \\
            \| \varepsilon_{h,k+N} - \delta \Bar{h}_{k+1} \|
        \end{bmatrix}}^2_{A_\delta^\top P_\mathrm{f} A_\delta}
    \end{split}
\end{equation*}
By construction $P_\mathrm{f}$ is selected such that $A_{\delta}^\top P_\mathrm{f} A_{\delta} - P_\mathrm{f} < -q I_2$. Therefore there exists a value of $\varphi$ small enough to have $-P_\mathrm{f} + (1 + \varphi^2)q I_2 + (1 + \varphi^2) A_{\delta}^\top P_\mathrm{f} A_{\delta} < 0$.

Combining all the computations, one obtains that
\begin{equation} \label{eq:deltaV-gamma}
    \begin{split}
        & V_{k+1} - V_k \leq - \| x^*_{0|k} - \bar{x}_{\infty} \|_Q^2 - \| u^*_{0|k} - \bar{u}_{\infty} \|_R^2 + \gamma(k)
    \end{split}
\end{equation}
where
\begin{equation*}
    \begin{split}
        & \gamma (k) = \sum_{i=1}^{N-1} \Big( 2 (x^*_{i|k} - \bar{x}_{\infty})^\top Q (\varepsilon_{k+i} - \delta \bar{x}_{k+1}) \\
        & + \| \varepsilon_{k+i} - \delta \bar{x}_{k+1} \|_Q^2 \Big) \\
        & + \sum_{i=1}^{N-1} \left( - 2 (u^*_{i|k} - \bar{u}_{\infty})^\top R \delta \bar{u}_{k+1} \right) + (N-1)\| \delta \bar{u}_{k+1} \|_R^2 \\
        & + \left(1 + \frac{1}{\varphi^2}\right) \|\varepsilon_{k+N} - \delta \bar{x}_{k+1} \|_Q^2  \\
        & + \left(1 + \frac{1}{\varphi^2}\right) \norm{\begin{bmatrix}
            \| \varepsilon_{c,k+N} - \delta \Bar{c}_{k+1} \| \\
            \| \varepsilon_{h,k+N} - \delta \Bar{h}_{k+1} \|
        \end{bmatrix}}^2_{A_\delta^\top P_\mathrm{f} A_\delta} \\
        & + \sum_{i=0}^{N-1} \left( 2 (x^*_{i|k+1} - \bar{x}_{k+1})^\top Q \delta \bar{x}_{k+1} \right. \\
        & + \left. 2 (u^*_{i|k+1} - \bar{u}_{k+1})^\top R \delta \bar{u}_{k+1} \right) \\
        & + 2 \begin{bmatrix}
            \| c^*_{N|k+1} - \bar{c}_{k+1} \| \\
            \| h^*_{N|k+1} - \bar{h}_{k+1} \|
        \end{bmatrix}^\top P_\mathrm{f} \begin{bmatrix}
            \| \delta \bar{c}_{k+1} \| \\
            \| \delta \bar{h}_{k+1} \|
        \end{bmatrix} \\
        & + \norm{\begin{bmatrix}
            \| \delta \bar{c}_{k+1} \| \\
            \| \delta \bar{h}_{k+1} \|
        \end{bmatrix}}_{P_\mathrm{f}}^2  + N \| \delta \bar{u}_{k+1} \|_R^2 + N \| \delta \bar{x}_{k+1} \|_Q^2
    \end{split}
\end{equation*}
In this expression: 
\begin{itemize}
    \item Terms $\varepsilon_{k+i}$, $\varepsilon_{c, k+N}$, $\varepsilon_{h, k+N}$ are present because $\hat{x}_{k+1} \neq x^*_{1|k}$. These terms have bounds related to the observer estimation error $\chi_k - \hat{\chi}_k$. In particular
    \begin{equation*}
        \begin{split}
            & \norm{\varepsilon_{k+1}} \stackrel{\eqref{eq:lyap_observer_constraint}}{\leq} L_{max} V_\mathrm{o} (\hat{\chi}_k, \chi_k) \stackrel{\eqref{eq:lyap_observer_pos_def}}{\leq} c_{\mathrm{o,u}} L_{max} \norm{\chi_k - \hat{\chi}_k}
        \end{split}
    \end{equation*}
    \begin{equation*}
        \begin{split}
            & \norm{\varepsilon_{k+i}} \stackrel{\eqref{eq:incremental_lyap_pos_def}}{\leq} \frac{1}{c_{\mathrm{s,l}}} V_\mathrm{s}(\Tilde{x}_{i-1|k+1}, x^*_{i|k}) \\
            & \stackrel{\eqref{eq:incremental_lyap_neg_def}}{\leq} \frac{\rho_\mathrm{s}^{i-1}}{c_\mathrm{s,l}} V_\mathrm{s}(\Tilde{x}_{0|k+1}, x^*_{1|k}) \stackrel{\eqref{eq:incremental_lyap_pos_def}}{\leq} \frac{c_{\mathrm{s,u}} \rho_\mathrm{s}^{i-1}}{c_{\mathrm{s,l}}} \norm{\varepsilon_{k+1}} \\
            & \leq \frac{c_{\mathrm{s,u}} \rho_\mathrm{s}^{i-1}}{c_{\mathrm{s,l}}} c_{\mathrm{o,u}} L_{max} \norm{\chi_k - \hat{\chi}_k}
        \end{split}
    \end{equation*}
    \item Terms $\delta \bar{x}_{k+1}$, $\delta \bar{u}_{k+1}$, $\delta \bar{c}_{k+1}$, $\delta \bar{h}_{k+1}$ are present because reference values $\bar{x}$ and $\bar{u}$ change at each time-step with the variation of the disturbance estimation $\hat{d}_k$. As already noted, these terms are bounded.
    \item Inputs are limited thanks to MPC constraint \eqref{eq:optimization_input_constraint}.
    \item States are limited in the set $\mathcal{X}^{MPC}$.
\end{itemize}

Moreover, since
\begin{equation*}
    \| x^*_{0|k} - \bar{x}_{\infty} \|_Q^2 + \| u^*_{0|k} - \bar{u}_{\infty} \|_R^2 \geq \lambda_{min} (Q) \| \hat{x}_k - \bar{x}_{\infty} \|^2
\end{equation*}
there exists constants $c = \lambda_{min} (Q)$ and $c_2 \geq 0$ and a $\mathcal{K}$-function $\tilde{\gamma}$ such that
\begin{equation}   \label{eq:bound-deltaV} 
    V_{k+1} - V_k \leq - c \| \hat{x}_k - \bar{x}_{\infty} \|^2 + \tilde{\gamma}(\norm{\chi_k - \hat{\chi}_k}) + c_2
\end{equation}

In view of \eqref{eq:lower_bound_Vk}-\eqref{eq:upper-bound-Vk}-\eqref{eq:bound-deltaV}, $V_k$ is an ISpS-Lyapunov function \cite{raimondo2009minmax}. Then the closed-loop system \eqref{eq:augmented_model}-\eqref{eq:reference_calculation}-\eqref{eq:mpc} is ISpS with respect to the observer estimation error $\chi - \hat{\chi}$.

\textbf{Part 3b:} To prove convergence, first note that
if $d_k \to \Bar{d}_\infty$ for $k \to \infty$, the observer estimation error $\chi - \hat{\chi}$ converges to 0 in view of Theorem \ref{th:observer}. Then, in view of \eqref{eq:bound-delta}, terms $\delta \bar{x}_k$, $\delta \bar{u}_k$, $\delta \bar{c}_k$ and $\delta \bar{h}_k$ converge to 0 for $k \to \infty$.

The bounds on $V_k$ and on $V_{k+1} - V_{k}$ are now studied for $k \to \infty$.

Consider the upper bound of $V_k$. If $d_k \to \Bar{d}_\infty$ and $y^0_k \to y^0_\infty$ for $k \to \infty$, in view of observer convergence, $\norm{\Bar{x}_k - \Bar{x}_\infty} \to 0$ for $k \to \infty$. Then there exist $\Bar{k} \in \mathbb{Z}_{\geq 0}$ such that $\norm{\Bar{x}_k - \Bar{x}_\infty} \leq \frac{\mu}{2}$ for all $k \geq \Bar{k}$.
Hence, for $k \geq \Bar{k}$, only cases 1 and 2 of the upper bound can appear. Then
\begin{equation}
    V_k \leq b \norm{\hat{x}_k - \bar{x}_\infty}^2 + \beta(k)
\end{equation}
where $\beta(k) \to 0$ for $k \to \infty$ for the convergence of $\delta \bar{x}_k$, $\delta \bar{u}_k$, $\delta \bar{c}_k$ and $\delta \bar{h}_k$.

Consider the bound on $V_{k+1} - V_{k}$ of Equation \eqref{eq:deltaV-gamma}. $\gamma(k) \to 0$ for $k \to \infty$ in view of the convergence of the observer estimation error and of $\delta \bar{x}_k$, $\delta \bar{u}_k$, $\delta \bar{c}_k$ and $\delta \bar{h}_k$.

Then for $k \geq \bar{k}$ the Lyapunov function $V_k$ is such that 
\begin{subequations} \label{eq:convergence-lyapunov}
    \begin{align}
        & a \norm{\hat{x}_k - \bar{x}_\infty}^2 \leq V_k \leq b \norm{\hat{x}_k - \bar{x}_\infty}^2 + \beta(k) \\
        & V_{k+1} - V_k \leq -c \norm{\hat{x}_k - \bar{x}_\infty}^2 + \gamma(k)
    \end{align}
\end{subequations}
for some $a, b, c > 0$, with $\beta(k) \to 0$ and $\gamma(k) \to 0$ for $k \to \infty$.

The fact that $\lim_{k \to \infty} \norm{\psi_k - \psi_\infty} = 0$ follows from \eqref{eq:convergence-lyapunov}, observer convergence proven in Theorem \ref{th:observer} and $\eqref{eq:eo-limit}$.

\subsection{Proof of Theorem \ref{th:offset-free}} \label{sec:proof-offset-free}

Following \cite{morari2012nonlinear_offset_free}, the proof can be derived as follows.

In view of Assumption \ref{ass:converging_plant} the output of the plant converges to a constant value $y_{\phi,\infty}$. Then it is sufficient to prove that $y_{\phi,\infty} = y^0_{\phi,\infty}$. 

In view of Assumption \ref{ass:converging_plant}, the observer states $\hat{x}$ and $\hat{d}$ converge to asymptotic values, denoted by $\hat{x}_\infty$ and $\hat{d}_\infty$.
By Corollary \ref{cor:observability} and Theorem \ref{th:observer}, the observer is nominally error free (i.e. it reaches steady state only if $y = \hat{y}$), and at steady state satisfies
\begin{subequations}
    \begin{align}
        \hat{x}_\infty &= f(\hat{x}_\infty, u_\infty)  \label{eq:observer-equilibrium-state} \\
        y_{\phi,\infty} &= g(\hat{x}_\infty) + \hat{d}_\infty \label{eq:observer-equilibrium-output}
    \end{align}
\end{subequations}
where $u_\infty$ is the input generated by the controller (Reference calculator + MPC), and is a function of $\hat{x}_\infty, \hat{d}_\infty, y^0_{\phi,\infty}$. From \eqref{eq:observer-equilibrium-state}, Assumption \ref{ass:setpoint_and_jacobian} and Theorem \ref{th:feasibility_stability} it follows that
\begin{equation} \label{eq:control-equilibrium}
    y^0_{\phi,\infty} = g(\hat{x}_\infty) + \hat{d}_\infty
\end{equation}
Hence, from a comparison of \eqref{eq:observer-equilibrium-output} and \eqref{eq:control-equilibrium}, it follows that $y_{\phi,\infty} = y^0_{\phi,\infty}$.
%\hfill $\square$

\printbibliography

\end{document}